\documentclass[%
superscriptaddress,
nofootinbib,
twocolumn,
amsmath,amssymb,
aps,
prd,
floatfix,
a4paper
]{revtex4-2}

\usepackage[colorlinks=true,linkcolor=blue,citecolor=blue,linktocpage=false]{hyperref}

\usepackage{newtxtext,newtxmath}
\usepackage{graphicx}
\usepackage{dcolumn}
\usepackage{bm}
\usepackage{enumitem}
\usepackage{placeins}
\usepackage{multirow}

\begin{document}

\preprint{}

\title{Renormalization-Free Galaxy Bias in Unified Lagrangian Perturbation Theory}

\author{Naonori Sugiyama}
\email{nao.s.sugiyama@gmail.com}
\affiliation{Independent Researcher, Tokyo, Japan}
\affiliation{National Astronomical Observatory of Japan, Mitaka, Tokyo 181-8588, Japan}
\thanks{Special Visiting Researcher (non-salaried)}

\date{\today}

\begin{abstract}
    We present a renormalization-free framework for modeling galaxy bias based on Unified Lagrangian Perturbation Theory (ULPT). In this approach, the galaxy density field is constructed entirely from Galileon-type operators, which also characterize the intrinsic nonlinear evolution of dark matter. This formulation ensures that the bias expansion is well defined at the field level, automatically satisfies the statistical conditions of vanishing ensemble and volume averages, and eliminates the need for any ad hoc renormalization procedures. We derive analytic expressions for the one-loop galaxy--galaxy and galaxy--matter power spectra and implement an efficient numerical algorithm using \texttt{FFTLog} and \texttt{FAST-PT}, enabling rapid and accurate evaluation of the full power spectrum. The resulting model requires only a minimal set of bias parameters, comprising three for correlation functions and four for power spectra. To assess its predictive accuracy, we perform joint fits to the halo--halo auto and halo--matter cross power spectra obtained from the \textit{Dark Emulator}, considering nine combinations of redshift and halo mass, with 100 cosmological models sampled for each combination. We find that a single set of bias parameters successfully and simultaneously reproduces both spectra with better than $\sim1\%$ accuracy up to $k \simeq 0.3\, h\,\mathrm{Mpc}^{-1}$ for typical linear bias values in the range $b_1 \sim 0.8$ to 2. For more strongly biased tracers with $b_1 \sim 3$, the agreement remains within $\sim1\%$ up to $k \simeq 0.2\, h\,\mathrm{Mpc}^{-1}$. We further confirm that the same bias parameters consistently describe the two-point correlation functions in configuration space down to $r \simeq 15\,h^{-1}\mathrm{Mpc}$ with comparable accuracy. Moreover, ULPT predicts the theoretical relation $b_{K^2}^{\mathrm{E}} = -\tfrac{3}{4} b_2^{\mathrm{E}}$ between second-order Eulerian local and tidal bias parameters, which is validated through comparison with empirical fitting formulas calibrated on $N$-body simulations. These findings demonstrate that the ULPT framework offers a physically interpretable, statistically consistent, and computationally efficient model for nonlinear galaxy bias, with promising applicability to other observables such as redshift-space distortions, bispectra, and density-field reconstruction. The numerical implementation developed in this work is publicly released as the open-source Python package \texttt{ulptkit} (\url{https://github.com/naonori/ulptkit}).
\end{abstract}

\maketitle

\section{Introduction}
\label{sec:intro}

The spatial clustering of galaxies encodes a wealth of cosmological information, offering critical insights into the physics of the early Universe, the nature of dark energy, and the total mass of neutrinos. However, accurately modeling galaxy clustering remains one of the central challenges in large-scale structure (LSS) analysis. This difficulty arises from the need to consistently incorporate several nonlinear effects that distort the observed distribution of galaxies. These include nonlinear gravitational evolution, redshift-space distortions (RSD)~\cite{Kaiser:1987qv}, and the impact of density-field reconstruction~\cite{Eisenstein:2006nk}. Furthermore, galaxies do not trace the underlying matter distribution directly but rather act as biased tracers, introducing additional complexity in the form of galaxy bias (for a review, see~\cite{Desjacques:2016bnm}). Each of these effects contributes in a distinct but interrelated manner, highlighting the need for a unified theoretical framework that can accurately capture their combined impact.

In pursuit of this goal, we recently proposed the \textit{Unified Lagrangian Perturbation Theory} (ULPT)~\cite{Sugiyama:inprep} as a systematic framework that consistently incorporates these nonlinear effects. ULPT reorganizes standard Lagrangian perturbation theory by explicitly decomposing the observed density field into two physically distinct components: the \textit{Jacobian deviation}, which captures the intrinsic nonlinear growth of matter fluctuations, and the \textit{displacement-mapping component}, which describes convective distortions induced by large-scale coherent flows. This decomposition provides a natural basis for implementing infrared (IR) safety, ensuring exact cancellation of long-wavelength contributions~\cite{Jain:1995kx,Scoccimarro:1995if,Kehagias:2013yd,Peloso:2013zw,Sugiyama:2013pwa,Sugiyama:2013gza,Blas:2013bpa,Blas:2015qsi,Lewandowski:2017kes} and enabling a consistent description of the nonlinear damping of baryon acoustic oscillations (BAO)~\cite{Sunyaev:1970eu,Peebles:1970ag,Sugiyama:2013gza,Senatore:2014via,Baldauf:2015xfa,Blas:2016sfa,Senatore:2017pbn,Ivanov:2018gjr,Lewandowski:2018ywf,Sugiyama:2020uil,Sugiyama:2024eye}. Importantly, this structure applies uniformly to pre- and post-reconstruction density fields, as well as to real and redshift space.

As a first step toward numerical applications of ULPT, we focus on the simplest case: dark matter clustering in real space before reconstruction. In this setting, we have developed a fast numerical algorithm to compute the one-loop matter power spectrum using FFT-based techniques~\cite{Sugiyama:2025myq}. The resulting ULPT predictions quantitatively match simulation-based emulators such as \textit{Dark Emulator}~\cite{Nishimichi:2018etk} and \textit{Euclid Emulator 2}~\cite{Euclid:2020rfv} with 2--3\% accuracy up to $k \simeq 0.4\, h\,\mathrm{Mpc}^{-1}$ for redshifts $z \gtrsim 0.5$. These results are achieved without introducing any free parameters. Each evaluation typically requires $1$--$2$ seconds per cosmological model. These results demonstrate that ULPT provides a computationally efficient and theoretically robust framework for modeling nonlinear matter clustering across a wide range of scales and redshifts.

In this paper, we extend the ULPT framework to describe biased tracers such as galaxies and dark matter halos, with the goal of constructing a renormalization-free model of nonlinear bias. The conventional approach to bias modeling, which relates the density fluctuations of galaxies or halos $\delta_g$ to the underlying matter density fluctuations $\delta_{\rm m}$, begins with the linear relation $\delta_g = b_1 \delta_{\rm m}$, originally proposed by Ref.~\cite{Kaiser:1984sw}. Higher-order corrections were subsequently introduced by Ref.~\cite{Fry:1992vr}, which formulated a local Taylor expansion of the galaxy density field in powers of the matter density contrast. This local bias model was later generalized to include nonlocal contributions, most notably from the tidal tensor $K_{ij}$, which encodes the effect of gravitational shear~\cite{McDonald:2009dh}.

Despite their success in capturing key features of galaxy clustering, bias models based on independent local and nonlocal operators, such as $\delta_{\rm m}^2$, $\delta_{\rm m}^3$, and $K_{ij}K^{ij}$, often exhibit unphysical behavior at the field level. Specifically, the modeled galaxy density field may fail to satisfy the requirement of vanishing ensemble and volume averages, leading to spurious constant offsets in the large-scale power spectrum. To correct for such artifacts, it is common to absorb the constant contribution to the power spectrum, which arises from stochastic noise terms~\cite{Dekel:1998eq}, into the $k = 0$ mode. These procedures are typically interpreted as a form of bias renormalization~\cite{McDonald:2006mx}.

An alternative but equivalent approach to expressing nonlocal bias terms, which are typically written in terms of the tidal tensor, is to employ Galileon-type operators~\cite{Chan:2012jj}, defined as nonlocal scalar invariants constructed from the gravitational or velocity potential. For example, the second-order Galileon operator is given by ${\cal G}_2 = -\tfrac{2}{3} \delta_{\rm m}^2 + K_{ij}K_{ij}$. These Galileon operators, by construction, individually satisfy the desired statistical properties of density fluctuations and have been shown to remain free from renormalization~\cite{Assassi:2014fva}. Consequently, contributions that require bias renormalization arise solely from terms that include local contributions from the matter density contrast, such as $\delta_{\rm m}^2$, $\delta_{\rm m}^3$, or $\delta_{\rm m} {\cal G}_2$.

Crucially, Ref.~\cite{Sugiyama:inprep} demonstrated that, within the ULPT framework, the intrinsic density fluctuations arising from the Jacobian deviation consist solely of Galileon-type operators. Motivated by this structure, we construct a bias model in which the nonlinear bias field is entirely described by a linear combination of the same Galileon operators that generate the dark matter fluctuations. The resulting expansion is well defined at the field level and does not require renormalization. This renormalization-free structure implies that the bias parameters retain their physical meaning and can be directly constrained from observations or simulations without additional regularization procedures. Moreover, the structure of the bias expansion mirrors that of the dark matter field, enabling the same computational pipeline to be applied with minimal modification. Specifically, the one-loop power spectrum for biased tracers can be evaluated by simply replacing the dark matter kernels with their bias-modified counterparts, without the need for any special operator-level manipulations or counterterm insertions.

To validate this approach, we evaluate the predictive accuracy of the ULPT bias model by comparing its one-loop predictions with the results from the \emph{Dark Emulator}, which provides simulation-calibrated halo--halo ($P_{\mathrm{hh}}$) and halo--matter ($P_{\mathrm{hm}}$) power spectra. Specifically, we compute these spectra across multiple halo mass bins and redshifts, using both a fiducial Planck 2015 cosmology and an extensive set of 100 cosmological models sampled from the emulator's comprehensive six-dimensional parameter space. We show that a single set of bias parameters simultaneously fits both $P_{\mathrm{hh}}$ and $P_{\mathrm{hm}}$ with accuracy at the 1\% level up to $k \simeq 0.3\, h\,\mathrm{Mpc}^{-1}$. Furthermore, we demonstrate that the same model accurately reproduces the corresponding two-point correlation functions ($\xi_{\mathrm{hh}}$ and $\xi_{\mathrm{hm}}$) down to $r \simeq 15\, h^{-1}\mathrm{Mpc}$.

Throughout this work, we adopt the fiducial cosmology implemented in the \textit{Dark Emulator} suite, which is consistent with the Planck 2015 best-fit $\Lambda$CDM model~\cite{Planck:2015fie}. The cosmological parameters are specified as follows: physical baryon density $\omega_b \equiv \Omega_b h^2 = 0.02225$, physical cold dark matter density $\omega_c \equiv \Omega_c h^2 = 0.1198$, dark energy density $\Omega_{\mathrm{de}} = 0.6844$, scalar spectral index $n_s = 0.9645$, amplitude of primordial curvature perturbations $\ln(10^{10} A_s) = 3.094$, dark energy equation-of-state parameter $w_0 = -1$, and total neutrino mass $\sum m_\nu = 0.06\,\mathrm{eV}$. The Hubble parameter is then determined to be $h = 0.6727$ from the flatness condition.

This paper is organized as follows. In Secs.~\ref{sec:ULPT_biased} and~\ref{sec:bias}, we present the ULPT formulation for biased tracers, introducing the structural decomposition of the density field and the construction of the renormalization-free bias model. Sec.~\ref{sec:power} provides a detailed derivation of the one-loop galaxy--galaxy and galaxy--matter power spectra within this framework, including an efficient numerical implementation using \texttt{FAST-PT}. In Sec.~\ref{sec:DarkEmu}, we describe the \textit{Dark Emulator}, which provides the simulation-calibrated reference spectra used for validation throughout this work. In Sec.~\ref{sec:vs_emu}, we validate the model against the \textit{Dark Emulator} outputs by performing joint fits to $P_{\mathrm{hh}}$ and $P_{\mathrm{hm}}$ across multiple redshifts and halo masses, and further confirm consistency in configuration space via the two-point correlation functions $\xi_{\mathrm{hh}}$ and $\xi_{\mathrm{hm}}$. Sec.~\ref{sec:compare_empirical} compares the bias parameters obtained from ULPT with empirical fitting formulas derived from $N$-body simulations. We conclude in Sec.~\ref{sec:conclusion} with a summary of our findings and a discussion of future prospects, including applications to redshift-space distortions, reconstruction, and extensions beyond the range of current emulators.

\section{Unified Lagrangian Framework for Biased Tracers}
\label{sec:ULPT_biased}

This section provides a concise overview of the ULPT formulation for biased tracers, as originally developed in Ref.~\cite{Sugiyama:inprep}. Section~\ref{sec:ULPT_biased_density} introduces the core equations that define the notation and physical context. Section~\ref{sec:ULPT_jacobian} then details the perturbative structure of the Jacobian deviation, which serves as the intrinsic source field in ULPT. This framework lays the foundation for the renormalization-free treatment of nonlinear bias presented in Sec.~\ref{sec:bias}.

\subsection{Density Contrast of Biased Tracers}
\label{sec:ULPT_biased_density}

We denote the galaxy (or halo) density field in Eulerian coordinates by $\rho_{\rm g}(\boldsymbol{x})$, with its density contrast defined as
\begin{equation}
    \rho_{\rm g}(\boldsymbol{x}) = \bar{\rho}_{\rm g} \left[ 1 + \delta_{\rm g}(\boldsymbol{x}) \right],
\end{equation}
where $\bar{\rho}_{\rm g}$ is the mean number density, and the subscript ``g'' denotes ``galaxy'' (or more generally, a biased tracer).

The mapping between Lagrangian coordinates $\boldsymbol{q}$ and Eulerian coordinates $\boldsymbol{x}$ is given by
\begin{equation}
    \boldsymbol{x} = \boldsymbol{q} + \boldsymbol{\Psi}(\boldsymbol{q}),
\end{equation}
where $\boldsymbol{\Psi}(\boldsymbol{q})$ is the displacement vector field. The Jacobian determinant of this transformation defines the volume element mapping:
\begin{equation}
    J(\boldsymbol{q}) = \det\left( \frac{\partial \boldsymbol{x}}{\partial \boldsymbol{q}} \right), \quad d^3x = J(\boldsymbol{q})\, d^3q.
\end{equation}

Let $\rho_{\rm b}(\boldsymbol{q})$ denote the biased density field defined in Lagrangian coordinates, related to its mean $\bar{\rho}_{\rm b}$ and the biased density contrast $\delta_{\rm b}(\boldsymbol{q})$ via
\begin{equation}
    \rho_{\rm b}(\boldsymbol{q}) = \bar{\rho}_{\rm b} \left[ 1 + \delta_{\rm b}(\boldsymbol{q}) \right].
\end{equation}
Assuming mass conservation under the coordinate transformation, the number of biased tracers is preserved:
\begin{equation}
    \rho_{\rm g}(\boldsymbol{x})\, d^3x = \rho_{\rm b}(\boldsymbol{q})\, d^3q.
\end{equation}
At the background level, the mean number density is independent of spatial coordinates. Therefore, mass conservation implies that the background densities are equal:
\begin{equation}
    \bar{\rho}_{\rm g} = \bar{\rho}_{\rm b}.
\end{equation}
This leads to the following relation between the Eulerian and Lagrangian density contrasts:
\begin{equation}
    \left[ 1 + \delta_{\rm g}(\boldsymbol{x}) \right]\, d^3x = \left[ 1 + \delta_{\rm b}(\boldsymbol{q}) \right]\, d^3q.
\end{equation}
Substituting $d^3x = J(\boldsymbol{q})\, d^3q$ into the above relation yields
\begin{equation}
    \delta_{\rm g}(\boldsymbol{q} + \boldsymbol{\Psi}(\boldsymbol{q})) = \frac{1 + \delta_{\rm b}(\boldsymbol{q})}{J(\boldsymbol{q})} - 1.
    \label{eq:delta_g_Lag}
\end{equation}

While the expression above defines the Eulerian density contrast in terms of Lagrangian variables, actual observations are made in Eulerian space. To express the density contrast explicitly in Eulerian coordinates while retaining the Lagrangian description, we employ the following identity:
\begin{equation}
    \delta_{\rm g}(\boldsymbol{x}) = \int d^3x' \, \delta_{\rm g}(\boldsymbol{x}') \, \delta_{\rm D}(\boldsymbol{x} - \boldsymbol{x}'),
\end{equation}
where $\delta_{\rm D}$ denotes the three-dimensional Dirac delta function, which enforces the coincidence of spatial positions in the integrand. We then change the integration variable using the Lagrangian-to-Eulerian mapping $\boldsymbol{x}' = \boldsymbol{q} + \boldsymbol{\Psi}(\boldsymbol{q})$, and substitute the volume element relation $d^3x' = J(\boldsymbol{q})\, d^3q$, together with the density contrast expression in terms of Lagrangian quantities from Eq.~\eqref{eq:delta_g_Lag}. This yields
\begin{equation}
    \delta_{\rm g}(\boldsymbol{x}) = \int d^3q \left[ \delta_{\rm J}(\boldsymbol{q}) + \delta_{\rm b}(\boldsymbol{q}) \right] \delta_{\rm D}(\boldsymbol{x} - \boldsymbol{q} - \boldsymbol{\Psi}(\boldsymbol{q})),
    \label{eq:delta_g}
\end{equation}
where we define the Jacobian deviation as
\begin{equation}
    \delta_{\rm J}(\boldsymbol{q}) \equiv 1 - J(\boldsymbol{q}).
    \label{eq:delta_J}
\end{equation}
In terms of the displacement field, the Jacobian deviation admits the exact expansion
\begin{align}
    \delta_{\rm J}(\boldsymbol{q})
    ={}& -\,\Psi_{i,i}(\boldsymbol{q})
    \nonumber\\
    & - \frac{1}{2}\!\left[
        \Psi_{i,i}(\boldsymbol{q})\,\Psi_{j,j}(\boldsymbol{q})
        - \Psi_{i,j}(\boldsymbol{q})\,\Psi_{j,i}(\boldsymbol{q})
      \right]
    \nonumber\\
    & - \frac{1}{6}\,\epsilon_{ijk}\,\epsilon_{lmn}\,
        \Psi_{i,l}(\boldsymbol{q})\,
        \Psi_{j,m}(\boldsymbol{q})\,
        \Psi_{k,n}(\boldsymbol{q}),
    \label{eq:dJ}
\end{align}
with indices $i,j,k,l,m,n\in\{x,y,z\}$ and implicit summation over repeated indices. Here $\epsilon_{ijk}$ denotes the Levi-Civita symbol, $\Psi_i$ is the $i$-th component of the displacement vector, and $\Psi_{i,j}\equiv \partial \Psi_i/\partial q_j$. This Jacobian deviation $\delta_{\rm J}$ can be interpreted as capturing the intrinsic density fluctuation that arises from the nonlinear deformation of volume elements under the coordinate transformation from Lagrangian to Eulerian space.

Taking the Fourier transform of Eq.~\eqref{eq:delta_g} yields
\begin{equation}
    \widetilde{\delta}_{\rm g}(\boldsymbol{k}) = \int d^3q\, e^{-i\boldsymbol{k} \cdot \boldsymbol{q}} e^{-i\boldsymbol{k} \cdot \boldsymbol{\Psi}(\boldsymbol{q})} \left[ \delta_{\rm J}(\boldsymbol{q}) + \delta_{\rm b}(\boldsymbol{q}) \right],
    \label{eq:delta_g_F}
\end{equation}
where we denote Fourier-transformed quantities with a tilde.

This expression makes explicit the structural decomposition of the density contrast into two physically distinct components:
\begin{itemize}
    \item The \textbf{Jacobian deviation} $\delta_{\rm J}$, which encodes intrinsic linear and nonlinear growth, including bias contributions via $\delta_{\rm b}$;
    \item The \textbf{displacement-mapping effect} $e^{-i\boldsymbol{k}\cdot\boldsymbol{\Psi}}$, which describes the nonlinear coordinate remapping induced by the Lagrangian-to-Eulerian transformation through the displacement field.
\end{itemize}

By expanding the exponential $e^{-i\boldsymbol{k}\cdot\boldsymbol{\Psi}}$ in Eq.~\eqref{eq:delta_g_F} and performing the inverse Fourier transform, the galaxy density contrast can be expressed entirely in terms of Eulerian coordinates, without explicit reference to the Lagrangian frame:
\begin{align}
    \delta_{\rm g}(\boldsymbol{x})
    &= \hspace{-0.5cm}\underbrace{\delta_{\rm J}(\boldsymbol{x}) + \delta_{\rm b}(\boldsymbol{x})}_{\text{Jacobian deviation with bias}} \nonumber \\
    &+ \underbrace{\sum_{n=1}^{\infty} \frac{(-1)^n}{n!}
    \partial_{i_1} \cdots \partial_{i_n} \left\{
        \Psi_{i_1}(\boldsymbol{x}) \cdots \Psi_{i_n}(\boldsymbol{x})
        \left[ \delta_{\rm J}(\boldsymbol{x}) + \delta_{\rm b}(\boldsymbol{x}) \right]
    \right\}}_{\text{Displacement-mapping effect}},
\label{eq:delta_g_decom}
\end{align}
where $\partial_i = \partial/\partial x_i$, and repeated indices are summed over.

In this formulation, the bias field $\delta_{\rm b}$ enters additively alongside the Jacobian deviation. As a result, the displacement-mapping effect acts uniformly on the combined term $[\delta_{\rm J} + \delta_{\rm b}]$ and does not alter its internal perturbative structure. This implies that the nonlinear structure of the bias field inherits the same perturbative characteristics as the Jacobian deviation.

\subsection{Perturbative Expansion of the Dark Matter Density Contrast}
\label{sec:ULPT_jacobian}

In this subsection, we present the perturbative expansion of the dark matter density contrast $\delta_{\rm m}$ up to third order within the ULPT framework, together with its constituent component, the Jacobian deviation $\delta_{\rm J}$. A detailed derivation is given in Ref.~\cite{Sugiyama:inprep}.

\subsubsection{Review of Standard Perturbation Theory}

In standard cosmological perturbation theory, any physical quantity $X$ is expanded perturbatively in terms of the linear matter density contrast, denoted by $\delta_{\rm m}^{(1)}$, where the subscript ``m'' stands for ``matter.'' The expansion is given by
\begin{equation}
    X = \sum_{n=1}^{\infty} X^{(n)},
\end{equation}
where $X^{(n)}$ represents the $n$-th order contribution and scales as $X^{(n)} = \mathcal{O}\big([\delta_{\rm m}^{(1)}]^n\big)$.

The $n$-th order contribution to the Fourier-transformed matter density contrast is expressed as
\begin{align}
    \widetilde{\delta}_{\rm m}^{(n)}(\boldsymbol{k})
    = \int_{\boldsymbol{k}_1 \cdots \boldsymbol{k}_n}
    F_n(\boldsymbol{k}_1, \ldots, \boldsymbol{k}_n)\,
    \widetilde{\delta}_{\rm m}^{(1)}(\boldsymbol{k}_1) \cdots \widetilde{\delta}_{\rm m}^{(1)}(\boldsymbol{k}_n),
\end{align}
where the integration measure is defined as
\begin{align}
    \int_{\boldsymbol{k}_1 \cdots \boldsymbol{k}_n}
    \equiv \int \frac{d^3k_1}{(2\pi)^3} \cdots \int \frac{d^3k_n}{(2\pi)^3}
    (2\pi)^3 \delta_{\rm D}(\boldsymbol{k} - \boldsymbol{k}_1 - \cdots - \boldsymbol{k}_n).
    \label{eq:measure}
\end{align}

By construction, the first-order kernel is unity: $F_1 = 1$. Higher-order kernels $F_n$ for $n \geq 2$ can be systematically derived using the well-established recursion relation~\cite{Bernardeau:2001qr}.

An important property of the perturbation theory kernels is their vanishing in the zero-mode limit:
\begin{equation}
    F_{n \geq 2}(\boldsymbol{k}_1, \ldots, \boldsymbol{k}_n) = 0,
    \quad \text{if } \boldsymbol{k}_1 + \cdots + \boldsymbol{k}_n = \boldsymbol{0}.
    \label{eq:Fn_property}
\end{equation}
This condition ensures that the $n$-th order contributions to the matter density contrast vanish under spatial and ensemble averaging:
\begin{equation}
    \int d^3x\, \delta_{\rm m}^{(n)}(\boldsymbol{x}) =
    \langle \delta_{\rm m}^{(n)}(\boldsymbol{x}) \rangle = 0
    \quad \text{for } n \geq 2.
\end{equation}
For the linear-order fluctuation ($n = 1$), the vanishing of these averages is not a consequence of the kernel structure, but instead follows from the statistical properties of the primordial fluctuations as predicted by inflationary theory. Taken together, these observations imply that the full matter density contrast satisfies
\begin{equation}
    \int d^3x\, \delta_{\rm m}(\boldsymbol{x}) =
    \langle \delta_{\rm m}(\boldsymbol{x}) \rangle = 0
\end{equation}
as a nonperturbative statistical property of the field.

\subsubsection{Jacobian Deviation and Its Statistical Properties}

In ULPT, the dark matter density contrast is expressed as a convolution of the Jacobian deviation $\delta_{\rm J}$ with a spatial Dirac delta function, which enforces mass conservation under the Lagrangian-to-Eulerian mapping:
\begin{equation}
    \delta_{\rm m}(\boldsymbol{x}) = \int d^3q\, \delta_{\rm J}(\boldsymbol{q})\,
    \delta_{\rm D}(\boldsymbol{x} - \boldsymbol{q} - \boldsymbol{\Psi}(\boldsymbol{q})).
\end{equation}
This formulation implies that $\delta_{\rm J}$ must itself satisfy the same statistical properties as $\delta_{\rm m}$:
\begin{equation}
    \int d^3q\, \delta_{\rm J}(\boldsymbol{q}) =
    \langle \delta_{\rm J}(\boldsymbol{q}) \rangle = 0.
    \label{eq:delta_J_prop}
\end{equation}

The $n$-th order contribution to the Jacobian deviation in Fourier space can be written as
\begin{align}
    \widetilde{\delta}_{\rm J}^{(n)}(\boldsymbol{k})
    = \int_{\boldsymbol{k}_1 \cdots \boldsymbol{k}_n}
    J_n(\boldsymbol{k}_1, \ldots, \boldsymbol{k}_n)\,
    \widetilde{\delta}_{\rm m}^{(1)}(\boldsymbol{k}_1) \cdots \widetilde{\delta}_{\rm m}^{(1)}(\boldsymbol{k}_n),
\end{align}
where the integration measure is defined as in Eq.~\eqref{eq:measure}.

By definition, the first-order kernel is unity, $J_1 = 1$, implying that $\delta_{\rm J}^{(1)} = \delta_{\rm m}^{(1)}$.  
Higher-order kernels $J_n$ for $n \geq 2$ are constrained by Eq.~\eqref{eq:delta_J_prop} to vanish in the zero-mode limit:
\begin{equation}
    J_{n \geq 2}(\boldsymbol{k}_1, \ldots, \boldsymbol{k}_n) = 0,
    \quad \text{if } \boldsymbol{k}_1 + \cdots + \boldsymbol{k}_n = \boldsymbol{0}.
    \label{eq:Jn_property}
\end{equation}

\subsubsection{Second- and Third-Order Contributions to the Jacobian Deviation}

The kernel functions $J_n$ at arbitrary order $n$ can be obtained by substituting the perturbative solutions of the displacement vector up to order $n$ into Eq.~\eqref{eq:dJ}. In this work, we focus on the solutions up to third order.

The $n$-th order perturbative solution of the displacement field in Fourier space is given by~\cite{Matsubara:2007wj}
\begin{align}
    \widetilde{\boldsymbol{\Psi}}^{(n)}(\boldsymbol{k})
    = \frac{i}{n!}
    \int_{\boldsymbol{k}_1 \cdots \boldsymbol{k}_n}
    \boldsymbol{L}_n(\boldsymbol{k}_1,\ldots,\boldsymbol{k}_n)\,
    \widetilde{\delta}_{\rm m}^{(1)}(\boldsymbol{k}_1)\cdots
    \widetilde{\delta}_{\rm m}^{(1)}(\boldsymbol{k}_n),
    \label{eq:psi_n}
\end{align}
where $\boldsymbol{L}_n$ denotes the $n$-th order kernel vector. It can be decomposed into longitudinal and transverse components as
\begin{align}
    \boldsymbol{L}_n(\boldsymbol{k}_1,\ldots,\boldsymbol{k}_n)
    & = \frac{1}{k_{1\cdots n}^2} \bigg[
        \boldsymbol{k}_{1\cdots n}\,
        S_n(\boldsymbol{k}_1,\ldots,\boldsymbol{k}_n) \nonumber \\
      & \quad \quad \quad  + \boldsymbol{k}_{1\cdots n} \times
        \boldsymbol{T}_n(\boldsymbol{k}_1,\ldots,\boldsymbol{k}_n)
    \bigg],
    \label{eq:Ln}
\end{align}
with $\boldsymbol{k}_{1\cdots n}=\boldsymbol{k}_1+\cdots+\boldsymbol{k}_n$. Here $S_n$ and $\boldsymbol{T}_n$ represent the longitudinal (scalar) and transverse (vector) components, respectively. In linear theory, $S_1=0$ and $\boldsymbol{T}_1=0$. All higher-order components ($n\geq2$) can be systematically computed using the recursion relations derived in Ref.~\cite{Matsubara:2015ipa}.

The $n$-th order kernel $\boldsymbol{L}_n$ can be expressed in terms of three geometric functions $U$, $V$, and $\boldsymbol{W}$ defined by
\begin{align}
    U(\boldsymbol{k}_1,\boldsymbol{k}_2)
    &= |\hat{\boldsymbol{k}}_1 \times \hat{\boldsymbol{k}}_2|^2
    = 1 - (\hat{\boldsymbol{k}}_1 \cdot \hat{\boldsymbol{k}}_2)^2,
    \label{eq:Udef} \\
    V(\boldsymbol{k}_1,\boldsymbol{k}_2,\boldsymbol{k}_3)
    &= \bigl| \hat{\boldsymbol{k}}_1 \cdot (\hat{\boldsymbol{k}}_2 \times \hat{\boldsymbol{k}}_3) \bigr|^2 \nonumber \\
    &= 1 - (\hat{\boldsymbol{k}}_1 \cdot \hat{\boldsymbol{k}}_2)^2
       - (\hat{\boldsymbol{k}}_2 \cdot \hat{\boldsymbol{k}}_3)^2
       - (\hat{\boldsymbol{k}}_3 \cdot \hat{\boldsymbol{k}}_1)^2 \nonumber \\
    &\quad + 2(\hat{\boldsymbol{k}}_1 \cdot \hat{\boldsymbol{k}}_2)
              (\hat{\boldsymbol{k}}_2 \cdot \hat{\boldsymbol{k}}_3)
              (\hat{\boldsymbol{k}}_3 \cdot \hat{\boldsymbol{k}}_1),
    \label{eq:Vdef} \\
    \boldsymbol{W}(\boldsymbol{k}_1,\boldsymbol{k}_2)
    &= (\hat{\boldsymbol{k}}_1 \times \hat{\boldsymbol{k}}_2)\,
       (\hat{\boldsymbol{k}}_1 \cdot \hat{\boldsymbol{k}}_2).
\end{align}
Here $U$ represents the squared norm of the cross product of two wavevectors, $V$ corresponds to the squared scalar triple product of three wavevectors, and $W$ denotes the cross product multiplied by the inner product of two wavevectors. These functions vanish when the total momentum is zero: for example, $U$ and $W$ vanish for collinear configurations, while $V=0$ for coplanar configurations. In addition, $W$ also vanishes when $\mathbf{k}_1$ and $\mathbf{k}_2$ are orthogonal.

Using these functions, the second- and third-order solutions for the displacement field can be written in a compact form~\cite{Matsubara:2015ipa}:
\begin{equation}
    S_2(\boldsymbol{k}_1,\boldsymbol{k}_2)
    = \frac{3}{7}\,U(\boldsymbol{k}_1,\boldsymbol{k}_2),
    \qquad
    \boldsymbol{T}_2(\boldsymbol{k}_1,\boldsymbol{k}_2)=0,
    \label{eq:L2}
\end{equation}
and
\begin{align}
    S_3(\boldsymbol{k}_1,\boldsymbol{k}_2,\boldsymbol{k}_3)
    &= \frac{5}{3}\,U(\boldsymbol{k}_1,\boldsymbol{k}_{23})\,S_2(\boldsymbol{k}_2,\boldsymbol{k}_3)
     - \frac{1}{3}\,V(\boldsymbol{k}_1,\boldsymbol{k}_2,\boldsymbol{k}_3), \nonumber \\
    \boldsymbol{T}_3(\boldsymbol{k}_1,\boldsymbol{k}_2,\boldsymbol{k}_3)
    &= \boldsymbol{W}(\boldsymbol{k}_1,\boldsymbol{k}_{23})\,
       S_2(\boldsymbol{k}_2,\boldsymbol{k}_3),
    \label{eq:L3}
\end{align}
where $\boldsymbol{k}_{23} \equiv \boldsymbol{k}_2 + \boldsymbol{k}_3$.

Substituting these second- and third-order displacement solutions into Eq.~\eqref{eq:dJ}, we obtain the explicit expressions for the second- and third-order Jacobian kernels,
\begin{align}
    J_2(\boldsymbol{k}_1,\boldsymbol{k}_2)
    &= -\frac{2}{7}\,U(\boldsymbol{k}_1,\boldsymbol{k}_2),
    \label{eq:J2} \\
    J_3(\boldsymbol{k}_1,\boldsymbol{k}_2,\boldsymbol{k}_3)
    &= -\frac{2}{21}\,U(\boldsymbol{k}_1,\boldsymbol{k}_{23})\,
                     U(\boldsymbol{k}_2,\boldsymbol{k}_3)
       + \frac{1}{9}\,V(\boldsymbol{k}_1,\boldsymbol{k}_2,\boldsymbol{k}_3),
    \label{eq:J3}
\end{align}
where in deriving Eq.~\eqref{eq:J3} we used the relation $S_2(\boldsymbol{k}_2,\boldsymbol{k}_3)=(3/7)U(\boldsymbol{k}_2,\boldsymbol{k}_3)$ from Eq.~\eqref{eq:L2}. It is worth emphasizing that up to third order, only the longitudinal components $S_2$ and $S_3$ contribute to $J_n$; the transverse component $\boldsymbol{T}_3$ does not appear explicitly in Eqs.~\eqref{eq:J2} and~\eqref{eq:J3}.

We further decompose the third-order contribution into two parts:
\begin{align}
    \delta_{\rm J}^{(3)}(\boldsymbol{q})
    &= \delta_{ {\rm J},U}^{(3)}(\boldsymbol{q})
     + \delta_{ {\rm J},V}^{(3)}(\boldsymbol{q}),
\end{align}
with
\begin{align}
    \delta_{ {\rm J},U}^{(3)}(\boldsymbol{q})
    &= -\frac{2}{21} \int \frac{d^3k_1}{(2\pi)^3}
                      \frac{d^3k_2}{(2\pi)^3}
                      \frac{d^3k_3}{(2\pi)^3}
       e^{i(\boldsymbol{k}_1 + \boldsymbol{k}_2 + \boldsymbol{k}_3)\cdot \boldsymbol{q}} \nonumber \\
    &\quad \times U(\boldsymbol{k}_1, \boldsymbol{k}_{23})
                 U(\boldsymbol{k}_2, \boldsymbol{k}_3)
        \widetilde{\delta}_{\rm m}^{(1)}(\boldsymbol{k}_1) 
        \widetilde{\delta}_{\rm m}^{(1)}(\boldsymbol{k}_2) 
        \widetilde{\delta}_{\rm m}^{(1)}(\boldsymbol{k}_3) , \\
        \delta_{ {\rm J},V}^{(3)}(\boldsymbol{q})
    &= \frac{1}{9} \int \frac{d^3k_1}{(2\pi)^3}
                    \frac{d^3k_2}{(2\pi)^3}
                    \frac{d^3k_3}{(2\pi)^3}
       e^{i(\boldsymbol{k}_1 + \boldsymbol{k}_2 + \boldsymbol{k}_3)\cdot \boldsymbol{q}} \nonumber \\
    &\quad \times V(\boldsymbol{k}_1, \boldsymbol{k}_2, \boldsymbol{k}_3)
        \widetilde{\delta}_{\rm m}^{(1)}(\boldsymbol{k}_1) 
        \widetilde{\delta}_{\rm m}^{(1)}(\boldsymbol{k}_2) 
        \widetilde{\delta}_{\rm m}^{(1)}(\boldsymbol{k}_3) .
\end{align}

Since the geometric kernels $U$ and $V$ vanish whenever the total momentum satisfies $\boldsymbol{k}_1 + \cdots + \boldsymbol{k}_n = \boldsymbol{0}$, each of $\delta_{\rm J}^{(2)}$, $\delta_{{\rm J},U}^{(3)}$, and $\delta_{{\rm J},V}^{(3)}$ independently satisfies the condition of vanishing spatial and ensemble averages, as required by Eq.~\eqref{eq:delta_J_prop}. As a result, the second- and third-order kernels $J_2$ and $J_3$ explicitly satisfy the constraint given in Eq.~\eqref{eq:Jn_property}.

\subsubsection{Galileon Operator Structure of the Jacobian Deviation}
\label{sec:ULPT_Galileon}

Throughout the remainder of this subsection, we omit the explicit dependence on the Eulerian coordinate $\boldsymbol{x}$ for notational simplicity.

The nonlinear structure of the Jacobian deviation $\delta_{\rm J}$ is closely related to a class of scalar invariants known as Galileon operators~\cite{Chan:2012jj,Assassi:2014fva}. The rescaled gravitational and velocity potentials are defined by
\begin{equation}
    \Phi_{\rm g} \equiv \partial^{-2} \delta_{\rm m}, \quad
    \Phi_{\rm v} \equiv -\frac{1}{Hf} \partial^{-2} \theta,
\end{equation}
where $\theta = \nabla \cdot \boldsymbol{v}$ is the velocity divergence, $H$ is the Hubble parameter, $f$ is the linear growth rate, and $\partial^{-2}$ denotes the inverse Laplacian operator.

From these potentials, the second- and third-order Galileon operators are constructed as
\begin{align}
    \mathcal{G}_2(\Phi_{\rm g}) &\equiv
    (\partial_{ij} \Phi_{\rm g})^2 - (\partial^2 \Phi_{\rm g})^2,
    \label{eq:G2_def} \\
    \mathcal{G}_3(\Phi_{\rm g}) &\equiv
    (\partial^2 \Phi_{\rm g})^3
    + 2\, \partial_{ij} \Phi_{\rm g} \, \partial_{jk} \Phi_{\rm g} \, \partial_{ki} \Phi_{\rm g} \nonumber \\
    &\quad - 3\, (\partial_{ij} \Phi_{\rm g})^2 \, \partial^2 \Phi_{\rm g},
    \label{eq:G3_def}
\end{align}
where $\partial_{ij} \equiv \partial_i \partial_j$.

In addition, we define the difference between the second-order Galileon operators constructed from the gravitational and velocity potentials as
\begin{equation}
    \Gamma_3 \equiv \mathcal{G}_2(\Phi_{\rm g}) - \mathcal{G}_2(\Phi_{\rm v}).
    \label{eq:Gamma3_def}
\end{equation}
The operator $\Gamma_3$ contributes only at third order or higher in perturbation theory.

Using these Galileon operators, the second- and third-order contributions to the Jacobian deviation can be written as
\begin{align}
    \delta_{\rm J}^{(2)} &= \frac{2}{7} \mathcal{G}_2^{(2)}, 
    \label{eq:delta_J2_G2} \\
    \delta_{\rm J}^{(3)} &= \frac{1}{6} \Gamma_3^{(3)} + \frac{1}{9} \mathcal{G}_3^{(3)},
    \label{eq:delta_J3_G3_Gamma3}
\end{align}
with the decomposition into $U$- and $V$-type contributions given by
\begin{align}
    \delta_{ {\rm J},U}^{(3)} &= \frac{1}{6} \Gamma_3^{(3)}, 
    \label{eq:delta_J3_Gamma3} \\
    \delta_{ {\rm J},V}^{(3)} &= \frac{1}{9} \mathcal{G}_3^{(3)}.
    \label{eq:delta_J3_G3}
\end{align}

To derive these relations, it is useful to note that when the Galileon operators defined in Eqs.~\eqref{eq:G2_def} and~\eqref{eq:G3_def} are transformed into Fourier space, they directly correspond to the geometric functions $U$ and $V$ introduced in Eqs.~\eqref{eq:Udef} and~\eqref{eq:Vdef}. Once this correspondence is recognized, Eqs.~\eqref{eq:delta_J2_G2} and~\eqref{eq:delta_J3_G3} follow in a straightforward manner. For Eq.~\eqref{eq:delta_J3_Gamma3}, the result can be obtained by substituting the perturbative solutions for $\delta_{\rm m}$ and $\theta$ up to second order into the definition of $\Gamma_3$ given in Eq.~\eqref{eq:Gamma3_def}.

Each of these Galileon-type operators individually satisfies the condition of vanishing ensemble average:
\begin{equation}
    \langle \mathcal{G}_2^{(2)} \rangle =
    \langle \Gamma_3^{(3)} \rangle =
    \langle \mathcal{G}_3^{(3)} \rangle = 0.
    \label{eq:gal_prop}
\end{equation}
Since the Jacobian deviation satisfies the condition of vanishing spatial and ensemble averages nonperturbatively, as shown in Eq.~\eqref{eq:delta_J_prop}, it is natural that its nonlinear structure is entirely composed of Galileon operators.

\subsubsection{Tidal Decomposition of Galileon Operators}
\label{sec:tidal_decomp}

An alternative and more physically intuitive way to express the Galileon operators is through their decomposition in terms of tidal fields~\cite{Desjacques:2016bnm}. The linear-order tidal tensor is defined as
\begin{equation}
    K^{(1)}_{ij} = \left( \frac{\partial_i \partial_j}{\partial^2} - \frac{1}{3} \delta^{\rm K}_{ij} \right)\delta_{\rm m}^{(1)},
\end{equation}
where $\delta^{\rm K}_{ij}$ is the Kronecker delta. Using this tensor, the second- and third-order Galileon operators can be written as
\begin{align}
    \mathcal{G}_2^{(2)} &= -\frac{2}{3}[\delta_{\rm m}^{(1)}]^2 + K^{(1)}_{ij} K^{(1)}_{ij}, \nonumber \\
    \mathcal{G}_3^{(3)} &= 2 K^{(1)}_{ij} K^{(1)}_{jk} K^{(1)}_{ki}
    - \delta_{\rm m}^{(1)} K^{(1)}_{ij} K^{(1)}_{ij} + \frac{2}{9} [\delta_{\rm m}^{(1)}]^3, \nonumber \\
    \Gamma_3^{(3)} &= \frac{8}{21} \left[ \frac{21}{8} Q_{\mathrm{td}}^{(3)}
    - \frac{2}{3}[\delta_{\rm m}^{(1)}]^3
    + \delta_{\rm m}^{(1)} K^{(1)}_{ij} K^{(1)}_{ij} \right],
    \label{eq:G_K_3}
\end{align}
with
\begin{equation}
    Q_{\mathrm{td}}^{(3)} = \frac{8}{21} K^{(1)}_{ij} \left( \frac{\partial_i \partial_j}{\partial^2} - \frac{1}{3} \delta^{\rm K}_{ij} \right) \left( [\delta_{\rm m}^{(1)}]^2 - \frac{3}{2} K^{(1)}_{kl} K^{(1)}_{kl} \right).
    \label{eq:O_td_def}
\end{equation}

As evident from the above expressions, local quantities such as $[\delta_{\rm m}^{(1)}]^2$, $[\delta_{\rm m}^{(1)}]^3$, and $\delta_{\rm m}^{(1)} K_{ij}^{(1)} K_{ij}^{(1)}$ appear only through specific combinations with the tidal tensor that are organized to form the Galileon operator structure. Within the ULPT framework, these combinations constitute the nonlinear components of the Jacobian deviation, and as such, they automatically satisfy the statistical condition of vanishing ensemble and volume averages.

\subsubsection{Displacement-Mapping Effects up to Third Order}
\label{sec:disp_mapping}

The second- and third-order contributions to the dark matter density contrast can be obtained from Eq.~\eqref{eq:delta_g_decom} as follows:
\begin{align}
    \delta_{\rm m}^{(2)} &= \delta_{\rm J}^{(2)} 
    - \partial_i \left[ \Psi^{(1)}_i \delta_{\rm J}^{(1)} \right], \\
    \delta_{\rm m}^{(3)} &= \delta_{\rm J}^{(3)} 
    - \partial_i \left[ \Psi^{(2)}_i \delta_{\rm J}^{(1)} \right]
    - \partial_i \left[ \Psi^{(1)}_i \delta_{\rm J}^{(2)} \right] \nonumber \\
    & \quad + \frac{1}{2} \partial_i \partial_j \left[ \Psi^{(1)}_i \Psi^{(1)}_j \delta_{\rm J}^{(1)} \right].
\end{align}

The displacement-mapping terms in the above expressions can be rewritten as
\begin{equation}
    -\partial_i \left( \Psi^{(1)}_i \delta_{\rm J}^{(1)} \right) = [\delta_{\rm m}^{(1)}]^2 - \Psi^{(1)}_i \partial_i \delta_{\rm m}^{(1)},
\end{equation}
and
\begin{align}
    & -\partial_i (\Psi^{(1)}_i \delta_{\rm J}^{(2)})
    -\partial_i (\Psi^{(2)}_i \delta_{\rm J}^{(1)})
    + \frac{1}{2}\partial_i\partial_j \left( \Psi^{(1)}_i \Psi^{(1)}_j \delta_{\rm J}^{(1)} \right) \nonumber \\
    &= [\delta_{\rm m}^{(1)}]^3 + \frac{4}{7} \delta_{\rm m}^{(1)} \mathcal{G}_2^{(2)} + \text{[shift-type terms]},
    \label{eq:delta3_mapping}
\end{align}
where ``shift-type terms'' refer to terms involving spatial derivatives of the Jacobian deviation, such as $\partial_i \delta_{\rm J}^{(1)}$ and $\partial_i \delta_{\rm J}^{(2)}$.

These expressions indicate that the displacement-mapping contributions contain both local terms involving $\delta_{\rm m}^{(1)}$, such as $[\delta_{\rm m}^{(1)}]^2$, $[\delta_{\rm m}^{(1)}]^3$, or $\delta_{\rm m}^{(1)} \mathcal{G}_2^{(2)}$, and nonlocal shift-type terms. The local terms arise from the displacement field through the identity $\nabla \cdot \boldsymbol{\Psi}^{(1)} = -\delta_{\rm m}^{(1)}$, and can thus be understood as being induced by the displacement vector. From this perspective, these local-looking terms are not independent but are part of the displacement-mapping structure.

Importantly, the local contributions appearing in the displacement-mapping terms are combined with the corresponding shift-type terms in such a way that the total expression satisfies the required statistical properties, namely, vanishing ensemble and volume averages. In contrast, the local terms such as $[\delta_{\rm m}^{(1)}]^2$, $[\delta_{\rm m}^{(1)}]^3$, or $\delta_{\rm m}^{(1)} \mathcal{G}_2^{(2)}$ do not satisfy these conditions on their own.

\subsubsection{Structural Summary of Dark Matter Fluctuations}

To conclude this subsection, we summarize the structural features of dark matter fluctuations within the ULPT framework as follows:
\begin{itemize}
    \item The Jacobian deviation $\delta_{\rm J}$ satisfies the statistical condition $\langle \delta_{\rm J}(\boldsymbol{q}) \rangle = \int d^3q\, \delta_{\rm J}(\boldsymbol{q}) = 0$.
    \item The perturbative contributions to $\delta_{\rm J}$ can be expressed entirely in terms of the geometric functions $U$ and $V$, which are directly related to Galileon operators.
    \item Each Galileon operator can be written as a specific combination of the local field $\delta_{\rm m}^{(1)}$ and the induced tidal tensor $K^{(1)}_{ij}$. As a result, the Jacobian deviation $\delta_{\rm J}$ contains local terms, but they appear only through such structured combinations and are therefore not treated as independent components.
    \item Displacement-mapping effects generate both local and shift-type terms, but the local terms appear only in specific combinations with shift-type terms, and are thus not treated as independent contributions in the perturbative expansion.

\end{itemize}

Although the decomposition of the dark matter density field can, in principle, depend on the choice of operator basis, we adopt a two-step procedure that emphasizes statistical consistency. We first decompose the density contrast into the Jacobian deviation and the displacement-mapping effect. The Jacobian deviation is then further expressed in terms of statistically well-defined components, such as $\delta_{\rm J}^{(2)}$, $\delta_{ {\rm J},U}^{(3)}$, and $\delta_{ {\rm J},V}^{(3)}$, which are equivalently represented by the Galileon operators $\mathcal{G}_2^{(2)}$, $\Gamma_3^{(3)}$, and $\mathcal{G}_3^{(3)}$. The simplicity of the geometric functions $U$ and $V$ in Fourier space reflects the natural emergence of this decomposition within the ULPT framework and provides a systematic foundation for the renormalization-free bias parameterization introduced in Sec.~\ref{sec:bias}.

\section{Renormalization-Free Bias Model}
\label{sec:bias}

\subsection{Conceptual Basis of the Bias Model}

The central idea of our bias model can be summarized as follows:
\begin{enumerate}

    \item \textbf{Separation of biased and unbiased components:}
    The ULPT framework provides a formulation that clearly separates the contribution directly affected by the biased fluctuation field, the \emph{Jacobian deviation} $\delta_{\rm J}$, from the remaining \emph{displacement-mapping} effect, which transports fields without modifying their internal structure. 
    In this formulation, the biased fluctuation $\delta_{\rm b}$ enters additively into $\delta_{\rm J}$, so that the displacement-mapping effect acts uniformly on the combined field $\delta_{\rm J}+\delta_{\rm b}$ while preserving the intrinsic structure of each bias contribution.

    \item \textbf{Inheritance of dark-matter properties:}
    Since biased tracers are physically generated from the underlying dark matter field, the biased fluctuation $\delta_{\rm b}$ is assumed to inherit, as much as possible, the nonlinear and statistical properties already satisfied by dark matter. In this way, the bias sector is placed on the same theoretical footing as the matter sector.

\end{enumerate}

Guided by these principles, in Sec.~\ref{sec:ULPT_jacobian}, we analyze the nonlinear structure of the Jacobian deviation $\delta_{\rm J}$ up to third order and show that it can be fully characterized by the perturbative components $\delta_{\rm J}^{(1)}$, $\delta_{\rm J}^{(2)}$, $\delta_{{\rm J},U}^{(3)}$, and $\delta_{{\rm J},V}^{(3)}$. These components are constructed solely from the two geometric functions $U$ and $V$ defined in Eqs.~\eqref{eq:Udef} and~\eqref{eq:Vdef}, where $U(\boldsymbol{k}_1,\boldsymbol{k}_2)=\lvert\hat{\boldsymbol{k}}_1\times\hat{\boldsymbol{k}}_2\rvert^{2}$ denotes the squared norm of the cross product of two wavevectors, and $V(\boldsymbol{k}_1,\boldsymbol{k}_2,\boldsymbol{k}_3)=\big\lvert\hat{\boldsymbol{k}}_1\cdot(\hat{\boldsymbol{k}}_2\times\hat{\boldsymbol{k}}_3)\big\rvert^{2}$ represents the squared scalar triple product of three wavevectors. Together, these functions provide a compact and well-structured basis for describing the intrinsic nonlinear evolution. The simplest bias model is then obtained by assigning a single bias parameter to each of these components. Because each operator individually satisfies the statistical property of vanishing ensemble and volume averages, the resulting bias model is inherently renormalization-free.

In this paper, we restrict our analysis to bias parameters associated with perturbative contributions up to third order, which are sufficient for one-loop power spectrum calculations. In future applications, however, it may be necessary to include additional higher-order bias parameters corresponding to the fourth- and fifth-order perturbative contributions relevant for two-loop calculations. A brief discussion of this possible extension is provided in Sec.~\ref{sec:HigherOrder}.

The bias operators adopted in this work correspond to Galileon-type operators, as described in Sec.~\ref{sec:ULPT_Galileon}. As shown in Eq.~\eqref{eq:G_K_3}, these can be expressed as specific combinations of higher local terms and tidal fields. In principle, one could alternatively adopt the conventional basis of higher local and tidal operators commonly used in standard bias expansions~\cite{Desjacques:2016bnm}. However, in that case, the well-structured mathematical properties encoded in the $U$ and $V$ functions are no longer preserved, and as a result, the individual operators no longer satisfy the statistical property of vanishing ensemble and volume averages. To restore this property, an explicit renormalization procedure is required. Moreover, such a basis involves a larger number of independent operators, thereby introducing additional bias parameters.

It should also be noted that our bias model is constructed solely from the nonlinear effects encapsulated in $\delta_{\rm J}$. However, scale-dependent contributions from higher-derivative bias terms not included in $\delta_{\rm J}$ may, in principle, arise. Such effects are beyond the scope of this work but could become relevant in more general bias models.

The primary objective of this work is to validate the minimal bias model within the ULPT framework, which achieves a compact and renormalization-free description with the smallest number of bias parameters. More general bias models, including additional operator contributions such as higher-order effects beyond third order or extra bias parameters at third order that may require renormalization, are regarded as natural extensions of this framework. Such generalizations would be worth pursuing in more detail within the ULPT framework if the minimal model is found to be insufficient for achieving the required accuracy in modeling observational data.

\subsection{Bias Parameterization}

We retain the commonly adopted linear bias relation at the nonperturbative level, in which the galaxy density contrast is proportional to the underlying dark matter density contrast. If we assume that the bias fluctuation $\delta_{\rm b}$ is proportional to the Jacobian deviation $\delta_{\rm J}$ nonperturbatively, we can write
\begin{equation}
    \delta_{\rm b}(\boldsymbol{q}) = b_1^{\rm u}\, \delta_{\rm J}(\boldsymbol{q}),
\end{equation}
where the superscript ``u'' indicates that the parameter is defined within the ULPT framework. Substituting this into Eq.~\eqref{eq:delta_g}, we find
\begin{equation}
    \delta_{\rm g}(\boldsymbol{x}) = \left( 1 + b_1^{\rm u} \right) \delta_{\rm m}(\boldsymbol{x}),
\end{equation}
which implies that $b^{\rm E}_1 = 1 + b_1^{\rm u}$ corresponds to the conventional Eulerian linear bias parameter. Equivalently, $b_1^{\rm u}$ represents the standard first-order Lagrangian bias.

Higher-order bias parameters are then introduced to capture deviations from this leading-order behavior. Up to third order, we parameterize the bias fluctuation in Lagrangian space as
\begin{align}
    \delta_{\rm b}(\boldsymbol{q})
    &= b_1^{\rm u}\, \delta_{\rm J}(\boldsymbol{q})
     + b_2^{\rm u}\, \delta_{\rm J}^{(2)}(\boldsymbol{q}) \nonumber \\
     &\quad + b_{3,U}^{\rm u}\, \delta_{ {\rm J},U}^{(3)}(\boldsymbol{q})
     + b_{3,V}^{\rm u}\, \delta_{ {\rm J},V}^{(3)}(\boldsymbol{q}).
    \label{eq:delta_b_sub}
\end{align}
Here, the subscripts ``1'', ``2'', and ``3'' on the bias parameters $b^{\rm u}_1$, $b^{\rm u}_2$, and $b^{\rm u}_3$ indicate their corresponding perturbative order in the expansion. The third-order contributions are further classified into $U$-type and $V$-type components, and accordingly the subscript ``3'' is supplemented by an additional label, resulting in $b_{3,U}^{\rm u}$ and $b_{3,V}^{\rm u}$. 

Although the first term $b_1^{\rm u}\delta_{\rm J}$ is defined nonperturbatively in Eq.~\eqref{eq:delta_b_sub}, in practical implementations it is truncated at third order,
\begin{equation}
    b_1^{\rm u}\, \delta_{\rm J} \approx b_1^{\rm u}\, \sum_{n=1}^{3} \delta_{\rm J}^{(n)}.
    \label{eq:b_u1_b1}
\end{equation}
Using this expression, Eq.~\eqref{eq:delta_b_sub} can be equivalently written as
\begin{align}
    \delta_{\rm b}(\boldsymbol{q})
    &= b_1^{\rm u}\, \delta_{\rm J}^{(1)}(\boldsymbol{q})
    + (b_1^{\rm u}+b_2^{\rm u})\, \delta_{\rm J}^{(2)}(\boldsymbol{q}) \nonumber \\
    &\quad + (b_1^{\rm u}+b_{3,U}^{\rm u})\, \delta_{ {\rm J},U}^{(3)}(\boldsymbol{q})
    + (b_1^{\rm u}+b_{3,V}^{\rm u})\, \delta_{ {\rm J},V}^{(3)}(\boldsymbol{q}).
    \label{eq:delta_b_alta}
\end{align}
While we do not adopt this form in the present work, one could, if preferred, redefine the bias parameters as
\(
b_2^{\rm u'} = b_1^{\rm u}+b_2^{\rm u},
\,
b_{3,U}^{\rm u'} = b_1^{\rm u}+b_{3,U}^{\rm u},
\,
b_{3,V}^{\rm u'} = b_1^{\rm u}+b_{3,V}^{\rm u},
\)
and work with the alternative parameter set $\{b_1^{\rm u}, b_2^{\rm u'}, b_{3,U}^{\rm u'}, b_{3,V}^{\rm u'}\}$. This redefinition is mathematically equivalent and may be useful in certain applications, although it does not affect the physical content of the model.

To account for additional small-scale or hidden effects not captured by large-scale density fields, we introduce a stochastic bias contribution to the bias fluctuation field $\delta_b$. Specifically, we add a stochastic field $\varepsilon(\boldsymbol{q})$, defined in Lagrangian space. This field represents random fluctuations in the galaxy--matter relation arising from unmodeled microscopic physics, environment-dependent processes, or residual contamination due to imperfect shot-noise subtraction in power spectrum measurements~\cite{Dekel:1998eq,Desjacques:2016bnm}.

We assume that the stochastic field $\varepsilon$ satisfies the following statistical properties. First, it has zero mean:
\begin{equation}
    \langle \varepsilon(\boldsymbol{q}) \rangle = 0.
    \label{eq:e0}
\end{equation}
Second, it is statistically independent of the deterministic components of the dark matter density field, namely the Jacobian deviation and the displacement field:
\begin{equation}
    \langle \delta_{\rm J}(\boldsymbol{q}) \, \varepsilon(\boldsymbol{q}') \rangle =
    \langle \Psi_i(\boldsymbol{q}) \, \varepsilon(\boldsymbol{q}') \rangle = 0.
    \label{eq:ejp0}
\end{equation}
Third, it is assumed to be spatially uncorrelated, obeying the white-noise condition
\begin{equation}
    \langle \varepsilon(\boldsymbol{q}) \, \varepsilon(\boldsymbol{q}') \rangle
    = N_{\varepsilon} \, \delta_{\rm D}(\boldsymbol{q} - \boldsymbol{q}'),
    \label{eq:e}
\end{equation}
where $N_{\varepsilon}$ is a constant characterizing the amplitude of stochasticity and is not constrained to be positive.

With the stochastic contribution included, the final expression for the bias fluctuation field in this work is given by
\begin{align}
    \delta_{\rm b}(\boldsymbol{q})
    &= b_1^{\rm u}\, \delta_{\rm J}(\boldsymbol{q})
     + b_2^{\rm u}\, \delta_{\rm J}^{(2)}(\boldsymbol{q}) \nonumber \\
     &\quad + b_{3,U}^{\rm u}\, \delta_{ {\rm J},U}^{(3)}(\boldsymbol{q})
     + b_{3,V}^{\rm u}\, \delta_{ {\rm J},V}^{(3)}(\boldsymbol{q})
     + \varepsilon(\boldsymbol{q}).
    \label{eq:delta_b}
\end{align}

Substituting Eq.~\eqref{eq:delta_b} into the Lagrangian expression for the galaxy density field in Eq.~\eqref{eq:delta_g}, we obtain the full field-level bias expansion:
\begin{align}
    \delta_{\rm g}(\boldsymbol{x}) &=
    b^{\rm E}_1\, \delta_{\rm m}(\boldsymbol{x}) \nonumber \\
    & \quad +
    b_2^{\rm u} \int d^3q\,  \delta_{\rm J}^{(2)}(\boldsymbol{q})\,
        \delta_{\rm D}(\boldsymbol{x} - \boldsymbol{q} - \boldsymbol{\Psi}(\boldsymbol{q})) \nonumber \\
    & \quad +
    b_{3,U}^{\rm u} \int d^3q\, \delta_{ {\rm J},U}^{(3)}(\boldsymbol{q})\,
        \delta_{\rm D}(\boldsymbol{x} - \boldsymbol{q} - \boldsymbol{\Psi}(\boldsymbol{q})) \nonumber \\
    & \quad +
    b_{3,V}^{\rm u} \int d^3q\, \delta_{ {\rm J},V}^{(3)}(\boldsymbol{q})\,
        \delta_{\rm D}(\boldsymbol{x} - \boldsymbol{q} - \boldsymbol{\Psi}(\boldsymbol{q})) \nonumber \\
    & \quad +
    \int d^3q\, \varepsilon(\boldsymbol{q})\,
        \delta_{\rm D}(\boldsymbol{x} - \boldsymbol{q} - \boldsymbol{\Psi}(\boldsymbol{q})),
    \label{eq:delta_g_field}
\end{align}
where we have used $b^{\rm E}_1 \equiv 1 + b_1^{\rm u}$ for convenience.

Each term on the right-hand side of Eq.~\eqref{eq:delta_g_field} is constructed to satisfy the statistical condition of vanishing ensemble and volume averages. Consequently, the galaxy density contrast obeys
\begin{equation}
    \langle \delta_{\rm g}(\boldsymbol{x}) \rangle = \int d^3x\, \delta_{\rm g}(\boldsymbol{x}) = 0,
    \label{eq:delta_g_ave}
\end{equation}
demonstrating that the model is statistically consistent and well defined at the field level. Here, the term ``field-level'' refers to the fact that the bias structure is formulated directly at the level of the density field itself, independently of any particular statistical observable.

The fact that our model is defined at the field level implies that the same perturbative treatment applied to the dark matter density field can be consistently extended to the galaxy density fluctuations. In particular, the power spectrum can be computed without the need for any renormalization procedure, simply by replacing the nonlinear kernels of the matter field with their bias-modified counterparts in Sec.~\ref{sec:power}.

An important consequence of this field-level formulation is that the bare bias parameters can be directly used in fitting procedures without requiring additional renormalization. This enables a consistent treatment across different statistical observables. Indeed, as we demonstrate in Sec.~\ref{sec:vs_emu}, the model successfully explains both the halo--halo and halo--matter statistics simultaneously using a single set of bias parameters.

\subsection{Generalization to Higher-Order Bias}
\label{sec:HigherOrder}

In this subsection we outline how the present bias construction extends to higher perturbative orders. A complete development is beyond the scope of this paper, but the pathway is straightforward and useful for future applications.

Our starting point is the Jacobian-deviation field $\delta_{\rm J}$ [Eq.~\eqref{eq:dJ}]. To obtain its higher-order contributions, one can substitute the higher-order Lagrangian displacement solutions directly into Eq.~\eqref{eq:dJ}. The relevant displacement kernels have been derived in the literature; see Ref.~\cite{Matsubara:2015ipa}. This immediately shows that the nonlinear functions characterizing $\delta_{\rm J}^{(n)}$ at any order $n$ are constructed from the same geometric objects that appear in the perturbative expansion of the displacement field itself, namely the scalar kernels $U$ and $V$ and, when relevant, the vector kernel $\boldsymbol{W}$.

As a concrete example, consider the fourth-order longitudinal component of the displacement, whose kernel $S_4$ can be expressed as~\cite{Matsubara:2015ipa}
\begin{align}
    S_4(\boldsymbol{k}_1,\boldsymbol{k}_2,\boldsymbol{k}_3,\boldsymbol{k}_4)
    & = \frac{28}{11}\, U(\boldsymbol{k}_1,\boldsymbol{k}_{234})\,S_3(\boldsymbol{k}_2,\boldsymbol{k}_3,\boldsymbol{k}_4)
    \nonumber \\
    & \quad - \frac{28}{11}\, \boldsymbol{W}(\boldsymbol{k}_1,\boldsymbol{k}_{234})\!\cdot\! \boldsymbol{T}_3(\boldsymbol{k}_2,\boldsymbol{k}_3,\boldsymbol{k}_4)
    \nonumber \\
    & \quad + \frac{17}{11}\, U(\boldsymbol{k}_{12},\boldsymbol{k}_{34})\, S_2(\boldsymbol{k}_1,\boldsymbol{k}_2)\, S_2(\boldsymbol{k}_3,\boldsymbol{k}_4)
    \nonumber \\
    & \quad - \frac{26}{11}\, V(\boldsymbol{k}_1,\boldsymbol{k}_2,\boldsymbol{k}_{34})\, S_2(\boldsymbol{k}_3,\boldsymbol{k}_4),
    \label{eq:S4}
\end{align}
where $\boldsymbol{k}_{ij\cdots} \equiv \boldsymbol{k}_i+\boldsymbol{k}_j+\cdots$ and $S_2$ and $S_3$ denote the lower-order longitudinal kernels. The fourth-order transverse (vector) component $\boldsymbol{T}_4$ is not required to construct $\delta_{\rm J}^{(4)}$, but the third-order transverse kernel $\boldsymbol{T}_3$ already enters through Eq.~\eqref{eq:S4}. This indicates that transverse contributions start to affect the nonlinear structure of $\delta_{\rm J}$ at fourth order.

By using Eqs.~\eqref{eq:L2} and~\eqref{eq:L3} to express $S_2$, $S_3$, and $\boldsymbol{T}_3$ in terms of $U$, $V$, and $\boldsymbol{W}$, one finds that the fourth-order Jacobian deviation is spanned by the following five nonlinear structures:
\begin{align}
    \delta_{\rm J}^{(4)} \in
    \Big\{\, &
    U(\boldsymbol{k}_1,\boldsymbol{k}_{234})\,V(\boldsymbol{k}_2,\boldsymbol{k}_3,\boldsymbol{k}_4), \nonumber \\
    & U(\boldsymbol{k}_1,\boldsymbol{k}_{234})\,U(\boldsymbol{k}_2,\boldsymbol{k}_{34})\,U(\boldsymbol{k}_3,\boldsymbol{k}_4), \nonumber \\
    & \boldsymbol{W}(\boldsymbol{k}_1,\boldsymbol{k}_{234})\!\cdot\!
    \boldsymbol{W}(\boldsymbol{k}_2,\boldsymbol{k}_{34})\,U(\boldsymbol{k}_3,\boldsymbol{k}_4), \nonumber \\
    & U(\boldsymbol{k}_{12},\boldsymbol{k}_{34})\,U(\boldsymbol{k}_1,\boldsymbol{k}_2)\,U(\boldsymbol{k}_3,\boldsymbol{k}_4), \nonumber \\
    & V(\boldsymbol{k}_1,\boldsymbol{k}_2,\boldsymbol{k}_{34})\,U(\boldsymbol{k}_3,\boldsymbol{k}_4)\,\Big\}.
    \label{eq:dJ4-basis}
\end{align}
Each of these functions vanishes when evaluated under the momentum-conservation constraint $\boldsymbol{k}_{1234}=\boldsymbol{0}$. As a result, each corresponding term independently satisfies the statistical conditions of vanishing ensemble and volume averages. This property provides the field-level reason why the bias construction based on $\delta_{\rm J}$ naturally avoids any ad hoc renormalization.

A minimal higher-order bias model at fourth order then represents the biased field $\delta_{\rm b}^{(4)}$ as a linear combination of the five independent structures in Eq.~\eqref{eq:dJ4-basis}, with one free bias parameter assigned to each structure. In this setup, there are five fourth-order bias parameters, and the model, by construction, preserves the vanishing-mean conditions term by term, thereby maintaining a renormalization-free formulation. The fourth-order biased contribution first appears in two-loop power-spectrum calculations and one-loop bispectrum calculations.

The same logic extends to fifth order (relevant for the two-loop power spectrum), where one builds the basis from the corresponding displacement kernels and includes transverse pieces as dictated by the Lagrangian solutions. A full enumeration and organization of the complete higher-order bias basis is left to future work.

\subsection{Relation to Existing Bias Models}
\label{sec:relation_standard}

In this subsection, we compare our ULPT-based bias model with the standard Eulerian bias expansion formulated in terms of the tidal tensor~\cite{Desjacques:2016bnm} (see also Refs.~\cite{Fujita:2020xtd,DAmico:2021rdb}).

Following Ref.~\cite{Desjacques:2016bnm}, the standard bias expansion up to second order is given by
\begin{equation}
    \delta_{\rm g}^{(1+2)}
    = b_1^{\rm E} \left[ \delta_{\rm m}^{(1)} + \delta_{\rm m}^{(2)} \right]
    + \frac{1}{2} b_2^{\rm E} \left[ \delta_{\rm m}^{(1)} \right]^2
    + b_{K^2}^{\rm E} K_{ij}^{(1)} K_{ij}^{(1)},
    \label{eq:delta_g_2_E}
\end{equation}
where we have suppressed the explicit Eulerian coordinate dependence for notational simplicity. Stochastic terms are also omitted. Here, $b_2^{\rm E}$ denotes the second-order local bias parameter, while $b_{K^2}^{\rm E}$ is the tidal bias parameter. The superscript ``E'' indicates that these parameters are defined in Eulerian space.

At third order, the expansion takes the form
\begin{align}
    \delta_{\rm g}^{(3)} &=
    b_1^{\rm E} \delta_{\rm m}^{(3)}
    + \frac{1}{3!} b_3^{\rm E} \left[ \delta_{\rm m}^{(1)} \right]^3
    + b_{K^3}^{\rm E} K_{ij}^{(1)} K_{jk}^{(1)} K_{ki}^{(1)} \nonumber \\
    &\quad + b_{\delta K^2}^{\rm E} \delta_{\rm m}^{(1)} K_{ij}^{(1)} K_{ij}^{(1)}
    + b_{\mathrm{td}}^{\rm E} Q_{\mathrm{td}}^{(3)} + \cdots,
    \label{eq:delta_g_3_E}
\end{align}
again omitting stochastic contributions.

In contrast, the ULPT-based bias expansion can be systematically derived by perturbatively expanding Eq.~\eqref{eq:delta_g_field} using Eq.~\eqref{eq:delta_g_decom}. Up to third order, the resulting expression is
\begin{align}
    \delta_{\rm g}^{(1+2)} &=
    b_1^{\rm E} \left[ \delta_{\rm m}^{(1)} + \delta_{\rm m}^{(2)} \right]+
    b_2^{\rm u} \delta_{\rm J}^{(2)}, \nonumber \\
    \delta_{\rm g}^{(3)} &=
    b_1^{\rm E} \delta_{\rm m}^{(3)}
    - b_2^{\rm u} \nabla \cdot \left[ \boldsymbol{\Psi}^{(1)}\, \delta_{\rm J}^{(2)} \right] \nonumber \\
    &\quad + b_{3,U}^{\rm u} \delta_{ {\rm J},U}^{(3)}
    + b_{3,V}^{\rm u} \delta_{ {\rm J},V}^{(3)},
    \label{eq:delta_g_3}
\end{align}
where all fields are expressed in Eulerian space.

Substituting the Galileon operator identities from Eq.~\eqref{eq:G_K_3}, the ULPT-based galaxy density contrast becomes
\begin{align}
    \delta_{\rm g}^{(1+2)}
    &= b_1^{\rm E} \left[ \delta_{\rm m}^{(1)} + \delta_{\rm m}^{(2)} \right]
    - \frac{4}{21} b_2^{\rm u} \left[ \delta_{\rm m}^{(1)} \right]^2
    + \frac{2}{7} b_2^{\rm u} K_{ij}^{(1)} K_{ij}^{(1)}, \\
    \delta_{\rm g}^{(3)}
    &= b_1^{\rm E} \delta_{\rm m}^{(3)}
    + \left( -\frac{8}{189} b_{3,U}^{\rm u} + \frac{2}{81} b_{3,V}^{\rm u} - \frac{4}{21} b_2^{\rm u} \right)
      \left[ \delta_{\rm m}^{(1)} \right]^3 \nonumber \\
    &\quad + \left( \frac{4}{63} b_{3,U}^{\rm u} - \frac{1}{9} b_{3,V}^{\rm u} + \frac{2}{7} b_2^{\rm u} \right)
      \delta_{\rm m}^{(1)} K_{ij}^{(1)} K_{ij}^{(1)} \nonumber \\
    &\quad + \frac{1}{6} b_{3,U}^{\rm u} Q_{\rm td}^{(3)}
      + \frac{2}{9} b_{3,V}^{\rm u} K_{ij}^{(1)} K_{jk}^{(1)} K_{ki}^{(1)} \nonumber \\
    &\quad - b_2^{\rm u} \boldsymbol{\Psi}^{(1)} \cdot \nabla \delta_{\rm J}^{(2)}.
    \label{eq:delta_g_3_U}
\end{align}

By comparing Eqs.~\eqref{eq:delta_g_2_E} and \eqref{eq:delta_g_3_E} with the ULPT expansion above, we obtain the following relations between the ULPT bias parameters and the conventional Eulerian bias parameters:

At second order, the Eulerian bias parameters are related to the ULPT coefficient $b_2^{\rm u}$ through the expressions
\begin{align}
    b_2^{\rm E} &= -\frac{8}{21} b_2^{\rm u}, \\
    b_{K^2}^{\rm E} &= \frac{2}{7} b_2^{\rm u},
    \label{eq:b2E}
\end{align}
indicating that both $b_2^{\rm E}$ and $b_{K^2}^{\rm E}$ originate from a single Lagrangian operator in the ULPT expansion. This structure leads to a specific theoretical prediction for the relation between the two Eulerian bias coefficients:
\begin{equation}
    b_{K^2}^{\rm E} = -\frac{3}{4} b_2^{\rm E}.
    \label{eq:b2Erelation}
\end{equation}
This prediction is a direct consequence of the Galileon operator basis adopted in ULPT, and can be tested by comparing with empirical fitting formulas calibrated on $N$-body simulations. We will quantitatively assess the consistency of this relation against existing fitting results in Sec.~\ref{sec:compare_empirical}.

At third order, we similarly find
\begin{align}
    b_3^{\rm E} &= \left( -\frac{16}{63} b_{3,U}^{\rm u} + \frac{4}{27} b_{3,V}^{\rm u} - \frac{8}{7} b_2^{\rm u} \right), \\
    b_{\delta K^2}^{\rm E} &= \frac{4}{63} b_{3,U}^{\rm u} - \frac{1}{9} b_{3,V}^{\rm u} + \frac{2}{7} b_2^{\rm u}, \\
    b_{K^3}^{\rm E} &= \frac{2}{9} b_{3,V}^{\rm u}, \\
    b_{\mathrm{td}}^{\rm E} &= \frac{1}{6} b_{3,U}^{\rm u}.
    \label{eq:b3E}
\end{align}

A key distinction from the standard bias model is that, at third order, the ULPT framework includes displacement-mapping contributions involving spatial derivatives of the Jacobian deviation $\delta_{\rm J}$. This results in the appearance of a shift-type term in the third-order bias expansion, as seen in Eq.~\eqref{eq:delta_g_3_U}, specifically $- b_2^{\rm u} \boldsymbol{\Psi}^{(1)} \cdot \nabla \delta_{\rm J}^{(2)}$. 

As demonstrated from Eq.~\eqref{eq:b2Erelation} to Eq.~\eqref{eq:b3E}, the bias parameters appearing in the standard bias expansion are not independent but are related through the bias parameters introduced in the ULPT framework. This property originates from the fact that the ULPT bias model inherits the same nonlinear structure and statistical properties that govern the dark matter field itself, thereby imposing stronger theoretical constraints than the standard bias model. As a consequence, the number of bias parameters in ULPT is smaller than in conventional formulations.

For example, when one imposes the condition that the volume and ensemble averages vanish for the second-order galaxy density fluctuation in the standard bias model [Eq.~\eqref{eq:delta_g_2_E}], the relation between $b_{2}^{\rm E}$ and $b_{K^2}^{\rm E}$ given by Eq.~\eqref{eq:b2Erelation} is required. In other words, this relation is not an arbitrary assumption but a direct consequence of the statistical constraints inherited from the underlying dark matter dynamics.

Conversely, starting from the perspective of the ULPT bias model makes it clear that adopting a more general standard bias model that violates these nonlinear properties of dark matter requires a well-motivated physical justification for such a choice.

\subsection{Perturbative Kernels for Biased Tracers}
\label{sec:kernels}

Within the framework of perturbation theory, the $n$th-order contribution to the density contrast of biased tracers (e.g., galaxies or haloes) in Fourier space can be expressed as
\begin{equation}
    \widetilde{\delta}_{\rm g}^{(n)}(\boldsymbol{k})
    = \int_{\boldsymbol{k}_1\cdots\boldsymbol{k}_n} F_{{\rm g}, n}(\boldsymbol{k}_1,\ldots,\boldsymbol{k}_n)\,
    \widetilde{\delta}_{\rm m}^{(1)}(\boldsymbol{k}_1)\cdots\widetilde{\delta}_{\rm m}^{(1)}(\boldsymbol{k}_n),
\end{equation}
where $F_{{\rm g}, n}$ denotes the $n$th-order kernel associated with the biased tracer.

From the bias expansion presented in Eq.~\eqref{eq:delta_g_3}, the first- and second-order galaxy kernels are given by
\begin{align}
    F_{{\rm g},1}(\boldsymbol{k}_1) &= b^{\rm E}_1, \nonumber \\
    F_{{\rm g},2}(\boldsymbol{k}_1, \boldsymbol{k}_2) &= b^{\rm E}_1 F_2(\boldsymbol{k}_1, \boldsymbol{k}_2)
    + b^{\rm u}_2\, J_2(\boldsymbol{k}_1, \boldsymbol{k}_2),
    \label{eq:Fg12}
\end{align}
where $F_2$ is the standard second-order SPT kernel, and $J_2$ is the ULPT-specific Jacobian kernel.

At third order, the galaxy kernel takes the form
\begin{align}
    F_{{\rm g},3}(\boldsymbol{k}_1, \boldsymbol{k}_2, \boldsymbol{k}_3)
    &= b^{\rm E}_1 F_3(\boldsymbol{k}_1, \boldsymbol{k}_2, \boldsymbol{k}_3) \nonumber \\
    &+ \frac{b_2^{\rm u}}{3} \left[
    \left( \frac{\boldsymbol{k}_{123} \cdot \boldsymbol{k}_1}{k_1^2} \right)
    J_2(\boldsymbol{k}_2, \boldsymbol{k}_3) + 2\,\text{perms.} \right] \nonumber \\
    &- \frac{2}{63}\, b_{3, U}^{\rm u}\,
    \left[ U(\boldsymbol{k}_1, \boldsymbol{k}_{23}) U(\boldsymbol{k}_2, \boldsymbol{k}_3)
    + 2\,\text{perms.} \right] \nonumber \\
    &+ \frac{1}{9}\,b_{3, V}^{\rm u}\, V(\boldsymbol{k}_1, \boldsymbol{k}_2, \boldsymbol{k}_3),
\end{align}
where the geometric functions $U$ and $V$ are defined in Eqs.~\eqref{eq:Udef} and \eqref{eq:Vdef}, and “perms.” denotes cyclic permutations over $\boldsymbol{k}_1$, $\boldsymbol{k}_2$, and $\boldsymbol{k}_3$.

However, in the one-loop power spectrum calculation (see Sec.~\ref{sec:power}), only a specific contraction of the third-order kernel is relevant:
\begin{align}
    F_{{\rm g},3}(\boldsymbol{k}, \boldsymbol{p}, -\boldsymbol{p})
    &= b^{\rm E}_1 F_3(\boldsymbol{k}, \boldsymbol{p}, -\boldsymbol{p}) \nonumber \\
    &- \frac{2}{63}\, b_{3, U}^{\rm u}\, \big[
    U(\boldsymbol{p}, \boldsymbol{k} - \boldsymbol{p}) U(\boldsymbol{k}, -\boldsymbol{p}) \nonumber \\
    &\hspace{1.7cm} +
    U(-\boldsymbol{p}, \boldsymbol{k} + \boldsymbol{p}) U(\boldsymbol{k}, \boldsymbol{p}) \big].
    \label{eq:Fg3}
\end{align}
In this expression, the terms proportional to $b_2^{\rm u}$ and $b_{3, V}^{\rm u}$ vanish due to the contraction structure. As a result, only the $b_{3, U}^{\rm u}$ term contributes to the one-loop correction. Since this paper focuses solely on the one-loop power spectrum, we henceforth simplify notation by omitting the subscript ``$U$'' and writing $b_{3, U}^{\rm u}$ as $b_3^{\rm u}$.

Furthermore, for clarity and brevity, we also omit the superscript ``E'' from the Eulerian linear bias parameter and write \( b_1^{\rm E} \equiv b_1 \) throughout the remainder of this paper.

\section{Power Spectrum of Biased Tracers in ULPT}
\label{sec:power}

In this section, we present the formulation and evaluation of the galaxy--galaxy auto power spectrum and the galaxy--matter cross power spectrum at one-loop order within the ULPT framework. The corresponding expressions for the nonlinear matter power spectrum have been established in Ref.~\cite{Sugiyama:2025myq}, to which we refer for technical details. In the case of galaxy clustering, the formulation remains structurally identical, with the only difference being the replacement of dark matter kernels with their bias-modified counterparts as defined in Eqs.~\eqref{eq:Fg12} and \eqref{eq:Fg3}.

Sections~\ref{sec:P_general}--\ref{sec:Conv} review the general structure of the ULPT power spectrum and the one-loop perturbative expansion scheme, as developed in Ref.~\cite{Sugiyama:2025myq}. In the following sections, \ref{sec:Source}, \ref{sec:stochastic}, and \ref{sec:ULPT_P}, we present our new results: the explicit derivation of the one-loop galaxy power spectra for biased tracers, incorporating nonlinear bias contributions within a renormalization-free framework. In particular, both the galaxy--galaxy auto power spectrum and the galaxy--matter cross power spectrum are consistently predicted using a common set of bias parameters. A comparison with simulation-based emulators is provided in Sec.~\ref{sec:vs_emu}.

Throughout this section, we frequently encounter Hankel transforms arising from statistical isotropy. For a spherically symmetric function \( f(r) \), the Hankel transform of order \( \ell \) is defined by
\begin{align}
    \tilde{f}_{\ell}(k) = (-i)^{\ell} (4\pi) \int_0^\infty dr\, r^2\, j_{\ell}(kr)\, f(r),
\end{align}
where \( j_{\ell}(x) \) denotes the spherical Bessel function of order \( \ell \). The corresponding inverse transform is
\begin{align}
    f(r) = i^{\ell} \int \frac{dk\, k^2}{2\pi^2} j_{\ell}(kr)\, \tilde{f}_{\ell}(k).
\end{align}
These transforms can be efficiently evaluated using FFTLog-based techniques~\cite{Hamilton:1999uv}, implemented in our work via the public package \texttt{mcfit}~\cite{mcfit}.

\subsection{General Structure}
\label{sec:P_general}

In the ULPT framework, the galaxy--galaxy auto power spectrum is expressed as
\begin{align}
    P_{\rm gg}(\boldsymbol{k}) = \int d^3r\, e^{-i\boldsymbol{k}\cdot\boldsymbol{r}}\,
    \left\langle e^{-i\boldsymbol{k}\cdot[\boldsymbol{\Psi}(\boldsymbol{q}) - \boldsymbol{\Psi}(\boldsymbol{q}')]}\,
    Y(\boldsymbol{q})\, Y(\boldsymbol{q}') \right\rangle,
    \label{eq:P_ULPT_general}
\end{align}
where \( \boldsymbol{r} = \boldsymbol{q} - \boldsymbol{q}' \) is the Lagrangian separation, and the composite field
\begin{align}
    Y(\boldsymbol{q}) = \delta_{\rm J}(\boldsymbol{q}) + \delta_{\rm b}(\boldsymbol{q})
\end{align}
represents the sum of the Jacobian deviation and the bias fluctuation field.

To evaluate Eq.~\eqref{eq:P_ULPT_general}, we decompose the ensemble average using the cumulant expansion. Defining
\begin{align}
    X \equiv -i\boldsymbol{k} \cdot \left[ \boldsymbol{\Psi}(\boldsymbol{q}) - \boldsymbol{\Psi}(\boldsymbol{q}') \right],
\end{align}
the expectation value can be written as~\cite{Scoccimarro:2004tg}
\begin{align}
    \left\langle e^{X} Y(\boldsymbol{q}) Y(\boldsymbol{q}') \right\rangle
    = \left\langle e^{X} \right\rangle
    \left[ \left\langle e^{X} Y Y' \right\rangle_{\rm c}
         + \left\langle e^{X} Y \right\rangle_{\rm c} \left\langle e^{X} Y' \right\rangle_{\rm c} \right],
    \label{eq:XYY}
\end{align}
where $Y' \equiv Y(\boldsymbol{q}')$, and $\langle\cdots\rangle_{\rm c}$ denotes the connected part.

The displacement-dependent factor,
\begin{align}
    \left\langle e^{X} \right\rangle
    = \exp\left[-\overline{\Sigma}(\boldsymbol{k}) + \Sigma(\boldsymbol{k}, \boldsymbol{r})\right],
\end{align}
is referred to as the \emph{displacement-mapping factor}. This factor is constructed solely from the displacement field and is statistically uncorrelated with the Jacobian deviation and the bias fluctuation field (collectively denoted by \( Y \)). Physically, it captures the nonlinear remapping of spatial positions induced by the Lagrangian-to-Eulerian coordinate transformation.

The exponent is defined through the cumulant expansion:
\begin{align}
    -\overline{\Sigma}(\boldsymbol{k}) + \Sigma(\boldsymbol{k}, \boldsymbol{r})
    = \sum_{m=2}^{\infty} \frac{1}{m!}
    \left\langle \left[ -i\boldsymbol{k} \cdot (\boldsymbol{\Psi}(\boldsymbol{q}) - \boldsymbol{\Psi}(\boldsymbol{q}')) \right]^m \right\rangle_{\rm c},
    \label{eq:Sigma_expand}
\end{align}
where the displacement variance is defined as
\begin{align}
    \overline{\Sigma}(\boldsymbol{k}) \equiv \Sigma(\boldsymbol{k}, \boldsymbol{r} = 0).
\end{align}

The remaining part of Eq.~\eqref{eq:XYY} involves correlations of the source field $Y$ and its interactions with the displacement field. We define the \emph{source correlation function} as
\begin{align}
    \xi_{\rm J,gg}(\boldsymbol{r})
    &\equiv
    \left\langle Y(\boldsymbol{q}) Y(\boldsymbol{q}') \right\rangle_{\rm c}
    + \left\langle (e^{X} - 1) Y Y' \right\rangle_{\rm c} \nonumber \\
   & \quad + \left\langle (e^{X} - 1) Y \right\rangle_{\rm c} \left\langle (e^{X} - 1) Y' \right\rangle_{\rm c},
\end{align}
where we have used the fact that $\langle Y \rangle = \langle Y' \rangle = 0$.

Combining the displacement-mapping factor and the source correlation function, the galaxy power spectrum can be written in compact form as
\begin{align}
    P_{\rm gg}(\boldsymbol{k}) =
    e^{-\overline{\Sigma}(\boldsymbol{k})}
    \int d^3r\, e^{-i\boldsymbol{k}\cdot\boldsymbol{r}}\,
    e^{\Sigma(\boldsymbol{k}, \boldsymbol{r})}\, \xi_{\rm J,gg}(\boldsymbol{r}).
    \label{eq:P_ULPT}
\end{align}

In cosmological perturbation theory, calculations are typically performed in Fourier space. Accordingly, the source correlation function appearing in the ULPT power spectrum [Eq.~\eqref{eq:P_ULPT}] is evaluated through its Fourier transform, namely the source power spectrum $P_{\rm J}$, as
\begin{equation}
    \xi_{\rm J,gg}(\boldsymbol{r}) = \int \frac{d^3k}{(2\pi)^3} e^{i\boldsymbol{k}\cdot\boldsymbol{r}} P_{\rm J}(\boldsymbol{k}).
\end{equation}
Here $P_{\rm J}$ corresponds to the power spectrum in the absence of displacement-induced remapping. With this expression, the full galaxy power spectrum can be written as
\begin{align}
    P_{\rm gg}(\boldsymbol{k}) = P_{\rm J,gg}(\boldsymbol{k}) + P_{\rm DM,gg}(\boldsymbol{k}),
    \label{eq:DMgg}
\end{align}
where the \emph{displacement-mapping (DM) correction} is given by
\begin{align}
    P_{\rm DM,gg}(\boldsymbol{k}) =
    \int d^3r\, e^{-i\boldsymbol{k}\cdot\boldsymbol{r}} \left[ e^{-\overline{\Sigma}(\boldsymbol{k}) + \Sigma(\boldsymbol{k}, \boldsymbol{r})} - 1 \right] \xi_{\rm J,gg}(\boldsymbol{r}).
    \label{eq:DM}
\end{align}

Throughout this work, we focus on real-space statistics, where statistical isotropy implies that all relevant quantities depend only on the magnitudes \( k = |\boldsymbol{k}| \) and \( r = |\boldsymbol{r}| \), and on the cosine angle \( \mu = \hat{\boldsymbol{k}} \cdot \hat{\boldsymbol{r}} \). For example, \( \Sigma(\boldsymbol{k}, \boldsymbol{r}) \) reduces to a function of three scalars: \( \Sigma(k, r, \mu) \). This symmetry substantially simplifies the numerical evaluation of the convolution integrals appearing in subsequent sections.

\subsection{Standard Perturbation Theory}
\label{sec:SPT}

In SPT, the galaxy power spectrum is computed as an expansion around the linear matter power spectrum, denoted by \( P_{\rm m}^{\rm (lin)} \), where the superscript ``lin'' indicates linear order. The SPT expression for the galaxy power spectrum at one-loop level takes the form
\begin{equation}
    P_{\rm gg}(k) = P_{\rm gg}^{\rm (lin)}(k) + P_{\rm gg}^{\text{(1-loop)}}(k),
\end{equation}
where the linear galaxy power spectrum is given by
\begin{equation}
    P_{\rm gg}^{\rm (lin)}(k) = b_1^2\, P_{\rm m}^{\rm (lin)}(k),
\end{equation}
as obtained from Eq.~\eqref{eq:Fg12}.

The one-loop correction is of order \( \mathcal{O}([P_{\rm m}^{\rm (lin)}]^2) \), and can be further decomposed into two distinct contributions: the so-called 22-type and 13-type terms. These correspond to the power spectrum arising from the auto-correlation of second-order density perturbations and the cross-correlation between first- and third-order perturbations, respectively. Explicitly,
\begin{equation}
    P_{\rm gg}^{\text{(1-loop)}}(k) =
    P_{\rm gg}^{(22)}(k) +
    P_{\rm gg}^{(13)}(k),
\end{equation}
with
\begin{align}
    P_{\rm gg}^{(22)}(k) &= 2 \int \frac{d^3p}{(2\pi)^3}
    \left[ F_{{\rm g},2}(\boldsymbol{k}, \boldsymbol{k} - \boldsymbol{p}) \right]^2 \nonumber \\
    &\quad \times P_{\rm m}^{\rm (lin)}(|\boldsymbol{k} - \boldsymbol{p}|)\, P_{\rm m}^{\rm (lin)}(p), \\
    P_{\rm gg}^{(13)}(k) &= 6 \int \frac{d^3p}{(2\pi)^3}
    F_{{\rm g},3}(\boldsymbol{k}, \boldsymbol{p}, -\boldsymbol{p})\, F_{{\rm g},1}(\boldsymbol{k}) \nonumber \\
    &\quad \times P_{\rm m}^{\rm (lin)}(k)\, P_{\rm m}^{\rm (lin)}(p).
\end{align}
The expressions above use the bias-modified kernels \( F_{{\rm g},n} \) defined in Sec.~\ref{sec:kernels}, thereby incorporating nonlinear galaxy bias consistently within the SPT formalism.

\subsection{One-Loop Expansion in ULPT}
\label{sec:Order}

In the ULPT framework, the one-loop galaxy power spectrum is computed by reproducing the SPT result at one-loop order and incorporating additional nonlinear corrections via the displacement-mapping factor. To implement this systematically, we expand the relevant quantities as follows:
\begin{align}
    \Sigma(k, r, \mu) &= \Sigma^{\rm (lin)}(k, r, \mu), \nonumber \\
    \xi_{\rm J,gg}(r) &= \xi_{\rm J,gg}^{\rm (lin)}(r) + \xi_{\rm J,gg}^{\text{(1-loop)}}(r), \nonumber \\
    P_{\rm J,gg}(k) &= P_{\rm J,gg}^{\rm (lin)}(k) + P_{\rm J,gg}^{\text{(1-loop)}}(k).
    \label{eq:SxiP}
\end{align}
The linear-order source correlation function and source power spectrum coincide with the standard linear SPT results:
\begin{align}
\xi_{\rm J,gg}^{\rm (lin)}(r) &= \xi_{\rm gg}^{\rm (lin)}(r) = b_1^2\, \xi_{\rm m}^{\rm (lin)}(r), \\
P_{\rm J,gg}^{\rm (lin)}(k) &= P_{\rm gg}^{\rm (lin)}(k) = b_1^2\, P_{\rm m}^{\rm (lin)}(k).
\end{align}

The DM contribution, as defined in Eq.~\eqref{eq:DM}, begins at one-loop order:
\begin{equation}
    P_{\rm DM,gg}(k) = P_{\rm DM,gg}^{\text{(1-loop)}}(k).
\end{equation}
Its explicit form is given by
\begin{align}
    P_{\rm DM,gg}^{\text{(1-loop)}}(k) & =
    \int d^3r\, e^{-i\boldsymbol{k}\cdot\boldsymbol{r}} \nonumber \\
    & \quad \times \left[ -\overline{\Sigma}^{\rm (lin)}(k) + \Sigma^{\rm (lin)}(k, r, \mu) \right]
    \xi_{\rm gg}^{\rm (lin)}(r),
    \label{eq:P_DM_1loop}
\end{align}
where \( \overline{\Sigma}^{\rm (lin)}(k) \equiv \Sigma^{\rm (lin)}(k, r{=}0, \mu) \) denotes the zero-separation limit of the displacement variance.

Combining this with the decomposition of the full power spectrum given in Eq.~\eqref{eq:DMgg}, we find that the one-loop source power spectrum satisfies
\begin{equation}
    P_{\rm J,gg}^{\text{(1-loop)}}(k) =
    P_{\rm gg}^{\text{(1-loop)}}(k) -
    P_{\rm DM,gg}^{\text{(1-loop)}}(k).
    \label{eq:PJ1loop}
\end{equation}

Thus, in practical implementations, the one-loop galaxy power spectrum in ULPT can be evaluated by computing only three quantities:
\begin{itemize}
    \item the linear displacement variance \( \Sigma^{\rm (lin)} \),
    \item the SPT one-loop galaxy power spectrum \( P_{\rm gg}^{\text{(1-loop)}} \), and
    \item the displacement-mapping correction \( P_{\rm DM,gg}^{\text{(1-loop)}} \).
\end{itemize}
There is no need to separately compute the one-loop source spectrum \( P_{\rm J,gg}^{\text{(1-loop)}} \), as it can be obtained via Eq.~\eqref{eq:PJ1loop}.

\subsection{Displacement-Mapping Factor}
\label{sec:DM}

We now present the explicit computation of the displacement-mapping factor within the linear approximation. Its exponent is given by
\begin{equation}
    \Sigma^{\rm (lin)}(k, r, \mu)
    = \int \frac{d^3p}{(2\pi)^3} e^{i\boldsymbol{p}\cdot\boldsymbol{r}}
    \left( \frac{\boldsymbol{k}\cdot\boldsymbol{p}}{p^2} \right)^2 P_{\rm m}^{\rm (lin)}(p),
    \label{eq:Sigma_lin}
\end{equation}
where \( \mu = \hat{\boldsymbol{k}} \cdot \hat{\boldsymbol{r}} \) denotes the cosine of the angle between the wavevector and the separation vector.

The angular integrals in Eq.~\eqref{eq:Sigma_lin} can be evaluated analytically, yielding
\begin{equation}
    \Sigma^{\rm (lin)}(k, r, \mu) = k^2 \left[ \sigma_0^2(r) + 2 \mathcal{L}_2(\mu) \sigma_2^2(r) \right],
\end{equation}
where \( \mathcal{L}_{\ell} \) is the Legendre polynomial of order \( \ell \), and the radial functions \( \sigma_\ell^2(r) \) are defined as
\begin{align}
    \sigma_{\ell}^2(r) = \frac{1}{3} i^{\ell} \int \frac{dk}{2\pi^2} j_{\ell}(kr)\, P_{\rm m}^{\rm (lin)}(k).
    \label{eq:sigma_l}
\end{align}

The zero-separation limit of the monopole term defines the linear displacement variance:
\begin{align}
    \bar{\sigma}^2 \equiv \sigma_0^2(0) = \frac{1}{3} \int \frac{dk}{2\pi^2} P_{\rm m}^{\rm (lin)}(k),
    \label{eq:sigma_bar}
\end{align}
which leads to the compact expression
\begin{align}
    \overline{\Sigma}^{\rm (lin)}(k) = k^2 \bar{\sigma}^2.
\end{align}

In numerical implementations, the evaluation of \( \sigma_\ell^2(r) \) at small separations becomes unstable when using FFTLog~\cite{Hamilton:1999uv,mcfit} due to the finite resolution and high-\(k\) cutoff of the linear matter power spectrum. In this work, we adopt \( P_{\rm m}^{\rm (lin)}(k) \) truncated at \( k_{\rm max} = 100\, h\,\mathrm{Mpc}^{-1} \), which induces artificial oscillations in \( \sigma_\ell^2(r) \) for \( r \lesssim 0.75\, h^{-1}\mathrm{Mpc} \).

To mitigate these numerical artifacts, we apply a smoothing procedure based on interpolation: for \( r < r_{\rm min} \), we interpolate smoothly between the analytic value at \( r = 0 \) and the numerically stable result at \( r = r_{\rm min} \), with \( r_{\rm min} = 0.75\, h^{-1}\mathrm{Mpc} \). A detailed discussion of this interpolation scheme in the context of the dark matter power spectrum can be found in Ref.~\cite{Sugiyama:2025myq}, where the same numerical settings and procedure were adopted. Since our primary analysis focuses on Fourier modes with \( k \lesssim 0.3\, h\,\mathrm{Mpc}^{-1} \), the impact of this small-scale interpolation on the final power spectrum is negligible.

\subsection{Convolution Integral}
\label{sec:Conv}

A central task in computing the ULPT galaxy power spectrum is the evaluation of convolution integrals that preserve the full exponential structure of the displacement-mapping factor. In Ref.~\cite{Sugiyama:2025myq}, we proposed an efficient and accurate approximation scheme based on this exponential representation.

The ULPT one-loop galaxy power spectrum is given by
\begin{align}
    P_{\rm gg}(k) &= 4\pi e^{-k^2\bar{\sigma}^2} \int dr\, r^2 \int \frac{d\mu}{2}\, e^{-ikr\mu} \nonumber \\
    &\quad \times e^{k^2\left[ \sigma_0^2(r) + 2 \mathcal{L}_2(\mu) \sigma_2^2(r) \right]} \xi_{\rm J,gg}(r),
\end{align}
where the source correlation function \( \xi_{\rm J,gg}(r) \) includes both linear and one-loop contributions, as defined in Eq.~\eqref{eq:SxiP}.

To isolate the convolution-induced corrections, we decompose the total power spectrum into a convolution-free (CF) term and a convolution-containing (CC) term:
\begin{equation}
    P_{\rm gg}(k) = P_{\rm CF,gg}(k) + P_{\rm CC,gg}(k),
\end{equation}
where the CF component is calculated as
\begin{align}
    P_{\rm CF,gg}(k) = e^{-k^2\bar{\sigma}^2} P_{\rm J,gg}(k),
\end{align}
and the CC component is given by
\begin{align}
    P_{\rm CC,gg}(k) &= 4\pi e^{-k^2\bar{\sigma}^2} \int dr\, r^2 \int \frac{d\mu}{2}\, e^{-ikr\mu} \nonumber \\
    &\quad \times \left[ e^{k^2\left[ \sigma_0^2(r) + 2 \mathcal{L}_2(\mu) \sigma_2^2(r) \right]} - 1 \right] \xi_{\rm J,gg}(r).
\end{align}
As noted in Eq.~\eqref{eq:SxiP}, the source power spectrum \( P_{\rm J,gg}(k) \) contains both the linear and one-loop contributions.

To accelerate the computation of \( P_{\rm CC,gg}(k) \), we expand the \( \mu \)-dependent exponential in Legendre polynomials and carry out the angular integration analytically:
\begin{equation}
    P_{\rm CC,gg}(k) = \sum_{n=0}^{\infty} P_{\rm CC,gg}^{[n]}(k),
\end{equation}
with the first few terms explicitly given by
\begin{align}
    P_{\rm CC,gg}^{[0]}(k) &= e^{-k^2 \bar{\sigma}^2} (4\pi) \int dr\, r^2 \left[ e^{k^2 \sigma_0^2(r)} - 1 \right] \nonumber \\
    &\quad \times j_0(kr) \xi_{\rm J,gg}(r), \\
    P_{\rm CC,gg}^{[n \geq 1]}(k) &= \frac{(2k^2)^n}{n!} e^{-k^2 \bar{\sigma}^2} (4\pi) \int dr\, r^2 e^{k^2 \sigma_0^2(r)} \nonumber \\
    &\quad \times I^{[n]}(k, r)\, \left[ \sigma_2^2(r) \right]^n \xi_{\rm J,gg}(r),
\end{align}
where the angular kernels \( I^{[n]}(k, r) \) are defined by
\begin{align}
    I^{[1]}(k, r) &= -j_2(kr), \\
    I^{[2]}(k, r) &= \frac{1}{5} j_0(kr) - \frac{2}{7} j_2(kr) + \frac{18}{35} j_4(kr), \\
    I^{[3]}(k, r) &= \frac{2}{35} j_0(kr) - \frac{291}{154} j_2(kr) + \frac{756}{2695} j_4(kr) - \frac{18}{77} j_6(kr).
\end{align}

Each term \( P_{\rm CC,gg}^{[n]}(k) \) corresponds to a Hankel-type transform and can be efficiently computed using FFT-based techniques. This expansion yields a numerically stable and computationally efficient method for evaluating displacement-induced convolution corrections in the ULPT power spectrum.

The accuracy and convergence of this expansion were quantitatively validated in Ref.~\cite{Sugiyama:2025myq} for the case of dark matter. At redshift \( z = 0 \), truncation at \( n = 3 \) achieves sub-percent precision across the range \( k \leq 0.4\, h\,\mathrm{Mpc}^{-1} \), with the maximum fractional error reaching only 0.025\%. To ensure comparable accuracy for biased tracers, we adopt the same truncation order \( n = 3 \) throughout this work.

Moreover, the convergence improves at higher redshifts. For instance, at \( z = 0.5 \), truncation at \( n = 2 \) already yields better than 0.4\% accuracy at \( k = 0.4\, h\,\mathrm{Mpc}^{-1} \), while at \( z = 1.0 \), the error further decreases to approximately 0.13\%. These results suggest that a truncation order of \( n = 2 \) is sufficient for most practical applications involving observational comparisons.

\subsection{Source Power Spectrum for Biased Tracers}
\label{sec:Source}

To complete the evaluation of the one-loop galaxy power spectrum in the ULPT framework, it remains to compute the one-loop source power spectrum and its Fourier counterpart, the source correlation function. We employ the \texttt{FAST-PT} algorithm~\cite{McEwen:2016fjn,Fang:2016wcf} to evaluate these quantities efficiently.

Following the decomposition in Eq.~\eqref{eq:P_DM_1loop}, the DM contribution is split into two components:
\begin{equation}
    P_{\rm DM,gg}^{(1\text{-loop})}(k) = P_{\rm DM,gg}^{(13)}(k) + P_{\rm DM,gg}^{(22)}(k),
\end{equation}
with
\begin{align}
    P_{\rm DM,gg}^{(13)}(k) &= -k^2 \bar{\sigma}^2 P_{\rm gg}^{\rm (lin)}(k), \\
    P_{\rm DM,gg}^{(22)}(k) &= \int \frac{d^3p}{(2\pi)^3}
    \left( \frac{\boldsymbol{k} \cdot \boldsymbol{p}}{p^2} \right)^2
    P_{\rm gg}^{\rm (lin)}(|\boldsymbol{k} - \boldsymbol{p}|)\, P_{\rm m}^{\rm (lin)}(p).
\end{align}

Since both terms depend solely on the linear galaxy power spectrum, they are unaffected by nonlinear bias contributions. Using \( P_{\rm gg}^{\rm (lin)}(k) = b_1^2\, P_{\rm m}^{\rm (lin)}(k) \), the DM terms can be rewritten as
\begin{align}
    P_{\rm DM,gg}^{(13)}(k) &= b_1^2\, P_{\rm DM,m}^{(13)}(k), \\
    P_{\rm DM,gg}^{(22)}(k) &= b_1^2\, P_{\rm DM,m}^{(22)}(k),
\end{align}
where
\begin{align}
    P_{\rm DM,m}^{(13)}(k) &= -k^2 \bar{\sigma}^2 P_{\rm m}^{\rm (lin)}(k), \\
    P_{\rm DM,m}^{(22)}(k) &= \int \frac{d^3p}{(2\pi)^3}
    \left( \frac{\boldsymbol{k} \cdot \boldsymbol{p}}{p^2} \right)^2
    P_{\rm m}^{\rm (lin)}(|\boldsymbol{k} - \boldsymbol{p}|)\, P_{\rm m}^{\rm (lin)}(p).
\end{align}

To obtain the one-loop source spectrum within ULPT, we subtract the DM contributions from the full SPT result:
\begin{align}
    P_{\rm J,gg}^{(13)}(k) &= P_{\rm gg}^{(13)}(k) - P_{\rm DM,gg}^{(13)}(k), \nonumber \\
    P_{\rm J,gg}^{(22)}(k) &= P_{\rm gg}^{(22)}(k) - P_{\rm DM,gg}^{(22)}(k).
\end{align}

The 13-type contribution is given by
\begin{equation}
    P_{\rm J,gg}^{(13)}(k) = b_1^2\, P_{\rm J,m}^{(13)}(k) + b_1\, b_{3}^{\rm u}\, P_{{\rm J}, b_1 b_3}^{(13)}(k),
\end{equation}
where each component is expressed as a single integral:
\begin{equation}
    P_{\rm J,X}^{(13)}(k) = \frac{k^3}{252(2\pi)^2} P_{\rm lin}(k) \int_0^\infty dr\, r^2\, Z_{{\rm J},X}(r)\, P_{\rm lin}(kr),
\end{equation}
with \( X = \{{\rm m}, b_1 b_3\} \). The kernel for the standard matter term~\cite{McEwen:2016fjn,Sugiyama:2025myq} is
\begin{align}
    Z_{\rm J,m}(r) &= \frac{12}{r^4} + \frac{10}{r^2} + 100 - 42r^2 \nonumber \\
    &\quad + \frac{3}{r^5}(7r^2 + 2)(r^2 - 1)^3 \ln\left| \frac{r+1}{r-1} \right|,
\end{align}
while the kernel for the nonlinear bias term is
\begin{align}
    Z_{ {\rm J},b_1 b_3}(r) &= \frac{12}{r^4} - \frac{44}{r^2} - 44 + 12r^2 \nonumber \\
    &\quad - \frac{6}{r^5}(r^2 - 1)^4 \ln\left| \frac{r+1}{r-1} \right|.
\end{align}

Although these integrals scale naively as \( \mathcal{O}(N^2) \), they can be efficiently recast as discrete convolutions using logarithmic variable transformations and evaluated in \( \mathcal{O}(N \log N) \) time via FFT-based methods implemented in \texttt{FAST-PT}.

The 22-type contribution is decomposed as
\begin{equation}
    P_{\rm J,gg}^{(22)}(k) = b_1^2\, P_{\rm J, m}^{(22)}(k) 
    + b_1 b_2^{\rm u}\, P_{{\rm J},b_1 b_2}^{(22)}(k) + (b_2^{\rm u})^2\, P_{{\rm J},b_2 b_2}^{(22)}(k),
\end{equation}
where each term is computed from its corresponding real-space correlation function:
\begin{equation}
    P_{{\rm J},X}^{(22)}(k) = 4\pi \int dr\, r^2\, j_0(kr)\, \xi_{{\rm J},X}^{(22)}(r),
\end{equation}
with \( X = \{{\rm m}, b_1 b_2, b_2 b_2\} \). The correlation function is expressed as~\cite{McEwen:2016fjn,Sugiyama:2025myq}
\begin{equation}
    \xi_{{\rm J},X}^{(22)}(r) = 2 \sum_{\alpha,\beta,\ell} c_{X,\alpha\beta\ell}\, J_{\alpha\beta\ell}(r),
\end{equation}
where
\begin{align}
    J_{\alpha\beta\ell}(r) &= \left[ i^{\ell} \int \frac{dk_1\, k_1^2}{2\pi^2}\, k_1^\alpha P^{(\rm lin)}_{\rm m}(k_1) j_\ell(k_1 r) \right] \nonumber \\
    &\quad \times \left[ i^{\ell} \int \frac{dk_2\, k_2^2}{2\pi^2}\, k_2^\beta P^{(\rm lin)}_{\rm m}(k_2) j_\ell(k_2 r) \right].
\end{align}
These nested Hankel transforms are evaluated using FFT-based techniques. The coefficients \( c_{X,\alpha\beta\ell} \) are listed in Table~\ref{tab:c_alpha_beta_l}.

To summarize, the source power spectrum for biased tracers in ULPT is given by
\begin{align}
    P_{\rm J}(k) &= b_1^2\, P_{\rm J,m}(k) + b_1 b_{3}^{\rm u}\, P_{{\rm J},b_1 b_3}^{(13)}(k) 
    \nonumber \\
    &\quad + b_1 b_2^{\rm u}\, P_{{\rm J},b_1 b_2}^{(22)}(k) + (b_2^{\rm u})^2\, P_{{\rm J},b_2 b_2}^{(22)}(k),
    \label{Eq:P_J}
\end{align}
where the matter source spectrum \( P_{\rm J,m}(k) \) includes linear and one-loop terms:
\begin{equation}
    P_{\rm J,m}(k) = P^{({\rm lin})}_{\rm m}(k) + P^{(13)}_{\rm J,m}(k) + P^{(22)}_{\rm J,m}(k).
\end{equation}

\renewcommand{\arraystretch}{1.6}
\begin{table}[!htbp]
    \centering
    \caption{
Coefficients \( c_{X,\alpha\beta\ell} \) used in the 22-type source correlation function \( \xi_{{\rm J},X}^{(22)}(r) \). Each coefficient controls the contribution of the corresponding term \( J_{\alpha\beta\ell}(r) \), which is constructed from Hankel transforms weighted by powers of the wavenumbers. The index \( X = \{{\rm m}, b_1 b_2, b_2 b_2\} \) labels the bias operator combinations.
}
\label{tab:c_alpha_beta_l}
    \begin{tabular}{ccc|ccc}
        \hline
        $\alpha$ & $\beta$ & $\ell$ & $c_{{\rm m}}$ & $c_{b_1b_2}$ & $c_{b_2b_2}$\\
        \hline
        0 & 0 & 0  & $\frac{242}{735}$  &  $- \frac{72}{245}$ &  $ \frac{32}{735}$\\
        0 & 0 & 2  & $\frac{671}{1029}$ &  $\frac{88}{343}$   &  $- \frac{64}{1029}$\\
        0 & 0 & 4  & $\frac{32}{1715}$  &  $\frac{64}{1715}$  &  $\frac{32}{1715}$\\
        1 & -1 & 1 & $\frac{27}{35}$    &  $- \frac{8}{35}$   &  0\\
        1 & -1 & 3 & $\frac{8}{35}$     &  $\frac{8}{35}$     &  0\\
        \hline
    \end{tabular}
\end{table}

\subsection{Stochastic Contributions}
\label{sec:stochastic}

In this subsection, we evaluate the contribution of stochastic bias to the galaxy power spectrum within the ULPT framework. The stochastic bias field, denoted by \( \varepsilon(\boldsymbol{q}) \), is assumed to be statistically uncorrelated with both the Jacobian deviation \( \delta_{\rm J} \) and the displacement field \( \boldsymbol{\Psi} \), and to have zero ensemble average, as expressed in Eqs.~\eqref{eq:e0} and \eqref{eq:ejp0}. Under these assumptions, the stochastic component contributes only to the galaxy auto power spectrum and does not affect the galaxy--matter cross power spectrum.

The stochastic contribution to the galaxy source correlation function is given by
\begin{align}
    \xi_{{\rm J},\varepsilon}(\boldsymbol{r}) &= \langle \varepsilon(\boldsymbol{q}) \varepsilon(\boldsymbol{q}') \rangle \nonumber \\
    &= N_{\varepsilon} \delta_{\rm D}(\boldsymbol{r}),
    \label{eq:Je}
\end{align}
where \( \boldsymbol{r} = \boldsymbol{q} - \boldsymbol{q}' \), and \( N_{\varepsilon} \) is the amplitude of the stochastic bias field defined in Eq.~\eqref{eq:e}.

Substituting this result into the general ULPT expression for the galaxy power spectrum [Eq.~\eqref{eq:P_ULPT}], we obtain
\begin{align}
    P_{\varepsilon}(\boldsymbol{k}) & =
    e^{-\overline{\Sigma}(\boldsymbol{k})}
    \int d^3r\, e^{-i\boldsymbol{k}\cdot\boldsymbol{r}} e^{\Sigma(\boldsymbol{k}, \boldsymbol{r})}
    N_{\varepsilon} \delta_{\rm D}(\boldsymbol{r}) \nonumber \\
    & = e^{-\overline{\Sigma}(\boldsymbol{k})} e^{\Sigma(\boldsymbol{k}, \boldsymbol{r} = \boldsymbol{0})}
    N_{\varepsilon} \nonumber \\
    & = N_{\varepsilon},
    \label{eq:P_ULPT_e}
\end{align}
where we have used the identity \( \overline{\Sigma}(\boldsymbol{k}) = \Sigma(\boldsymbol{k}, \boldsymbol{r} = \boldsymbol{0}) \).

As a result, the stochastic contribution to the galaxy auto power spectrum reduces to a constant \( N_{\varepsilon} \), independent of the displacement-mapping factor. Although the stochastic field $\varepsilon(\boldsymbol{q})$ is defined in Lagrangian space, this behavior is fully consistent with the standard treatment of stochastic terms in Eulerian space, where such contributions are typically modeled as scale-independent white-noise components characterized solely by a constant amplitude~\cite{Dekel:1998eq,Desjacques:2016bnm}.

\subsection{Galaxy Auto and Galaxy--Matter Cross Power Spectra}
\label{sec:ULPT_P}

In this subsection, we present the final expressions for the galaxy auto power spectrum and the galaxy--matter cross power spectrum at one-loop order within the ULPT framework.

Each constituent term in the galaxy power spectrum is computed using the following expression:
\begin{align}
    P_{\rm X}(k) &= 4\pi\, e^{-k^2\bar{\sigma}^2} \int dr\, r^2 \int \frac{d\mu}{2}\, e^{-ikr\mu} \nonumber \\
    &\quad \times e^{k^2\left[ \sigma_0^2(r) + 2 \mathcal{L}_2(\mu)\, \sigma_2^2(r) \right]} \, \xi_{{\rm J},X}(r),
\end{align}
where \( X = \{ {\rm m}, b_1 b_2, b_2 b_2, b_1 b_3, \varepsilon \} \). For \( X = \{ {\rm m}, b_1 b_2, b_2 b_2, b_1 b_3 \} \), the source correlation functions \( \xi_{{\rm J},X}(r) \) are the inverse Fourier transforms of the corresponding terms defined in Eq.~\eqref{Eq:P_J}, while \( \xi_{{\rm J},\varepsilon}(r) \) is given by Eq.~\eqref{eq:Je}.

By collecting all relevant contributions, the one-loop galaxy auto power spectrum is given by
\begin{align}
    P_{\rm gg}(k)
    &= b_1^2\, P_{\rm m}(k)
    + b_1 b_2^{\rm u}\, P_{b_1 b_2}(k) \nonumber \\
    &\quad + (b_2^{\rm u})^2\, P_{b_2 b_2}(k)
    + b_1 b_{3}^{\rm u}\, P_{b_1 b_3}(k)
    + N_{\varepsilon}.
\end{align}

The galaxy--matter cross power spectrum is defined as
\begin{align}
    P_{\rm gm}(\boldsymbol{k})
    &= \int d^3r\, e^{-i\boldsymbol{k}\cdot\boldsymbol{r}} \left\langle
    e^{-i\boldsymbol{k}\cdot[\boldsymbol{\Psi}(\boldsymbol{q}) - \boldsymbol{\Psi}(\boldsymbol{q}')]} \right. \nonumber \\
    &\quad \left. \times
    \left[\delta_{\rm J}(\boldsymbol{q}) + \delta_{\rm b}(\boldsymbol{q})\right] \delta_{\rm J}(\boldsymbol{q}')
    \right\rangle.
\end{align}
This expression shares the same structural form as the auto spectrum, but involves only one biased density field. Proceeding analogously, we obtain the corresponding one-loop galaxy--matter cross power spectrum:
\begin{align}
    P_{\rm gm}(k)
    = b_1\, P_{\rm m}(k)
    + \frac{1}{2} b_2^{\rm u}\, P_{b_1 b_2}(k)
    + \frac{1}{2} b_{3}^{\rm u}\, P_{b_1 b_3}(k).
\end{align}

\begin{figure}[!t]
    \centering
    \includegraphics[width=\columnwidth]{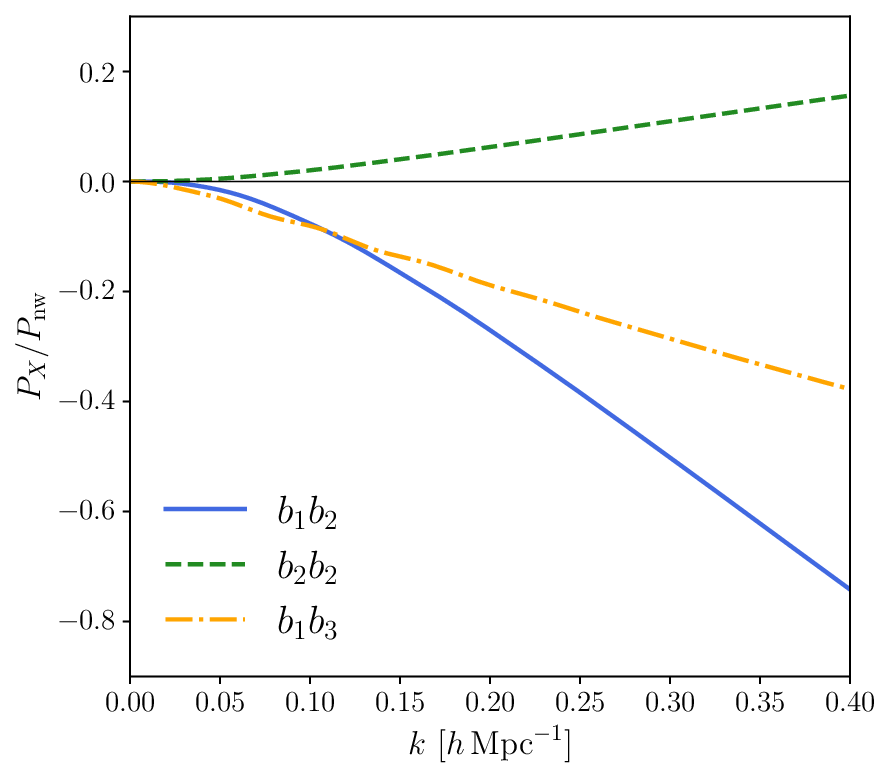}
\caption{
Nonlinear bias correction terms \( P_X \) for \( X = \{ b_1 b_2,\, b_2 b_2,\, b_1 b_3 \} \) at redshift \( z = 0 \), normalized by the no-wiggle linear matter power spectrum \( P_{\rm nw}(k) \). All correction terms smoothly approach zero in the large-scale limit \( k \to 0 \), as expected for physically consistent nonlinear contributions. This infrared behavior highlights the renormalization-free nature of the ULPT bias expansion.
}
    \label{fig:P_ULPT_bias}
\end{figure}

Figure~\ref{fig:P_ULPT_bias} illustrates the nonlinear bias correction terms \( P_X \) for \( X = \{ b_1 b_2,\, b_2 b_2,\, b_1 b_3 \} \) at redshift \( z = 0 \), normalized by the no-wiggle linear matter power spectrum \( P_{\rm nw}(k) \) that excludes baryon acoustic oscillations (BAO)~\cite{Eisenstein:1997ik,Hamann:2010pw,Chudaykin:2020aoj}.

We observe that the \( b_1 b_2 \) and \( b_1 b_3 \) terms are negative across the full \( k \)-range shown, while the \( b_2 b_2 \) term remains strictly positive. However, since the nonlinear bias parameters \( b_2^{\rm u} \) and \( b_3^{\rm u} \) can take either sign, the net contribution of these terms to the galaxy power spectrum depends on the specific parameter values.

A particularly important feature is that all correction terms smoothly approach zero in the large-scale limit \( k \to 0 \), as expected for nonlinear contributions. This infrared behavior reflects a key aspect of the ULPT framework: as shown in Eqs.~\eqref{eq:delta_g_field} and~\eqref{eq:delta_g_ave}, the galaxy density contrast is constructed to be statistically consistent at the field level, ensuring both vanishing ensemble and volume averages. Consequently, no renormalization procedure is required to absorb constant offsets in the power spectrum.

\section{Dark Emulator}
\label{sec:DarkEmu}

\begin{table}[!htbp]
\centering
\caption{Parameter ranges for \textit{Dark Emulator}}
\begin{tabular}{lc}
\hline
$\omega_b$ & $[0.02114, 0.02336]$ \\
$\omega_c$ & $[0.10782, 0.13178]$ \\
$\Omega_{\Lambda}$ & $[0.54752, 0.82128]$ \\
$\ln(10^{10} A_s)$ & $[2.4752, 3.7128]$ \\
$n_s$ & $[0.9163, 1.0127]$ \\
$w$ & $[-1.2, -0.8]$ \\
$\sum m_\nu$ [eV] & fixed (0.06 eV) \\
\hline
\end{tabular}
\label{tab:param_ranges}
\end{table}

For comparison with our ULPT predictions, we employ the public version of \textit{Dark Emulator}~\cite{Nishimichi:2018etk}, which provides real-space halo--halo (\( P_{\rm hh} \)) and halo--matter (\( P_{\rm hm} \)) power spectra, along with their corresponding two-point correlation functions: the halo--halo auto-correlation function and the halo--matter cross-correlation function. These quantities are available across a broad range of cosmological models. The parameter ranges covered by the emulator are summarized in Tab.~\ref{tab:param_ranges}.

The emulator is constructed from a suite of $N$-body simulations covering 101 flat \( w \)CDM cosmologies, each simulated with a single realization. The only exception is the fiducial cosmology, which is consistent with the Planck 2015 best-fit $\Lambda$CDM parameters~\cite{Planck:2015fie} and is realized with multiple simulations to suppress sample variance.

Specifically, the halo--halo power spectrum \( P_{\rm hh} \) is derived from simulations with a box size of \( (2\, h^{-1} \mathrm{Gpc})^3 \), while the halo--matter cross power spectrum \( P_{\rm hm} \) is obtained from simulations with \( (1\, h^{-1} \mathrm{Gpc})^3 \) volumes. For the fiducial cosmology, 14 realizations are used to compute \( P_{\rm hh} \), and 28 realizations for \( P_{\rm hm} \).

The emulator provides halo clustering statistics for halos with masses in the range \( 10^{12} \lesssim M \lesssim 10^{14}\, h^{-1} M_\odot \), evaluated over 21 redshift snapshots from \( z = 0 \) to \( z = 1.48 \). These outputs serve as a benchmark for validating theoretical models of halo bias and nonlinear clustering on large scales.

\section{Validation of ULPT Predictions with Dark Emulator}
\label{sec:vs_emu}

In this section, we assess the predictive accuracy of the ULPT framework by comparing its one-loop results with simulation-based outputs from \textit{Dark Emulator}. The comparison allows us to evaluate its performance across a wide range of mass scales, redshifts, and cosmological parameter choices.

Section~\ref{sec:analysis} outlines the setup of our comparison, including the definition of the target observables and the methodology used to estimate bias parameters. In Section~\ref{sec:fid}, we focus on the fiducial cosmology consistent with the Planck 2015 $\Lambda$CDM parameters, for which multiple realizations are available in the emulator. This reduces statistical fluctuations in the emulator outputs and enables a more stable and reliable comparison with the ULPT predictions. We then extend our analysis in Section~\ref{sec:100} to a suite of 100 cosmological models randomly sampled from the emulator's six-dimensional parameter space, in order to test the robustness of our model across broader cosmological variations.

\subsection{Analysis Methodology}
\label{sec:analysis}

\subsubsection{Observables and Fitting Strategy}

\begin{table}[t]
\centering
\caption{Linear Eulerian bias parameter \( b_1 \) predicted by the \textit{Dark Emulator} for the redshift and halo mass bins considered in this work. These values are used as reference benchmarks, while \( b_1 \) is treated as a free parameter in the fitting analysis.}
\label{tab:linear_bias_emu}
\begin{tabular}{cc|@{\hspace{12pt}}c}
\hline\hline
$z$ & $\log_{10}(M/M_\odot)$ 
& $b_1$\\
\hline
0.0 & 12.5 & $0.83$  \\
0.0 & 13.0 & $1.02$  \\
0.0 & 13.5 & $1.34$  \\
0.5 & 12.5 & $1.13$  \\
0.5 & 13.0 & $1.45$  \\
0.5 & 13.5 & $2.04$  \\
1.0 & 12.5 & $1.59$  \\
1.0 & 13.0 & $2.10$  \\
1.0 & 13.5 & $3.08$  \\
\hline\hline
\end{tabular}
\end{table}

In this work, we jointly fit the halo--halo auto power spectrum and the halo--matter cross power spectrum provided by \textit{Dark Emulator}, adopting the one-loop ULPT predictions as our theoretical framework. Since the bias model employed in ULPT is defined at the field level, it should, in principle, be applicable to any statistical measure derived from the underlying density field. To test this consistency explicitly, we simultaneously fit $P_{\rm hh}$ and $P_{\rm hm}$, which exhibit distinct dependences on the bias parameters. This joint analysis plays a critical role in validating the field-level formulation of the bias expansion.

In addition to the power spectra, we perform the same analysis using the halo--halo auto correlation function and the halo--matter cross correlation function, thereby further assessing the consistency of ULPT across both configuration and Fourier space.

The comparison is carried out at three redshifts, \( z = 0.0,\, 0.5,\, \text{and } 1.0 \), and for three halo mass bins at each redshift, defined by \( \log_{10}(M/M_\odot) = 12.5,\, 13.0,\, \text{and } 13.5 \). This yields a total of nine halo samples, spanning a representative range in both redshift and mass. These combinations enable a systematic investigation of the redshift evolution and mass dependence of the halo bias, as well as the predictive accuracy of the ULPT model across different halo populations.

As a reference, Table~\ref{tab:linear_bias_emu} presents the values of the linear Eulerian bias parameter \( b_1 \) predicted by the \textit{Dark Emulator} for the redshift and halo mass bins considered in this work. While \( b_1 \) is treated as a free parameter in our fitting analysis, these values serve as useful benchmarks for the expected bias amplitude across the sampled halo populations. As shown in the table, the emulator predicts \( b_1 \) values ranging from approximately $0.8$ to $3.0$, reflecting the wide range covered by our selection of redshifts and halo masses.

\subsubsection{Covariance Matrix}
\label{sec:cov}

In this analysis, we consider only the Gaussian contributions to the covariance matrix.

For the halo--halo auto power spectrum \( P_{\rm hh}(k) \), the diagonal Gaussian covariance is given by
\begin{equation}
    \mathrm{Cov}\left[ P_{\rm hh}(k), P_{\rm hh}(k') \right]
    = \frac{2}{N_{\rm hh}(k)}\, \delta^{\rm K}_{k,k'} \left[ P_{\rm hh}(k) + \frac{1}{\bar{n}} \right]^2,
\end{equation}
where \( N_{\rm hh}(k) = 4\pi k^2 \Delta k V_{\rm hh}/(2\pi)^3 \) is the number of modes in a shell of width \( \Delta k = 0.01\, h\, \mathrm{Mpc}^{-1} \), and \( V_{\rm hh} = (2\, h^{-1} \mathrm{Gpc})^3 \) is the simulation volume. The symbol \( \delta^{\rm K}_{k,k'} \) denotes the Kronecker delta, which equals unity when \( k = k' \) and vanishes otherwise. The parameter \( \bar{n} \) denotes the mean halo number density. As the \textit{Dark Emulator} does not provide exact values of \( \bar{n} \) for each mass bin, we adopt a fiducial value of \( \bar{n} = 10^{-4}\, (h^{-1} \mathrm{Mpc})^{-3} \), which lies within the emulator's validated range~\cite{Nishimichi:2018etk}.

For the halo--matter cross power spectrum \( P_{\rm hm}(k) \), the Gaussian covariance takes the form
\begin{align}
    &\mathrm{Cov}\left[ P_{\rm hm}(k), P_{\rm hm}(k') \right] \nonumber \\
    &\quad = \frac{1}{N_{\rm hm}(k)}\, \delta^{\rm K}_{k,k'} \left\{
        \left[ P_{\rm hh}(k) + \frac{1}{\bar{n}} \right] P_{\rm m}(k)
        + \left[ P_{\rm hm}(k) \right]^2
    \right\},
\end{align}
where \( N_{\rm hm}(k) = 4\pi k^2 \Delta k V_{\rm hm}/(2\pi)^3 \) and \( V_{\rm hm} = (1\, h^{-1} \mathrm{Gpc})^3 \).

For simplicity, we neglect the cross-covariance between \( P_{\rm hh} \) and \( P_{\rm hm} \), assuming the two to be uncorrelated. All required power spectra are computed from the \textit{Dark Emulator} outputs.

For the correlation functions, the corresponding covariance matrices are computed as Hankel transforms of the power spectra. For the halo--halo case, the Gaussian covariance is given by
\begin{align}
    \mathrm{Cov}\left[ \xi_{\rm hh}(r), \xi_{\rm hh}(r') \right]
    & = \frac{2}{V_{\rm hh}} \int \frac{dk\, k^2}{2\pi^2} j_0(kr)\, j_0(kr') \nonumber \\
    & \quad \times
    \left\{ \left[ P_{\rm hh}(k) \right]^2 + \frac{2}{\bar{n}}\, P_{\rm hh}(k) \right\} \nonumber \\
    & \quad + \frac{2}{V_{\rm hh}} \frac{\delta^{\rm K}_{r,r'}}{4\pi r^2 \Delta r}
    \left( \frac{1}{\bar{n}} \right)^2,
\end{align}
where \( \Delta r = 5\, h^{-1}\, \mathrm{Mpc} \) denotes the bin width in configuration space. The Kronecker delta \( \delta^{\rm K}_{r,r'} \) is unity when \( r = r' \) and vanishes otherwise. 
For the halo--matter case:
\begin{align}
    \mathrm{Cov}\left[ \xi_{\rm hm}(r), \xi_{\rm hm}(r') \right]
    & = \frac{1}{V_{\rm hm}} \int \frac{dk\, k^2}{2\pi^2} j_0(kr)\, j_0(kr') \nonumber \\
    & \quad \times
    \left\{
        \left[ P_{\rm hh}(k) + \frac{1}{\bar{n}} \right] P_{\rm m}(k)
        + \left[P_{\rm hm}(k)\right]^2
    \right\}.
\end{align}
Even under the Gaussian assumption, these correlation-function covariances exhibit non-negligible off-diagonal correlations between radial bins.

We note, however, that our analysis neglects non-Gaussian contributions, particularly those arising from the connected four-point function (trispectrum). As a result, the statistical uncertainties derived from this covariance model are likely to be underestimated. In particular, previous studies have shown that the cumulative signal-to-noise ratio of the power spectrum tends to saturate at \( k \gtrsim 0.3\, h\, \mathrm{Mpc}^{-1} \) due to non-Gaussian effects~\cite{Takahashi:2009bq}, suggesting that the neglected trispectrum terms may be relevant for the scales probed in this work.

Additional uncertainties arise from the choice of \( \bar{n} \), which directly enters the covariance but cannot be accurately determined from the emulator outputs for each halo sample. Variations in \( \bar{n} \) can lead to systematic shifts in the estimated error amplitudes.

Moreover, the omission of the cross-covariance between \( P_{\rm hh} \) and \( P_{\rm hm} \) may affect the joint constraints on bias parameters, particularly in simultaneous fits involving both statistics.

Despite these simplifications, we emphasize that the main purpose of this work is to determine whether a common set of ULPT bias parameters can accurately describe the \textit{Dark Emulator} predictions across redshift and mass bins. A comprehensive treatment of statistical errors, including non-Gaussian contributions, sample variance, and cross-covariances, is beyond the scope of the present study.

\subsubsection{Estimation of Bias Parameters from MCMC Analysis}

We estimate the bias parameters by performing a Markov Chain Monte Carlo (MCMC) analysis using the publicly available \texttt{MontePython}~\cite{Brinckmann:2018cvx} code.

The fitting procedure is based on minimizing the standard chi-squared statistic,
\begin{equation}
    \chi^2 = \left( \mathbf{D} - \mathbf{M} \right)^{\rm T} \mathbf{C}^{-1} \left( \mathbf{D} - \mathbf{M} \right),
    \label{eq:chi2_def}
\end{equation}
where \( \mathbf{D} \) is the data vector from \textit{Dark Emulator}, \( \mathbf{M} \) is the ULPT prediction, and \( \mathbf{C} \) is the covariance matrix.

For joint fits to both the halo--halo and halo--matter statistics, we define the total chi-squared as
\begin{equation}
    \chi^2_{\rm total} = \chi^2_{\rm hh} + \chi^2_{\rm hm}.
    \label{eq:chi2_total_def}
\end{equation}

Throughout the analysis, the cosmological parameters are fixed to those adopted in the emulator. Only the bias parameters defined within the ULPT framework are allowed to vary. For power-spectrum-based analyses, the free parameters include the Eulerian linear bias \( b_1 \), which is related to the ULPT parameter via \( b_1 = 1 + b_1^{\rm u} \), the second- and third-order nonlinear bias parameters \( b_2^{\rm u} \) and \( b_3^{\rm u} \), and the stochastic amplitude \( N_{\varepsilon} \), yielding a total of four parameters.

In contrast, for analyses based on two-point correlation functions, the constant stochastic contribution \( N_{\varepsilon} \) does not contribute to the signal. In this case, the parameter space is reduced to three: \( b_1 \), \( b_2^{\rm u} \), and \( b_3^{\rm u} \).

\subsection{Fiducial Cosmology}
\label{sec:fid}

\subsubsection{Power Spectrum Comparison}
\label{sec:fid_pk}

\begin{figure*}[!h]
    \centering
    \includegraphics[width=\textwidth]{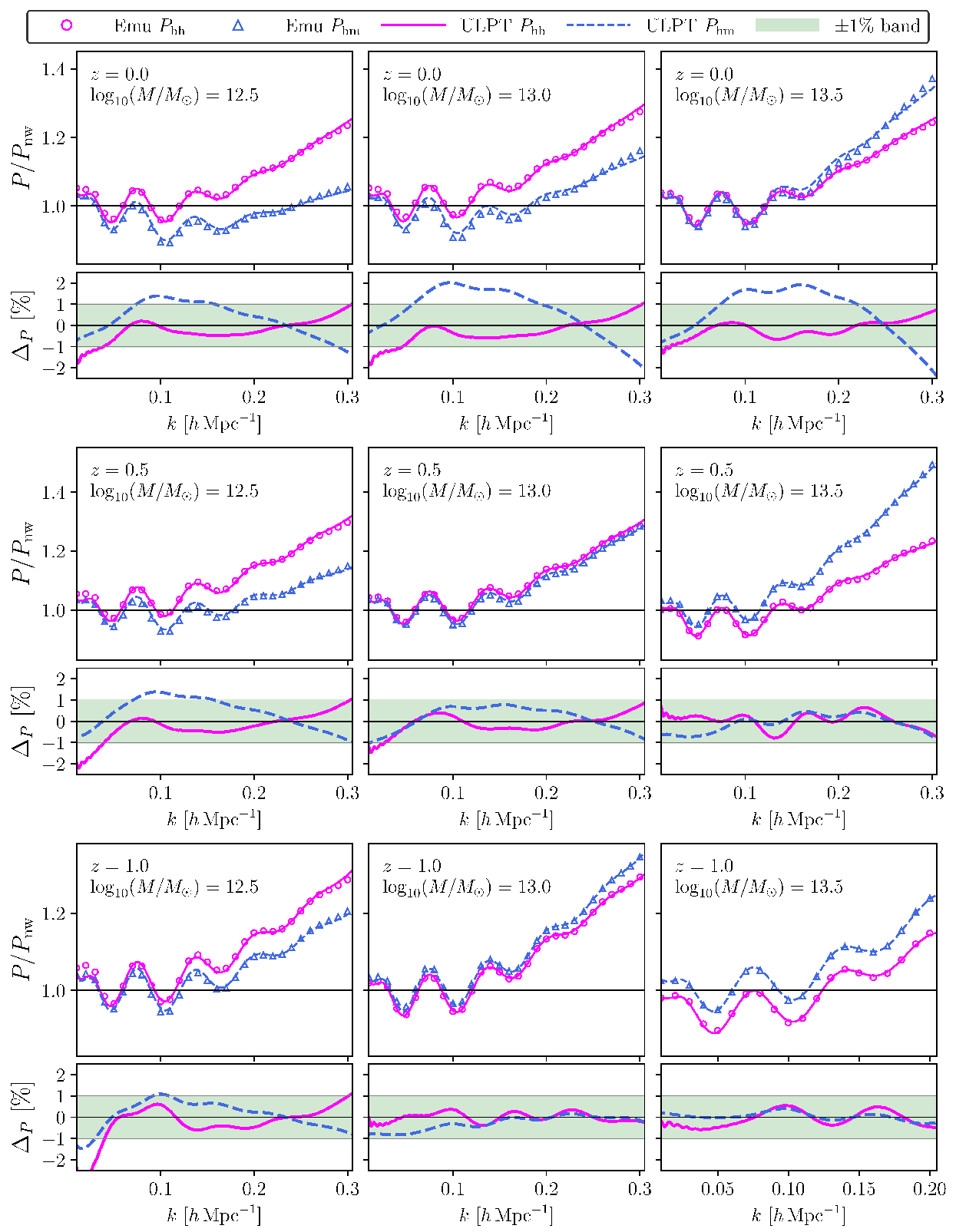}
\caption{
Comparison between the ULPT predictions and \textit{Dark Emulator} outputs for the halo--halo auto power spectrum \( P_{\rm hh}(k) \) and the halo--matter cross power spectrum \( P_{\rm hm}(k) \), shown across all redshift and halo mass bins. In each panel, the upper sub-panel displays the ratios \( P_{\rm hh}/(b_1^2 P_{\rm nw}) \) and \( P_{\rm hm}/(b_1 P_{\rm nw}) \), where \( P_{\rm nw} \) denotes the no-wiggle linear matter power spectrum. The lower sub-panel shows the relative deviation between the ULPT and emulator predictions, defined as \( \Delta_{P}[\%] = 100 \times (P_{\rm ULPT} - P_{\rm Emu})/P_{\rm Emu} \). Magenta and blue lines denote \( P_{\rm hh} \) and \( P_{\rm hm} \), respectively; solid and dashed curves indicate ULPT fits, while points represent emulator data. In nearly all cases, ULPT achieves better than 1\% accuracy up to \( k = 0.3\, h\,\mathrm{Mpc}^{-1} \), except for the highest redshift and mass bin (\( z = 1.0 \), \( \log_{10}(M/M_\odot) = 13.5 \)), where sub-percent agreement extends up to \( k = 0.2\, h\,\mathrm{Mpc}^{-1} \).
}
    \label{fig:Phh_Phm}
\end{figure*}

\begin{table*}[t]
\centering
\caption{
Mean values and marginalized 1$\sigma$ credible intervals for the bias parameters \( (b_1,\, b_2^{\rm u},\, b_3^{\rm u}) \) and the stochastic amplitude \( N_{\varepsilon} \), obtained from the joint fit to the halo--halo and halo--matter power spectra at each redshift \( z \) and halo mass \( \log_{10}(M/M_\odot) \). Also listed are the reduced minimum chi-squared values \( \chi^2_{\rm min}/\mathrm{dof} \). The notation \( x^{+\Delta_+}_{-\Delta_-} \) denotes the 68\% credible interval around the mean, and numbers in parentheses indicate the best-fit values of each parameter. The fitting range is \( 0.01 \leq k \leq 0.3\, h\,\mathrm{Mpc}^{-1} \), except for the highest redshift and mass bin (\( z = 1.0 \), \( \log_{10}(M/M_\odot) = 13.5 \)), where \( k_{\rm max} = 0.2\, h\,\mathrm{Mpc}^{-1} \) is adopted.
}
\label{tab:bias_summary}
\begin{tabular}{cc|@{\hspace{12pt}}r@{\hspace{15pt}}r@{\hspace{15pt}}r@{\hspace{15pt}}r@{\hspace{15pt}}c}
\hline\hline
$z$ & $\log_{10}(M/M_\odot)$ 
& \multicolumn{1}{c}{$b_1$} & \multicolumn{1}{c}{$b^{\rm u}_2$} &  \multicolumn{1}{c}{$b^{\rm u}_3$} & \multicolumn{1}{c}{$N_{\varepsilon}$} & $\chi^2_{\rm min}/{\rm dof}$ \\
\hline
0.0 & 12.5 & $0.83\,(0.83)^{+0.007}_{-0.007}$ & $1.74\, (1.74)^{+0.11}_{-0.11}$ & $-0.16\, (-0.17)^{+0.23}_{-0.24}$ & $139.49\,(137.84)^{+47.19}_{-48.56}$ & 0.06 \\
0.0 & 13.0 & $1.02\, (1.02)^{+0.007}_{-0.007}$ & $1.91\, (1.91)^{+0.11}_{-0.11}$ & $-0.47\, (-0.51)^{+0.23}_{-0.23}$ & $158.91\,(153.47)^{+53.13}_{-48.46}$ & 0.19 \\
0.0 & 13.5 & $1.34\, (1.34)^{+0.008}_{-0.008}$ & $1.33\, (1.32)^{+0.13}_{-0.13}$ & $-0.47\, (-0.45)^{+0.28}_{-0.28}$ & $-0.30\, (5.09)^{+45.02}_{-41.82}$ & 0.34 \\
0.5 & 12.5 & $1.11\, (1.11)^{+0.008}_{-0.008}$ & $2.20\,(2.23)^{+0.26}_{-0.24}$ & $-0.80\,(-0.85)^{+0.49}_{-0.54}$ & $97.29\,(92.19)^{+51.95}_{-47.30}$ & 0.07 \\
0.5 & 13.0 & $1.44\,(1.44)^{+0.010}_{-0.009}$ & $1.47\,(1.44)^{+0.30}_{-0.28}$ & $-0.82\,(-0.77)^{+0.59}_{-0.60}$ & $63.72\,(69.59)^{+43.16}_{-33.82}$ & 0.05 \\
0.5 & 13.5 & $2.04\,(2.04)^{+0.016}_{-0.017}$ & $-2.45\,(-2.44)^{+0.58}_{-0.59}$ & $2.17\,(2.17)^{+1.15}_{-1.20}$ & $-1032.85\,(-1019.52)^{+150.24}_{-100.14}$ & $0.08$ \\
1.0 & 12.5 & $1.54\,(1.54)^{+0.012}_{-0.013}$ & $1.42\,(1.59)^{+0.74}_{-0.58}$ & $-0.63\,(-0.93)^{+1.19}_{-1.42}$ & $99.84\,(100.84)^{+40.77}_{-25.03}$ & $0.06$ \\
1.0 & 13.0 & $2.10\,(2.10)^{+0.018}_{-0.018}$ & $-2.73\,(-2.61)^{+1.03}_{-1.01}$ & $2.14\,(1.90)^{+1.92}_{-2.19}$ & $-323.66\,(-295.17)^{+126.55}_{-58.45}$ & $0.02$ \\
1.0 & 13.5 & $3.10\,(3.10)^{+0.037}_{-0.047}$ & $-11.23\,(-11.36)^{+2.61}_{-2.07}$ & $6.41\,(6.71)^{+3.80}_{-5.00}$ & $-3818.06\,(-3780.50)^{+1295.61}_{-637.34}$ & $0.06$ \\
\hline\hline
\end{tabular}
\end{table*}

\begin{table*}[t]
\centering
\caption{
Mean values and marginalized 1$\sigma$ credible intervals for the bias parameters \( (b_1,\, b_2^{\rm u},\, b_3^{\rm u}) \), obtained from the joint fit to the halo--halo and halo--matter correlation functions \( \xi_{\rm hh}(r) \) and \( \xi_{\rm hm}(r) \) at each redshift \( z \) and halo mass \( \log_{10}(M/M_\odot) \). The reduced minimum chi-squared values \( \chi^2_{\rm min}/\mathrm{dof} \) are also listed. The notation \( x^{+\Delta_+}_{-\Delta_-} \) indicates the 68\% credible interval around the mean. The numbers in parentheses indicate the best-fit values of each parameter. The fitting range is fixed to \( 15 \leq r \leq 200\, h^{-1} \mathrm{Mpc} \) for all cases. The results are found to be in full agreement with those from the power spectrum analysis shown in Table~\ref{tab:bias_summary}, within the 1$\sigma$ credible intervals.
}
\label{tab:bias_summary_xi}
\begin{tabular}{cc|@{\hspace{12pt}}r@{\hspace{15pt}}r@{\hspace{15pt}}r@{\hspace{15pt}}c}
\hline\hline
$z$ & $\log_{10}(M/M_\odot)$ 
& \multicolumn{1}{c}{$b_1$} & \multicolumn{1}{c}{$b_2^{\rm u}$} & \multicolumn{1}{c}{$b_3^{\rm u}$} 
& $\chi_{\rm min}^2/{\rm dof}$ \\
\hline
0.0 & 12.5 & $0.85\,(0.84)^{+0.01}_{-0.02}$ & $0.88\,(1.21)^{+1.29}_{-1.17}$ & $0.04\,(-0.09)^{+0.51}_{-0.55}$ & $0.01$\\
0.0 & 13.0 & $1.04\,(1.03)^{+0.01}_{-0.02}$ & $0.98\,(0.92)^{+1.19}_{-1.20}$ & $-0.25\,(-0.40)^{+0.47}_{-0.52}$ & $0.02$\\
0.0 & 13.5 & $1.35\,(1.35)^{+0.01}_{-0.03}$ & $1.01\,(0.67)^{+1.29}_{-1.58}$ & $-0.28\,(-0.44)^{+0.47}_{-0.53}$ & $0.02$\\
0.5 & 12.5 & $1.13\,(1.13)^{+0.01}_{-0.03}$ & $0.56\,(0.40)^{+1.90}_{-2.11}$ & $-0.32\,(-0.66)^{+1.12}_{-1.16}$ & $0.01$\\
0.5 & 13.0 & $1.45\,(1.46)^{+0.02}_{-0.03}$ & $0.61\,(-0.43)^{+2.24}_{-2.91}$ & $-0.07\,(-0.44)^{+1.08}_{-1.16}$ & $0.01$\\
0.5 & 13.5 & $2.03\,(2.05)^{+0.04}_{-0.05}$ & $-1.15\,(-2.68)^{+1.61}_{-3.46}$ & $0.46\,(0.59)^{+1.16}_{-1.35}$ & $0.02$\\
1.0 & 12.5 & $1.56\,(1.57)^{+0.02}_{-0.04}$ & $-0.63\,(-2.39)^{+2.75}_{-4.07}$ & $0.89\,(0.54)^{+2.25}_{-2.24}$ & $0.01$\\
1.0 & 13.0 & $2.09\,(2.12)^{+0.03}_{-0.05}$ & $-0.79\,(-4.21)^{+3.27}_{-6.42}$ & $1.76\,(1.72)^{+2.21}_{-2.51}$ & $0.01$\\
1.0 & 13.5 & $3.05\,(3.07)^{+0.06}_{-0.05}$ & $-6.10\,(-8.03)^{+2.09}_{-4.89}$ & $1.78\,(2.15)^{+2.81}_{-3.04}$ & $0.04$\\
\hline\hline
\end{tabular}
\end{table*}

We evaluate the performance of ULPT by jointly fitting the halo--halo auto power spectrum \( P_{\rm hh}(k) \) and the halo--matter cross power spectrum \( P_{\rm hm}(k) \), using one-loop ULPT predictions. Figure~\ref{fig:Phh_Phm} compares the ULPT best-fit results with the \textit{Dark Emulator} outputs across all redshift and halo mass bins. In each panel, the upper sub-panel shows the ratios \( P_{\rm hh}/(b_1^2 P_{\rm nw}) \) and \( P_{\rm hm}/(b_1 P_{\rm nw}) \), where \( P_{\rm nw} \) denotes the no-wiggle linear matter power spectrum. These ratios asymptotically approach unity on large scales. This behavior is consistent with expectations from linear theory. The lower sub-panels display the relative difference between the ULPT predictions and emulator results, defined as
\begin{equation}
    \Delta_{P}\,[\%] = 100\times\left( \frac{P_{\rm ULPT}-P_{\rm Emu}}{P_{\rm Emu}} \right).
    \label{eq:Delta_P}
\end{equation}

The fitting range is set to \( 0.01 \leq k \leq 0.3\, h\,\mathrm{Mpc}^{-1} \) in all cases, except for the highest redshift and mass bin (\( z = 1.0 \), \( \log_{10}(M/M_\odot) = 13.5 \)), for which we adopt a more conservative upper bound of \( k_{\rm max} = 0.2\, h\,\mathrm{Mpc}^{-1} \).

For almost all samples, ULPT reproduces both \( P_{\rm hh}(k) \) and \( P_{\rm hm}(k) \) at the sub-percent level, showing no significant systematic deviations. This result demonstrates that ULPT can simultaneously and accurately describe both auto and cross power spectra over a wide range of halo masses and redshifts, using only four free parameters.

The inferred bias parameters are summarized in Table~\ref{tab:bias_summary}, together with their marginalized \( 1\sigma \) uncertainties and the reduced minimum chi-squared values. Since the \textit{Dark Emulator} predictions for the fiducial cosmology are averaged over multiple realizations, the statistical uncertainties are small. Consequently, the best-fit models typically achieve reduced chi-squared values well below unity. Notably, the best-fit values of \( b_1 \) are in excellent agreement with those independently predicted by the emulator (see Table~\ref{tab:linear_bias_emu}), further validating the robustness of our fitting procedure.

\subsubsection{Hybrid Analysis with Emulator Matter Spectrum at \( z = 0 \)}
\label{sec:hybrid}

\begin{figure*}[!t]
    \centering
    \includegraphics[width=\textwidth]{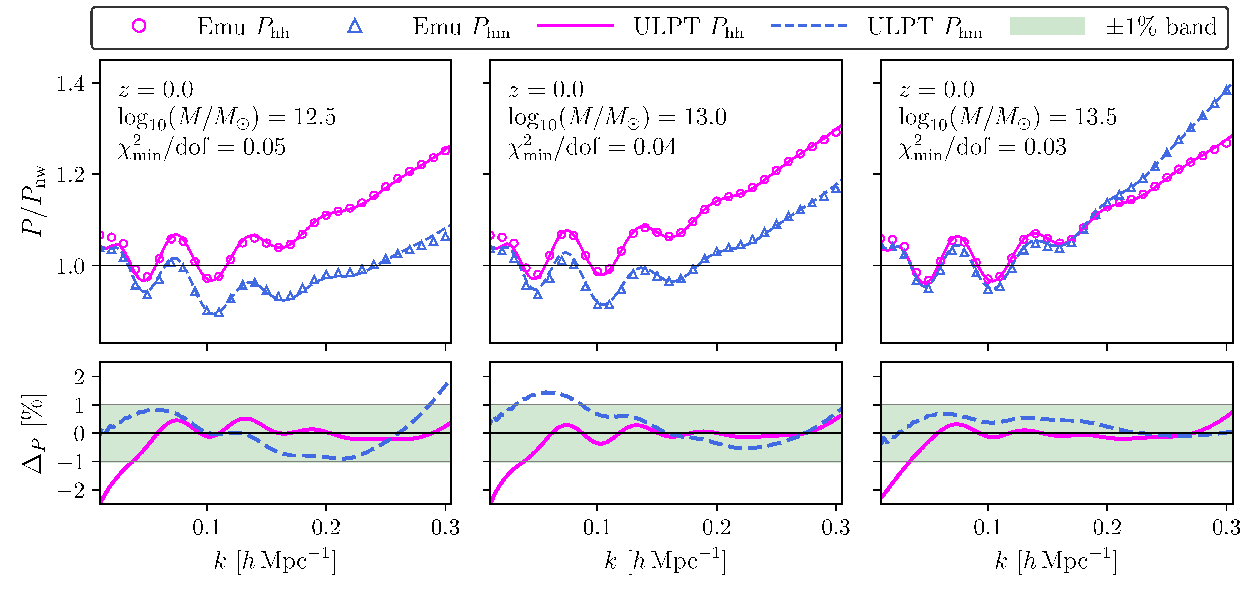}
\caption{
Same as Fig.~\ref{fig:Phh_Phm}, but for the hybrid analysis at \( z = 0 \), in which the matter power spectrum \( P_{\rm m} \) from the emulator is substituted for the ULPT prediction. The residuals for both \( P_{\rm hh}(k) \) and \( P_{\rm hm}(k) \) remain within 1\% over the entire fitting range, with \( P_{\rm hh}(k) \) achieving sub-0.5\% accuracy across most scales. The reduced chi-squared values are significantly improved, indicating that the dominant source of discrepancy between the ULPT and emulator predictions at \( z = 0 \) originates from the nonlinear matter power spectrum.
}
    \label{fig:Phh_Phm_z0}
\end{figure*}

In the analysis presented in Sec.~\ref{sec:fid_pk}, the halo power spectra were computed entirely within the ULPT framework, including both the nonlinear matter power spectrum and the associated bias terms. However, as shown in Ref.~\cite{Sugiyama:2025myq}, the one-loop ULPT prediction for the nonlinear matter power spectrum deviates from the \textit{Dark Emulator}, which is calibrated against high-resolution $N$-body simulations, by approximately 5\% at \( z = 0 \), 3\% at \( z = 0.5 \), and 2\% at \( z = 1 \) up to \( k = 0.4\, h\,\mathrm{Mpc}^{-1} \). These theoretical uncertainties in the matter spectrum can propagate into the halo power spectrum and degrade the accuracy of bias parameter estimation.

At \( z = 0 \), for \( \log_{10}(M/M_\odot) = 13.0 \) and 13.5, the agreement between ULPT and emulator predictions, particularly for \( P_{\rm hm}(k) \), deteriorates to the 2\% level. The corresponding reduced chi-squared values, \( \chi^2_{\rm min}/{\rm dof} = 0.19 \) and 0.34, are substantially higher than those at other redshifts and halo masses, indicating a decline in fit quality.

To isolate the impact of the matter spectrum discrepancy, we repeat the analysis at \( z = 0 \), where the theoretical error is most pronounced. In this test, we retain the ULPT bias expansion but substitute the ULPT matter power spectrum with that from the emulator under the fiducial cosmology. This hybrid approach removes the perturbative uncertainty in \( P_{\rm m} \) while preserving the field-level consistency of the ULPT bias model.

The results of this test are presented in Fig.~\ref{fig:Phh_Phm_z0}. The agreement improves markedly: residuals between ULPT and emulator predictions for both \( P_{\rm hh}(k) \) and \( P_{\rm hm}(k) \) fall within 1\% across the full fitting range. Notably, \( P_{\rm hh}(k) \) achieves sub-0.5\% accuracy over most scales, clearly exceeding the 1\% precision threshold. Correspondingly, the reduced chi-squared values improve to 0.04 and 0.03 for the two previously problematic cases, indicating a substantial enhancement in fit quality. These results strongly suggest that the primary source of error in the standard ULPT prediction at \( z = 0 \) arises from inaccuracies in the nonlinear matter power spectrum rather than the halo bias treatment.

These findings, in turn, reinforce the robustness of the ULPT bias model itself. They confirm that the remaining discrepancies are attributable mainly to limitations in the perturbative matter modeling.

\subsubsection{$\chi^2$ statistics}
\label{sec:chi2_statistics}

We provide a detailed discussion of the interpretation of the minimum reduced chi-squared values, $\chi^2_{\rm min}/{\rm dof}$, presented in Secs.~\ref{sec:fid_pk} and~\ref{sec:hybrid}. Although the precise interpretation depends on the number of degrees of freedom (dof), a reduced chi-squared value close to unity is generally considered to indicate a statistically reasonable fit, whereas significantly larger values suggest that the model is statistically disfavored and may be rejected at a given significance level.

As discussed in Sec.~\ref{sec:cov}, several sources of uncertainty are associated with the covariance matrix adopted in this paper. First, the non-Gaussian contribution to the covariance is neglected. Second, the cross-covariances between $P_{\rm hh}$ and $P_{\rm hm}$ are ignored. Third, the shot noise is modeled by fixing the halo number density to a fiducial value of $\bar{n} = 10^{-4}\,(h^{-1}\mathrm{Mpc})^{-3}$ for all halo samples. Among these, the first two effects, if included, would increase the statistical errors, while the third effect can either increase or decrease the errors depending on the actual number density. These considerations imply that the statistical errors used in this paper are generally underestimated.

Since the $\chi^2$ statistic defined in Eq.~\eqref{eq:chi2_def} is computed using the inverse covariance matrix, underestimated statistical errors directly lead to overestimated $\chi^2$ values. As a result, the $\chi^2$ values quoted in this paper represent more stringent tests of the model than would be the case with the true covariance.

It is also important to recall that, in the fiducial cosmology, the outputs of $P_{\rm hh}$ and $P_{\rm hm}$ from the \textit{Dark Emulator} represent averages over multiple realizations. In such cases, multiplying the computed $\chi^2$ values by the number of realizations used in the averaging allows one to recover the statistically appropriate values. However, since the number of realizations differs between $P_{\rm hh}$ (28) and $P_{\rm hm}$ (14), a well-defined total $\chi^2$ cannot be recovered exactly. As an approximate treatment, we adopt the arithmetic mean of the two realization numbers, $(28+14)/2$, corresponding to $21$ realizations, to obtain a pseudo-restored value of $\chi^2$. \footnote{Strictly speaking, since the cross-covariance between $P_{\rm hh}$ and $P_{\rm hm}$ is ignored and the total $\chi^2$ is defined through Eq.~\eqref{eq:chi2_total_def}, a well-defined total $\chi^2$ could be obtained by multiplying $\chi^2_{\rm hh}$ and $\chi^2_{\rm hm}$ by 28 and 14, respectively. However, the implementation used in this work only outputs the total $\chi^2$, and we therefore adopted the simplified approach described above.}

As an example, consider the fiducial analysis in Sec.~\ref{sec:fid_pk}, which uses the wavenumber range $0.01 \leq k \leq 0.3\,h\, \mathrm{Mpc}^{-1}$ with a bin width of $\Delta k = 0.01\,h\, \mathrm{Mpc}^{-1}$ and four bias parameters. In this case, the number of degrees of freedom is given by ${\rm dof} = 2 \times 30 - 4 = 56$. The restored chi-squared value is then computed as $\chi^2_{\rm restored} = 21 \times 56 \times \left(\chi^2_{\rm min}/{\rm dof}\right)$, where a factor of $21$ accounts for the number of realizations. If $\chi^2_{\rm min}/{\rm dof} \lesssim 0.07$, the corresponding $p$-value exceeds $0.01$, indicating that the fit is not rejected at the 1\% significance level. As shown in Table~\ref{tab:bias_summary}, the vast majority of halo samples satisfy this criterion.

An exception is the case with $z=0.5$ and $\log_{10}(M/M_\odot)=13.5$, where $\chi^2_{\rm min}/{\rm dof}=0.08$. This slight excess is well within the tolerance, given the systematic tendency for our $\chi^2$ values to be overestimated due to the underestimated statistical errors. In contrast, at $z=0.0$, a clearly larger $\chi^2_{\rm min}/{\rm dof}$ is observed, but as discussed in Sec.~\ref{sec:hybrid}, this is not due to a breakdown of the bias model but rather due to the accuracy limits of the theoretical prediction for dark matter, which is addressed separately.

Finally, it is important to emphasize that the $\chi^2_{\rm min}$ values in this work should not be interpreted as absolute statistical standards because of the multiple uncertainties involved in their calculation. Rather, they serve as useful \textit{relative} indicators, for example, when comparing different halo samples or examining the scale dependence of the fits as discussed in Sec.~\ref{sec:kmax}.

\subsubsection{Choice of Fitting Scale \( k_{\rm max} \)}
\label{sec:kmax}

\begin{figure}[!t]
    \centering
    \includegraphics[width=\columnwidth]{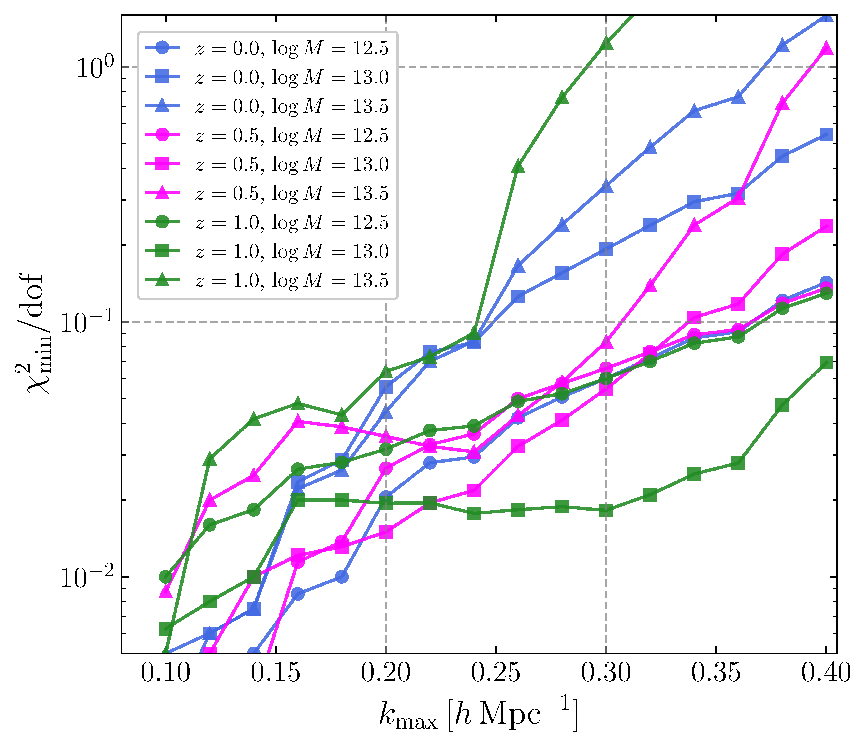}
\caption{
Minimum reduced chi-squared values obtained from joint fits to \( P_{\rm hh}(k) \) and \( P_{\rm hm}(k) \), plotted as a function of the maximum wavenumber \( k_{\rm max} \). Each line corresponds to a different combination of redshift and halo mass. A lower \( \chi^2_{\rm min}/\mathrm{dof} \) indicates better agreement between ULPT predictions and \textit{Dark Emulator} outputs.
}
    \label{fig:chi2}
\end{figure}

To complement the analysis presented in Sec.~\ref{sec:fid_pk}, we provide here a detailed discussion of how the maximum fitting wavenumber \( k_{\rm max} \) was determined.

We perform a series of joint fits to \( P_{\rm hh}(k) \) and \( P_{\rm hm}(k) \) over the range \( 0.01 \leq k \leq k_{\rm max} \), varying \( k_{\rm max} \) from 0.1 to 0.4 \( h\,\mathrm{Mpc}^{-1} \) in steps of 0.02. The resulting minimum reduced chi-squared values, \( \chi^2_{\rm min}/\mathrm{dof} \), are plotted in Fig.~\ref{fig:chi2} as a function of \( k_{\rm max} \).

As shown in the figure, the scale at which ULPT achieves good agreement with the \textit{Dark Emulator} predictions, indicated by a substantially sub-unity value of \( \chi^2_{\rm min}/\mathrm{dof} \), depends on both redshift and halo mass. 

To define a conservative yet broadly applicable benchmark, we adopt \( k_{\rm max} = 0.3\, h\,\mathrm{Mpc}^{-1} \) as the default fitting scale. This choice corresponds to the typical upper limit at which ULPT maintains high accuracy across a wide range of redshifts and halo masses. Indeed, for nearly all samples considered in this work, the reduced chi-squared remains below 0.1 with this cutoff.

There are, however, three notable exceptions: the cases at \( z = 0 \) with \( \log_{10}(M/M_\odot) = 13.0 \) and 13.5, and the case at \( z = 1.0 \) with \( \log_{10}(M/M_\odot) = 13.5 \). In the first two, the reduced chi-squared slightly exceeds 0.1 but remains comfortably below unity. As discussed in Sec.~\ref{sec:hybrid}, these discrepancies are significantly reduced by replacing the ULPT matter power spectrum with the emulator prediction, suggesting that the dominant uncertainty originates from the nonlinear matter spectrum.

The remaining outlier at \( z = 1.0 \), \( \log_{10}(M/M_\odot) = 13.5 \), corresponds to a highly biased halo population with \( b_1 \simeq 3 \), where the larger bias leads to an earlier breakdown of the ULPT prediction. For this case, we adopt a more conservative cutoff of \( k_{\rm max} = 0.2\, h\,\mathrm{Mpc}^{-1} \).

Taken together, these results indicate that ULPT is reliably applicable up to \( k_{\rm max} = 0.3 \, h\,\mathrm{Mpc}^{-1} \) for halo samples with linear bias in the range \( 0.8 \lesssim b_1 \lesssim 2 \). For more strongly biased tracers with \( b_1 \sim 3 \), the valid fitting range is reduced to \( k_{\rm max} \simeq 0.2\, h\,\mathrm{Mpc}^{-1} \).

We emphasize that our choice of \( k_{\rm max} = 0.3 \, h\,\mathrm{Mpc}^{-1} \) is motivated by the goal of identifying a universal scale limit at which ULPT yields accurate predictions across redshifts and halo masses. In practice, the valid range may extend to smaller scales (higher \( k \)) depending on the specific case. Notably, this scale also marks the domain where simultaneous fits to both \( P_{\rm hh}(k) \) and \( P_{\rm hm}(k) \) remain robust. If only \( P_{\rm hh}(k) \) is considered, the fitting accuracy may improve even further.

\subsubsection{Correlation Function Comparison}
\label{sec:fid_cf}

\begin{figure*}[!t]
    \centering
    \includegraphics[width=\textwidth]{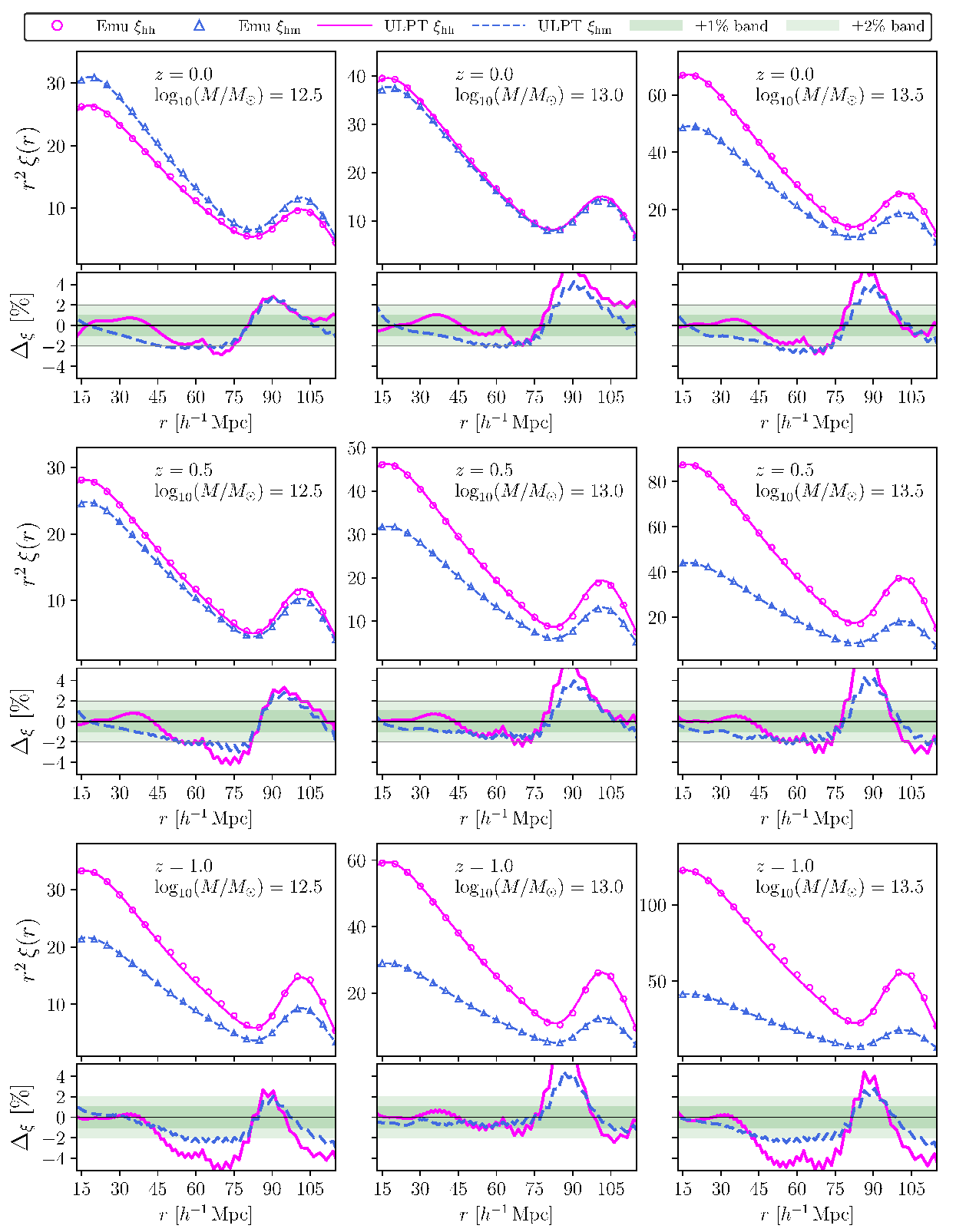}
\caption{
Comparison between the ULPT predictions and \textit{Dark Emulator} outputs for the halo--halo auto correlation function \( \xi_{\rm hh}(r) \) and the halo--matter cross correlation function \( \xi_{\rm hm}(r) \), shown for all redshift and halo mass bins. In each panel, the upper sub-panel displays the correlation functions themselves, while the lower sub-panel shows the relative deviation between the ULPT and emulator predictions, defined as \( \Delta_{\xi}[\%] = 100 \times (\xi_{\rm ULPT} - \xi_{\rm Emu}) / \xi_{\rm Emu} \). Magenta and blue colors denote \( \xi_{\rm hh} \) and \( \xi_{\rm hm} \), respectively; solid and dashed lines indicate the ULPT predictions, while points represent emulator data. ULPT reproduces both correlation functions to within 1\% accuracy over the range \( 15 \leq r \lesssim 45\, h^{-1} \mathrm{Mpc} \), with only modest deviations (2--4\%) at larger separations.
}
    \label{fig:Xihh_Xihm}
\end{figure*}

\begin{figure*}[!t]
    \centering
    \includegraphics[width=\textwidth]{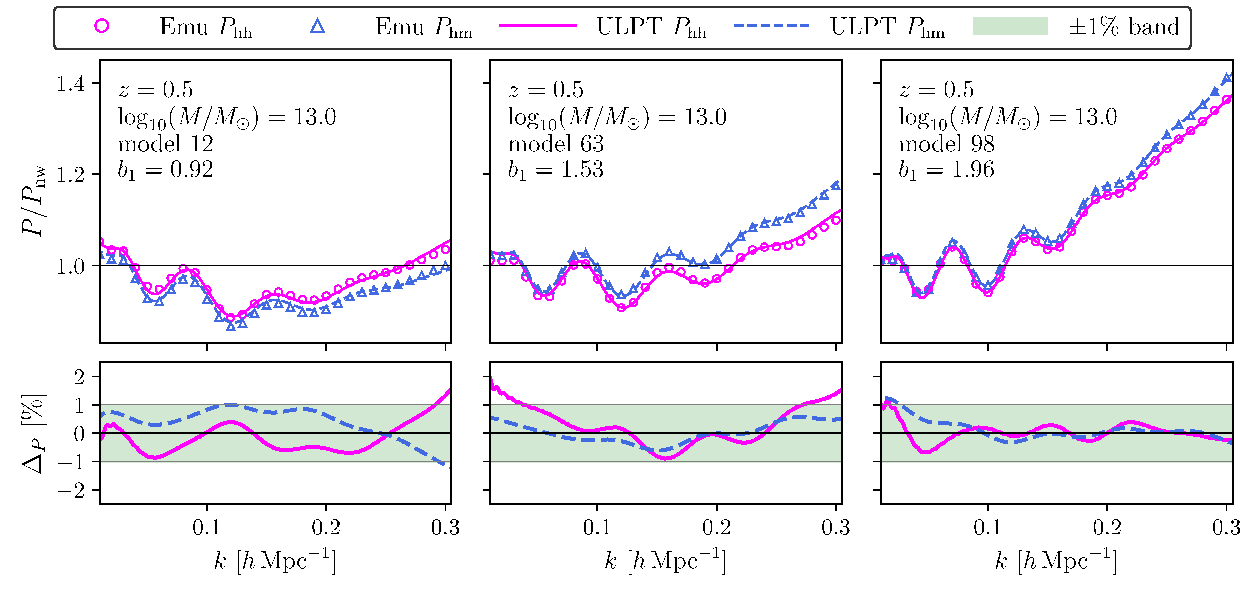}
\caption{
Comparison of the halo--halo power spectrum \( P_{\rm hh}(k) \) and halo--matter power spectrum \( P_{\rm hm}(k) \) for three cosmological models at \( z = 0.5 \) and \( \log_{10}(M/M_\odot) = 13.0 \). These models are selected from the 100-model sample to represent linear bias values of approximately \( b_1 = 1.0 \), 1.5, and 2.0. The corresponding cosmological parameters for models 12, 63, and 98 are provided in Table~\ref{tab:100param}. Although the redshift and halo mass are fixed, the nonlinear structures of both \( P_{\rm hh}(k) \) and \( P_{\rm hm}(k) \) vary significantly due to cosmological dependence. In all three cases, ULPT predictions agree with the emulator outputs to within sub-percent accuracy over the entire fitting range.
}
\label{fig:Phh_Phm_cosmo}
\end{figure*}

We conclude our analysis by examining the consistency between ULPT and emulator predictions in configuration space. Specifically, we perform a joint fit to the halo--halo auto-correlation function \( \xi_{\rm hh}(r) \) and the halo--matter cross-correlation function \( \xi_{\rm hm}(r) \), both computed via inverse Hankel transforms of the corresponding one-loop ULPT power spectra:
\begin{equation}
    \xi_X(r) = \int \frac{dk\, k^2}{2\pi^2} j_0(kr)\, P_X(k),
\end{equation}
where \( X = \{{\rm hh}, {\rm hm}\} \).

Figure~\ref{fig:Xihh_Xihm} presents the comparison for all nine redshift and mass bin combinations, using a common fitting range of \( 15 \leq r \leq 200\, h^{-1} \mathrm{Mpc} \). The lower panel in each subplot shows the relative deviation:
\begin{equation}
    \Delta_{\xi}[\%] = 100 \times \left( \frac{\xi_{\rm ULPT} - \xi_{\rm Emu}}{\xi_{\rm Emu}} \right).
    \label{eq:Delta_xi}
\end{equation}
The choice of \( r_{\rm min} = 15\, h^{-1} \mathrm{Mpc} \) satisfies \( \pi/k_{\rm max} < r_{\rm min} < 2\pi/k_{\rm max} \) for \( k_{\rm max} = 0.3\, h\,\mathrm{Mpc}^{-1} \), matching the small-scale limit used in the power spectrum analysis.

Across nearly the entire fitting range, ULPT successfully reproduces both \( \xi_{\rm hh}(r) \) and \( \xi_{\rm hm}(r) \) within approximately 1--2\% accuracy for all nine redshift and mass bins. In a few cases, localized deviations at the 4\% level are observed around \( r \simeq 90\, h^{-1} \mathrm{Mpc} \) (positive) and \( r \simeq 60\, h^{-1} \mathrm{Mpc} \) (negative). At smaller scales \( (r \lesssim 45\, h^{-1} \mathrm{Mpc}) \), the agreement remains consistently better than 1\% in all cases. Given that statistical uncertainties grow at larger separations, these moderate deviations beyond \( r \gtrsim 60\, h^{-1} \mathrm{Mpc} \) have negligible impact on the overall fit quality.

The best-fit bias parameters from the correlation function analysis are listed in Table~\ref{tab:bias_summary_xi}, along with their marginalized 1$\sigma$ uncertainties. Since the constant stochastic amplitude \( N_\varepsilon \) does not contribute to the correlation function, only three parameters are fitted: \( b_1 \), \( b_2^{\rm u} \), and \( b_3^{\rm u} \). In all nine cases, the reduced chi-squared values are well below unity, indicating excellent fit quality. However, the parameter constraints are generally weaker than those obtained from the power spectrum analysis (Table~\ref{tab:bias_summary}). This difference likely arises from the fact that scale-dependent nonlinear bias terms are more prominent in Fourier space, providing greater sensitivity to the parameters.

Nevertheless, the bias parameters inferred from the correlation functions remain fully consistent with those from the power spectrum analysis within 1$\sigma$ uncertainties. This agreement confirms the internal consistency of the ULPT framework across both Fourier and configuration space statistics.

\subsection{Validation across 100 Cosmologies}
\label{sec:100}

\begin{figure*}[!h]
    \centering
    \includegraphics[width=\textwidth]{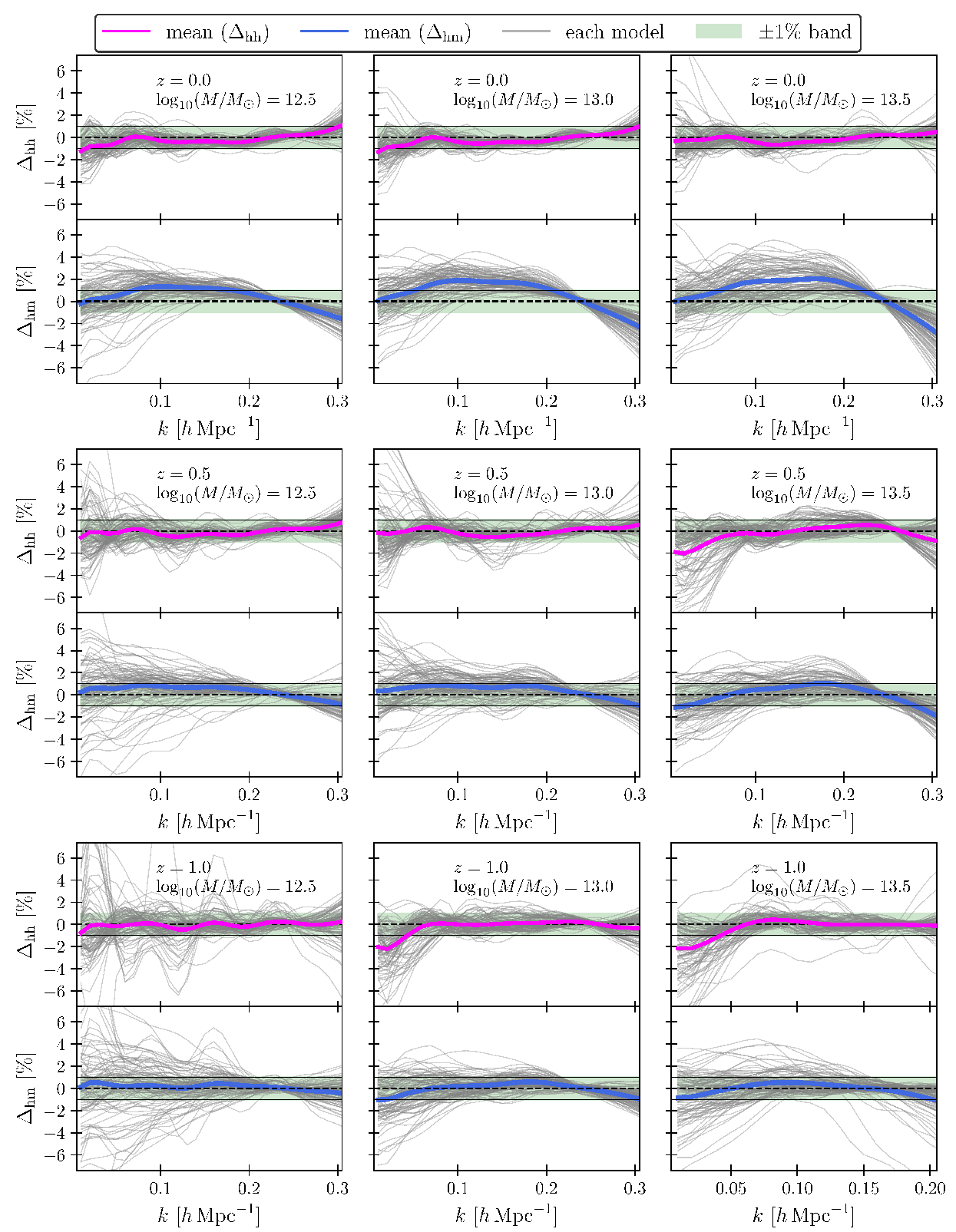}
\caption{
Relative deviation between ULPT and \textit{Dark Emulator} predictions for the halo--halo power spectrum \( P_{\rm hh}(k) \) and the halo--matter power spectrum \( P_{\rm hm}(k) \), evaluated across 100 randomly sampled cosmologies within the emulator's validated parameter range (see Table~\ref{tab:param_ranges}). The relative deviation is defined by Eq.~\eqref{eq:Delta_P} and denoted as \( \Delta_{\rm hh} \) for halo--halo and \( \Delta_{\rm hm} \) for halo--matter. Each gray line corresponds to one cosmological model. The solid magenta and blue lines show the means of \( \Delta_{\rm hh} \) and \( \Delta_{\rm hm} \), respectively, averaged over the 100 models.
}
\label{fig:Phh_Phm_emu}
\end{figure*}

In this subsection, we extend the validation of ULPT by performing joint fits to \( P_{\rm hh}(k) \) and \( P_{\rm hm}(k) \) for 100 cosmological models randomly sampled from within the parameter ranges covered by Dark Emulator (see Table~\ref{tab:param_ranges}). The full set of cosmological parameters used in this analysis is listed in Table~\ref{tab:100param} in Appendix~\ref{ap:100param}.

\subsubsection{Demonstration with Example Cosmologies}

Before presenting the full statistical evaluation across the entire ensemble, we begin by illustrating how variations in cosmological parameters affect the nonlinear halo power spectra. Figure~\ref{fig:Phh_Phm_cosmo} shows the predicted \( P_{\rm hh}(k) \) and \( P_{\rm hm}(k) \) at fixed redshift \( z = 0.5 \) and halo mass \( \log_{10}(M/M_\odot) = 13.0 \), for three representative models selected to span a range of linear bias values: \( b_1 \approx 1.0 \), 1.5, and 2.0. Although the redshift and halo mass are fixed, the nonlinear structures in both spectra differ appreciably, reflecting the sensitivity of halo clustering to the underlying cosmology.

In each of these cases, the ULPT predictions closely match those from the \textit{Dark Emulator}, achieving sub-percent accuracy across the entire fitting range. These results highlight the robustness of the ULPT framework under substantial cosmological variations.

\subsubsection{Statistical Assessment Across 100 Cosmologies}
\label{sec:ensemble_stats}

\begin{figure*}[!t]
    \centering
    \includegraphics[width=\textwidth]{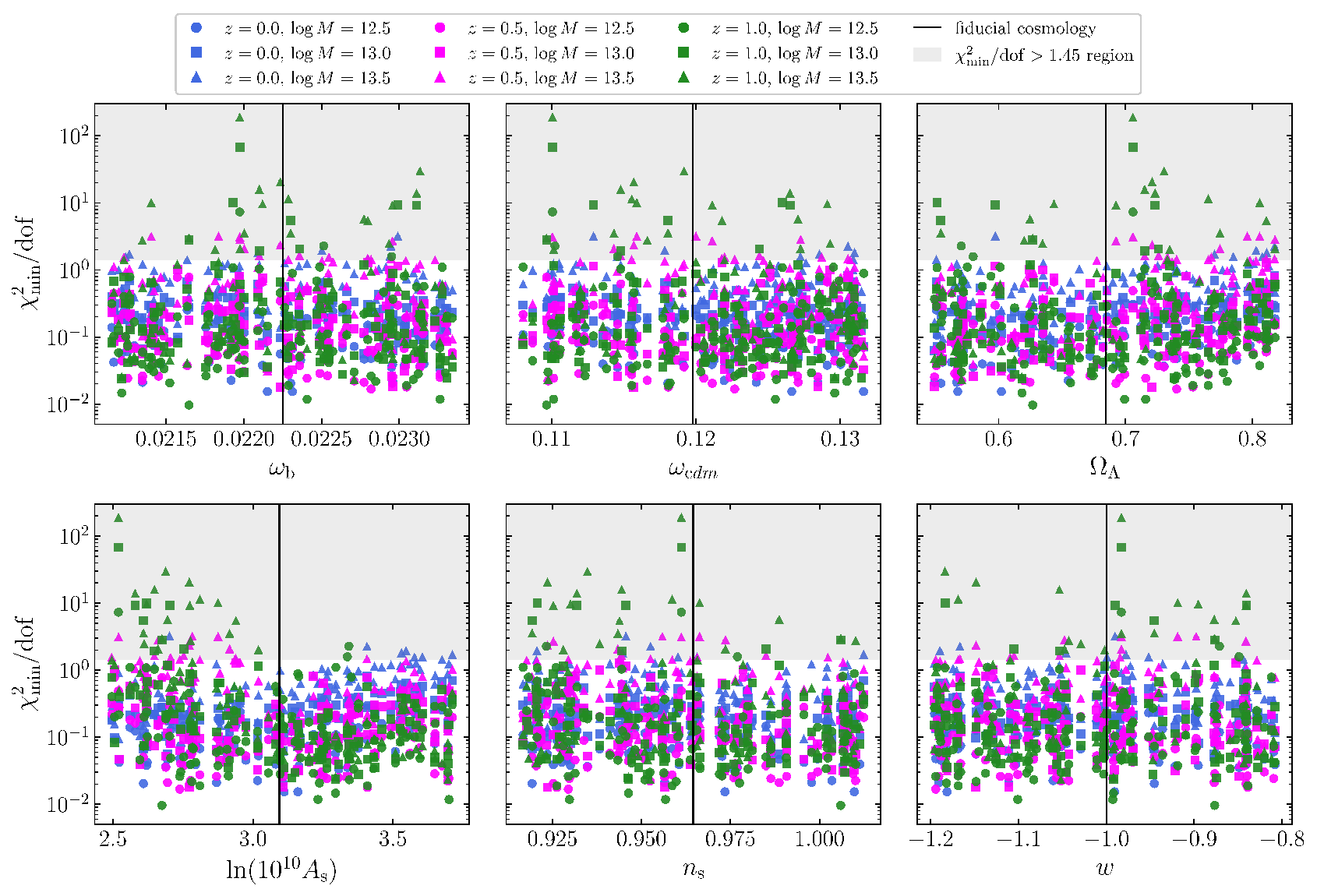}
    \caption{
Dependence of \( \chi^2_{\rm min}/{\rm dof} \) on the six standard cosmological parameters, evaluated across 900 cases.
These 900 realizations correspond to three redshifts (\(z=0,\,0.5,\,1.0\)) and three halo masses (\(\log_{10}M/h^{-1}M_\odot = 12.5,\,13.0,\,13.5\)),
each sampled over 100 randomly selected cosmological parameter sets within the emulator's validated range.
The shaded gray region indicates statistically significant deviations (\( \chi^2_{\rm min}/{\rm dof}>1.45 \)).
Vertical black lines mark the fiducial parameter values.
Colors denote redshifts ($z=0$: blue, $z=0.5$: red, $z=1$: green), and marker shapes correspond to halo masses
(\(\log_{10}M/h^{-1}M_\odot = 12.5\): circles, $13.0$: squares, $13.5$: triangles).
Statistically significant deviations cluster at lower values of \( \ln(10^{10}A_s) \), reflecting enhanced bias in cosmologies with smaller curvature amplitudes.
}
    \label{fig:params_chi2}
\end{figure*}

We now turn to a comprehensive statistical assessment of our ULPT predictions. Figure~\ref{fig:Phh_Phm_emu} shows the relative deviations between ULPT and emulator predictions for both halo--halo and halo--matter power spectra, as defined in Eq.~\eqref{eq:Delta_P}, evaluated across the full set of 100 cosmological models. Each gray line corresponds to one cosmological realization, while the solid magenta and blue lines indicate the ensemble averages of the relative deviations, \( \Delta_{\rm hh} \) for the halo--halo case and \( \Delta_{\rm hm} \) for the halo--matter case, respectively.

Although the relative deviations occasionally exceed \(5\%\) for individual models, typically at specific wavenumbers or within certain redshift--mass bins, the ensemble-averaged deviations remain fully consistent with those obtained for the fiducial cosmology. In almost all cases, ULPT reproduces both \( P_{\rm hh}(k) \) and \( P_{\rm hm}(k) \) with better than \(1\%\) accuracy across the entire fitting range.

\medskip
\noindent
\textbf{Origin of the observed scatter.}
The scatter seen in Fig.~\ref{fig:Phh_Phm_emu} reflects several effects. The first and dominant contribution is the \emph{statistical noise intrinsic to the emulator itself}. For the fiducial cosmology, Dark Emulator predictions are based on averages over multiple $N$-body realizations, whereas for all other cosmological models they rely on single realizations. Consequently, because each cosmological model is effectively evaluated using a different realization, the emulator predictions include not only the genuine physical dependence on cosmological parameters but also statistical fluctuations associated with realization-to-realization variance.

The second contribution arises from the \emph{dependence of the valid fitting range on the halo bias strength}. As discussed in Sec.~\ref{sec:kmax}, the maximum wavenumber up to which the ULPT bias model remains accurate depends on the halo sample properties, primarily characterized by the linear bias amplitude \( b_1 \). More strongly biased halos, corresponding to larger \( b_1 \), tend to reach the breakdown of the ULPT model at larger scales, leading to smaller effective \( k_{\rm max} \). Because the bias strength varies with cosmology, redshift, and halo mass, this effect naturally introduces additional scatter among models. Furthermore, even for fixed cosmological parameters, stochastic fluctuations in individual $N$-body realizations can alter the effective halo bias and the resulting sample properties.

\medskip
\noindent
\textbf{Comparison with the dark matter case.}
Before quantifying this scatter, it is instructive to recall our previous analysis of the dark matter power spectrum~\cite{Sugiyama:2025myq}, in which ULPT predictions were compared with both Dark Emulator~\cite{Nishimichi:2018etk} and Euclid Emulator 2~\cite{Euclid:2020rfv} across 100 cosmologies. In that study, ULPT provides a fully deterministic prediction that depends solely on cosmological parameters, without any nuisance parameters. Therefore, any difference between the two emulator results can be entirely attributed to the statistical fluctuations inherent to each emulator.

The analysis showed that ULPT agrees with both emulators at the \(2\)--\(3\%\) level up to \( k \simeq 0.4\,h\,\mathrm{Mpc}^{-1} \) for \( z \geq 0.5 \) when averaged over 100 cosmologies. In the case of Dark Emulator, however, deviations from the mean for individual cosmologies can reach up to \( \sim 5\% \), which originates from the fact that each nonfiducial cosmology is based on a single $N$-body realization. Repeating the comparison with Euclid Emulator 2, which has smaller statistical uncertainties than Dark Emulator, reduces the scatter to \(1\)--\(2\%\) without changing the mean offset. This clearly demonstrates that the observed dispersion arises from emulator-side statistical noise rather than a breakdown of perturbation theory. These statistical fluctuations at the dark matter level naturally propagate to the halo case, accounting for the majority of the model-by-model scatter seen in Fig.~\ref{fig:Phh_Phm_emu}.

\medskip
\noindent
\textbf{Quantifying emulator noise.}
To assess this effect more quantitatively, we use the minimum \( \chi^2 \) statistic as a diagnostic. For our baseline fits up to \( k_{\rm max}=0.3\,h\,\mathrm{Mpc}^{-1} \) with a bin width of \( \Delta k = 0.01 \) and four free bias parameters, the number of degrees of freedom is ${\rm dof}=56$. For the single case of \( z=1.0 \) and \( \log_{10}(M/M_\odot)=13.5 \), we adopt \( k_{\rm max}=0.2 \), yielding ${\rm dof}=36$. For these degrees of freedom, a $p$-value threshold of $0.01$ corresponds to \( \chi^2_{\rm min}/{\rm dof}\simeq 1.48 \) and \( \simeq 1.61 \), respectively. For simplicity, we adopt \( \chi^2_{\rm min}/{\rm dof}<1.45 \) as a uniform criterion: models exceeding this threshold are classified as showing statistically significant deviations between the ULPT and emulator predictions.

Across three redshift bins and three halo mass bins, we evaluate 100 cosmologies for each combination, yielding a total of 900 cases. Among these, only 50 cases (\( 5.6\% \)) exceed the significance threshold, while the remaining \( 94.4\% \) show no statistically significant deviation. This confirms that the vast majority of the fluctuations seen in Fig.~\ref{fig:Phh_Phm_emu} originate from statistical scatter rather than from any systematic failure of the ULPT framework.

\subsubsection{Dependence on Cosmological Parameters and Bias Strength}
\label{sec:param_dependence}

\begin{figure}[!t]
    \centering
    \includegraphics[width=\columnwidth]{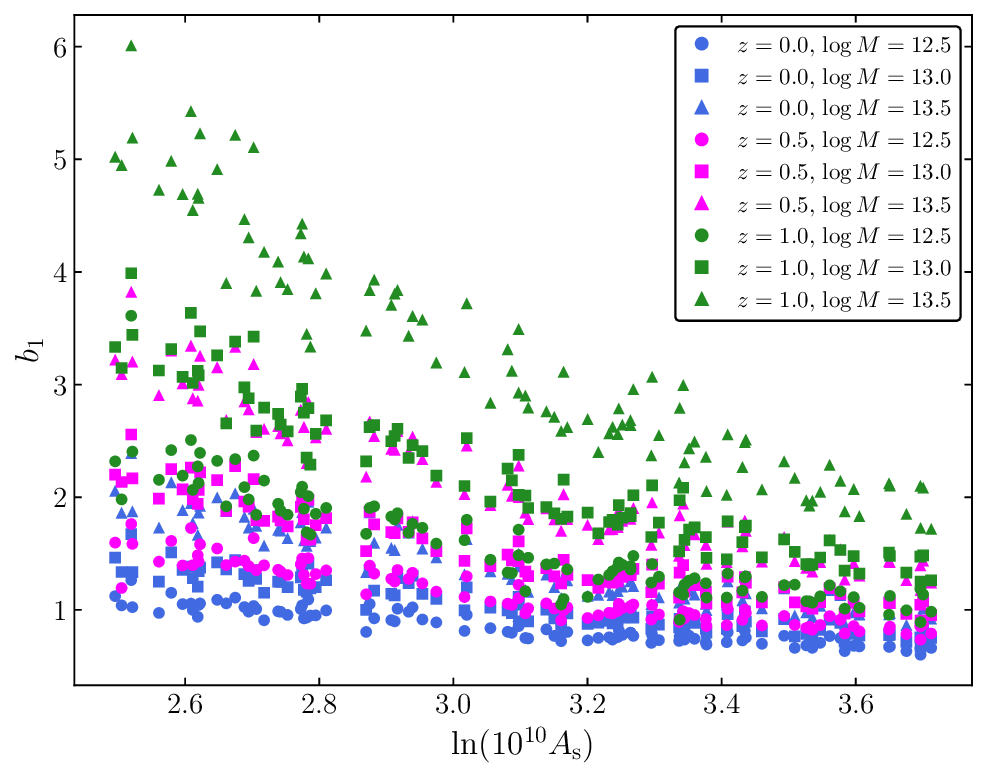}
    \caption{
    Correlation between the amplitude of primordial curvature perturbations \( \ln(10^{10}A_s) \) and the linear bias \( b_1 \) predicted by Dark Emulator. Lower curvature amplitudes correspond to larger \( b_1 \), reaching values up to \( b_1 \simeq 6 \) within the emulator’s parameter coverage.
    }
    \label{fig:As_b1}
\end{figure}

\begin{figure}[!t]
    \centering
    \includegraphics[width=\columnwidth]{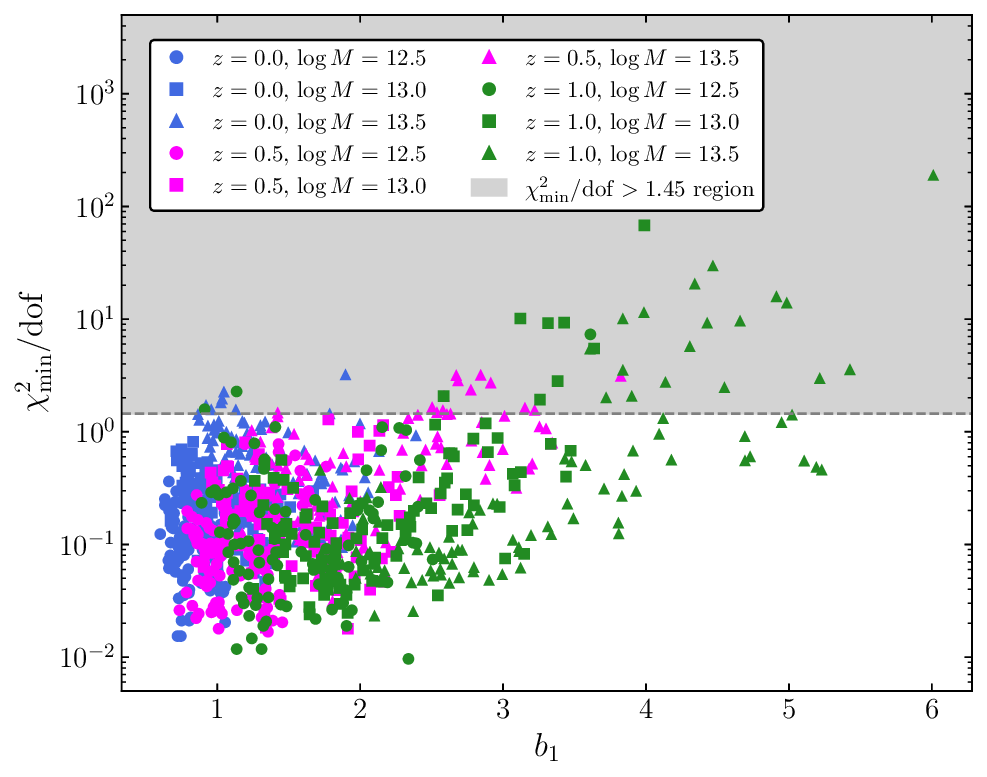}
    \caption{
    Relationship between linear bias \( b_1 \) and \( \chi^2_{\rm min}/{\rm dof} \). Larger \( b_1 \) values correlate with greater discrepancies between ULPT and emulator predictions, particularly for \( b_1 \gtrsim 3 \), where several cases exhibit large deviations (\( \chi^2_{\rm min}/{\rm dof} \gtrsim 10 \)). Nevertheless, some high-bias cases are well fit, indicating that this is a statistical trend rather than an absolute threshold.
    }
    \label{fig:b1_chi2}
\end{figure}

To further investigate the origin of the \(5.6\%\) of cases in which ULPT and Dark Emulator predictions show statistically significant discrepancies, we examine their dependence on cosmological parameters. Figure~\ref{fig:params_chi2} shows the distribution of \( \chi^2_{\rm min}/{\rm dof} \) for all 900 cases as a function of the six standard cosmological parameters. The shaded gray region corresponds to \( \chi^2_{\rm min}/{\rm dof}>1.45 \), indicating statistically significant deviations. Vertical black lines mark the fiducial parameter values. Different colors represent redshifts (\(z=0\): blue, \(z=0.5\): red, \(z=1\): green), while different marker shapes indicate halo masses (\(\log_{10}M/h^{-1}M_\odot = 12.5\): circles, 13.0: squares, 13.5: triangles).

For the baryon density \( \omega_{\rm b} \), cold dark matter density \( \omega_{\rm c} \), dark energy density \( \Omega_\Lambda \), scalar spectral index \( n_{\rm s} \), and dark energy equation-of-state parameter \( w \), the deviations are broadly scattered across parameter space without any noticeable systematic trend relative to the fiducial cosmology. In contrast, for the amplitude of primordial curvature perturbations, \( \ln(10^{10}A_s) \), nearly all statistically significant deviations are concentrated at lower amplitudes than the fiducial value.

This trend can be physically interpreted as follows. Smaller curvature amplitudes correspond to reduced fluctuation amplitudes, making halo formation rarer and thereby increasing the linear bias \( b_1 \). Figure~\ref{fig:As_b1} shows the correlation between \( \ln(10^{10}A_s) \) and the linear bias parameter \( b_1 \) calculated by Dark Emulator. As expected, lower values of \( A_s \) lead to larger \( b_1 \), with some cases reaching \( b_1 \simeq 6 \) within the parameter range covered by the emulator. Because larger bias amplifies nonlinear effects and narrows the validity range of the ULPT bias model, it naturally increases the likelihood of statistically significant deviations.

\medskip
\noindent
\textbf{Impact of linear bias.}
Figure~\ref{fig:b1_chi2} shows the relationship between \( b_1 \) and \( \chi^2_{\rm min}/{\rm dof} \). A clear trend emerges: larger \( b_1 \) systematically correlates with larger discrepancies between ULPT and Dark Emulator predictions. In particular, for \( b_1 \gtrsim 3 \), several cases exhibit \( \chi^2_{\rm min}/{\rm dof} \gtrsim 10 \), indicating a breakdown of the ULPT bias model in this regime. Nonetheless, some high-bias cases are still well described by ULPT, suggesting that this represents a statistical trend rather than an absolute threshold.

Motivated by this observation, we restrict our analysis to cases with \( b_1 \leq 3 \). This selection reduces the total number of samples from 900 to 825. Among these, only 21 cases (\( 2.5\% \)) exceed the statistical threshold for significant deviation, roughly half the fraction of the original sample. We therefore conclude that approximately half of the significant outliers arise from enhanced bias effects associated with low curvature amplitude.

\medskip
\noindent
\textbf{Residual deviations and interpretation.}
The remaining \(2.5\%\) of outliers likely originate from multiple factors. First, as discussed in Sec.~\ref{sec:chi2_statistics}, the covariance matrix adopted in this analysis tends to underestimate uncertainties, leading to systematically inflated \( \chi^2 \) values. Some samples flagged as statistically significant may in fact be statistically consistent when the true covariance is taken into account. Second, as noted in Sec.~\ref{sec:hybrid}, at \( z=0 \) the ULPT prediction for the matter power spectrum itself is less accurate, increasing the mismatch with the emulator. This issue can be mitigated by employing a hybrid approach that combines the ULPT bias formulation with emulator outputs for the matter power spectrum.

Future improvements to ULPT and its associated analysis framework, together with emulators that exhibit lower statistical noise in the halo power spectrum, should enable a more complete understanding of these residual deviations. Importantly, the present analysis demonstrates that most discrepancies across cosmologies do not indicate a breakdown of ULPT but rather reflect the statistical limitations of emulator-based predictions. Furthermore, roughly half of the statistically significant deviations arise from enhanced bias in cosmologies with lower curvature amplitudes. Conversely, when $\ln(10^{10} A_s)$ is close to its fiducial Planck 2015 value, ULPT maintains its predictive accuracy across parameter space.

\medskip

\noindent
In summary, being free from statistical noise and inherently stable across cosmological models, ULPT offers a particularly robust framework for MCMC-based parameter inference, which requires repeated evaluations over a wide parameter range. This highlights one of its key advantages for precision cosmological analysis.

\section{Comparison with Empirical Fitting Formulas for Eulerian Bias}
\label{sec:compare_empirical}

In this section, we compare the bias parameters \( (b_1,\, b_2^{\rm u},\, b_3^{\rm u}) \), obtained from the joint analysis of the halo--halo auto and halo--matter cross power spectra within the ULPT framework, with empirical fitting formulas for Eulerian bias coefficients.

As discussed in Sec.~\ref{sec:relation_standard}, the second-order Eulerian bias parameters, namely the local bias \( b^{\rm E}_2 \) and the tidal bias \( b^{\rm E}_{K^2} \), originate from a single Lagrangian operator characterized by the coefficient \( b_2^{\rm u} \). Their explicit relations, given by \( b_2^{\rm E} = -\tfrac{8}{21} b_2^{\rm u} \) and \( b_{K^2}^{\rm E} = \tfrac{2}{7} b_2^{\rm u} \) in Eq.~\eqref{eq:b2E}, lead to the theoretical consistency condition \( b_{K^2}^{\rm E} = -\tfrac{3}{4} b_2^{\rm E} \) shown in Eq.~\eqref{eq:b2Erelation}. This provides a concrete and testable prediction that can be directly compared against simulation-based fitting results.

Lazeyras et al.~\cite{Lazeyras:2015lgp} proposed fitting formulas for the second- and third-order local Eulerian bias parameters using response function techniques and separate universe simulations. These methods probe the nonlinear response of halo abundance to long-wavelength density perturbations. The fitting formula for the second-order local bias is expressed as a function of the linear bias \( b_1 \):
\begin{equation}
    b_2^{\rm E}(b_1) = 0.412 - 2.143\, b_1 + 0.929\, b_1^2 + 0.008\, b_1^3.
\end{equation}

Although a fitting formula for the third-order local Eulerian bias parameter \( b_3^{\rm E} \) also exists, we do not include it in this comparison. As shown in Eq.~\eqref{eq:b3E}, the ULPT expression for \( b_3^{\rm E} \) depends on both \( b_{3,{\rm U}}^{\rm u} \) and \( b_{3,{\rm V}}^{\rm u} \), while only the former contributes to the one-loop power spectrum. The latter remains unconstrained, making a meaningful comparison infeasible.

Modi et al.~\cite{Modi:2016dah} provided a fitting formula for the tidal bias \( b^{\rm E}_{K^2} \), derived from three independent estimators: Fourier-space cross-correlations, real-space measurements of the density PDF, and the peak-background split applied to small-volume simulations. Their estimator isolates the tidal component using orthogonal polynomial decomposition:
\begin{equation}
    b_{K^2}^{\rm E}(b_1) = 1.03 - 0.615\, b_1 + 0.188\, b_1^2 - 0.072\, b_1^3.
\end{equation}

A notable feature of both studies is that the bias parameters were directly measured from $N$-body simulations without renormalization or theoretical priors. As a result, these can be directly compared to the bare bias parameters predicted by ULPT.

\begin{figure}[!t]
    \centering
    \includegraphics[width=\columnwidth]{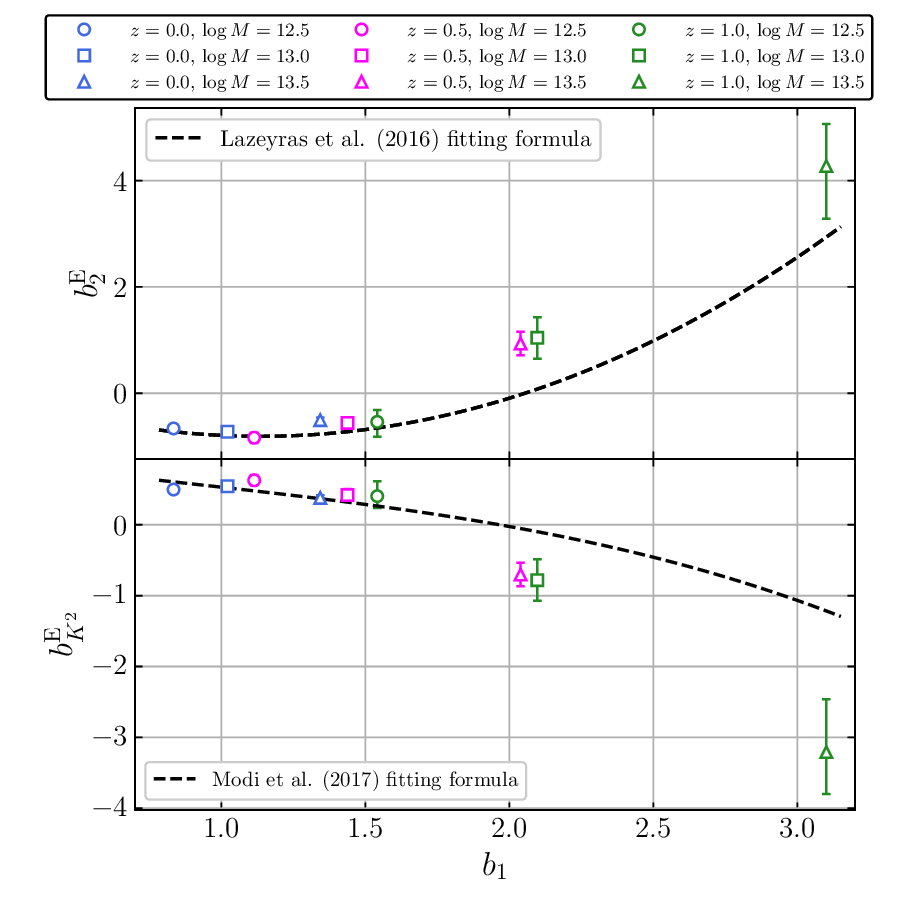}
\caption{
    Comparison between the Eulerian bias parameters derived from ULPT and the empirical fitting formulas as functions of the linear bias \( b_1 \). The top panel shows the second-order local bias \( b^{\rm E}_2 \), and the bottom panel shows the tidal bias \( b^{\rm E}_{K^2} \). Points represent ULPT predictions based on the fiducial cosmology using Eq.~\eqref{eq:b2E}, and black dashed lines indicate the empirical formulas. The ULPT results match the fitting relations in the low-bias regime ($b_1\lesssim 1.5$) and accurately capture key nonlinear features: both \( b^{\rm E}_2 \) and \( b^{\rm E}_{K^2} \) undergo sign reversals near \( b_1 \sim 2.0 \), and their magnitudes increase rapidly with increasing \( b_1 \), with \( b^{\rm E}_2 \) growing positively and \( b^{\rm E}_{K^2} \) becoming increasingly negative. This agreement highlights the predictive power and internal consistency of the ULPT framework.
}
    \label{fig:fitting_bias}
\end{figure}

Table~\ref{tab:bias_summary} summarizes the ULPT results for \( b_1 \), \( b_2^{\rm u} \), and \( b_3^{\rm u} \) obtained from fits to the fiducial cosmology. Figure~\ref{fig:fitting_bias} compares the corresponding Eulerian bias parameters, computed using Eq.~\eqref{eq:b2E}, with the empirical formulas from Refs.~\cite{Lazeyras:2015lgp, Modi:2016dah}. The ULPT results are plotted as data points, while the empirical formulas are shown as black dashed curves. All values of \( b_1 \) correspond to the posterior means from the fits.

The top panel of Fig.~\ref{fig:fitting_bias} shows the relation between \( b^{\rm E}_2 \) and \( b_1 \). The ULPT predictions exhibit excellent agreement with the fitting formula for \( b_1 \lesssim 1.5 \), and successfully reproduce the sign change near \( b_1 \sim 2.0 \), as well as the steep growth in the high-bias regime, consistent with the nonlinear behavior encoded in the empirical fit.

The bottom panel displays the relation between \( b^{\rm E}_{K^2} \) and \( b_1 \). Again, the ULPT results are in good agreement with the empirical fit for \( b_1 \lesssim 1.5 \), with deviations gradually increasing at higher values. The steep decline and eventual sign reversal near \( b_1 \sim 2.0 \) are also well captured.

Together, these results demonstrate that the ULPT predictions simultaneously reproduce the behaviors of both \( b^{\rm E}_2 \) and \( b^{\rm E}_{K^2} \), while satisfying the theoretical relation in Eq.~\eqref{eq:b2Erelation}. This consistency provides strong evidence for the internal coherence of the ULPT framework. The agreement with independently derived simulation-based bias parameters further supports the validity and robustness of the bias modeling approach implemented in ULPT.

\section{Conclusion}
\label{sec:conclusion}

In this work, we have developed a renormalization-free framework for modeling galaxy bias based on Unified Lagrangian Perturbation Theory (ULPT). In this formulation, the bias field is constructed entirely from Galileon-type operators that also characterize the intrinsic nonlinear structure of the dark matter field. As a result, the bias expansion is well defined at the field level, inherently satisfies the statistical conditions of vanishing ensemble and volume averages, and eliminates the need for any ad hoc renormalization procedures. Consequently, all bias parameters can be interpreted as physically meaningful quantities.

Within this framework, we derived analytic expressions for the one-loop galaxy--galaxy and galaxy--matter power spectra, incorporating nonlinear bias effects in a fully renormalization-free manner. We implemented an efficient numerical algorithm based on \texttt{FAST-PT} and \texttt{FFTLog}, which enables rapid and accurate evaluation of the full power spectrum.

A key feature of our approach is that it constitutes a minimal bias model defined at the field level. For power spectrum analyses, only four parameters are required: the linear bias $b_1$, two nonlinear Lagrangian bias coefficients $b^{\rm u}_2$ and $b^{\rm u}_3$, and the stochastic amplitude $N_{\varepsilon}$. For correlation functions, where the stochastic contribution does not affect the signal, the number of parameters reduces to three. Because the model is constructed directly at the field level, it is naturally compatible with a wide range of statistical observables. As a representative test case, we examined whether the model can simultaneously describe the halo--halo auto and halo--matter cross power spectra, $P_{\mathrm{hh}}$ and $P_{\mathrm{hm}}$, using a common set of bias parameters.

To validate the model's predictive performance, we first examined the fiducial cosmology in the \textit{Dark Emulator}, whose outputs are averaged over multiple $N$-body realizations and thus have reduced statistical noise. ULPT predictions were found to reproduce both the halo--halo and halo--matter power spectra, \( P_{\mathrm{hh}}(k) \) and \( P_{\mathrm{hm}}(k) \), with better than 1\% accuracy up to \( k \simeq 0.3\,h\,\mathrm{Mpc}^{-1} \) for tracers with typical linear bias values (\( b_1 \sim 0.8\!-\!2 \)), and within 1\% up to \( k \simeq 0.2\,h\,\mathrm{Mpc}^{-1} \) for more strongly biased halos (\( b_1 \sim 3 \)). In configuration space, the same bias parameters consistently reproduce the two-point correlation functions, \( \xi_{\mathrm{hh}}(r) \) and \( \xi_{\mathrm{hm}}(r) \), down to \( r \simeq 15\,h^{-1}\mathrm{Mpc} \), demonstrating the model's internal consistency across Fourier and real-space statistics.

We then extended this validation to a broader parameter space by performing a statistical assessment across 100 cosmologies sampled from the six-dimensional parameter range of the emulator. In contrast to the fiducial case, each cosmological model is based on a single $N$-body realization and therefore carries residual statistical fluctuations from the underlying simulations. Across three redshifts and three halo mass bins, we jointly compared ULPT predictions for \( P_{\mathrm{hh}} \) and \( P_{\mathrm{hm}} \) with emulator results for a total of 900 cases. Over 94\% of all cases show no statistically significant deviation, confirming that the small scatter among individual cosmologies primarily reflects emulator-side statistical noise rather than a limitation of ULPT itself. Roughly half of the remaining significant outliers are associated with cosmologies having lower primordial curvature amplitudes, where enhanced halo bias amplifies nonlinear effects and reduces the valid scale range of the ULPT model. This large-scale stability highlights a key advantage of ULPT: being free from statistical noise, it provides a robust and stable framework for MCMC-based cosmological parameter inference across a wide parameter space, demonstrating its strong potential for precision cosmological analysis.

We also showed that the bias parameters inferred from ULPT naturally satisfy theoretical relations between second-order Eulerian bias coefficients. In particular, ULPT predicts that the local quadratic bias $b_2^{\mathrm{E}}$ and the tidal bias $b_{K^2}^{\mathrm{E}}$ are not independent but instead obey the relation $b_{K^2}^{\mathrm{E}} = -\tfrac{3}{4} b_2^{\mathrm{E}}$, as a direct consequence of the underlying Galileon operator structure. This relation arises from the fact that both coefficients originate from a single second-order Lagrangian operator $b_2^{\rm u}$ in the ULPT expansion. The predicted values of $b_2^{\mathrm{E}}$ and $b_{K^2}^{\mathrm{E}}$ obtained via this mapping are in good agreement with empirical fitting formulas calibrated on $N$-body simulations. These findings further support the physical interpretability and predictive reliability of the renormalization-free bias model.

Looking ahead, the field-level nature of ULPT provides a promising foundation for future developments. The framework can be systematically extended to incorporate additional observables such as the bispectrum, redshift-space distortions, and the effects of reconstruction. Moreover, because the formalism is not tied to any specific emulator, it can be generalized beyond the parameter space covered by current simulation-based models. Potential applications include cosmologies with dynamical dark energy, primordial non-Gaussianity, modified gravity theories, and other extensions of the standard model. These directions will be explored in future work.

\begin{acknowledgments}
N.S. acknowledges financial support from JSPS KAKENHI Grant No.~25K07343, administratively hosted by the National Astronomical Observatory of Japan. N.S. is grateful to the developers of publicly available Python packages used in this work, including \texttt{mcfit}~\cite{mcfit} for Hankel transforms and \texttt{FAST-PT}~\cite{McEwen:2016fjn,Fang:2016wcf} for efficient evaluation of one-loop corrections. The nonlinear reference spectra were obtained from \textit{Dark Emulator}~\cite{Nishimichi:2018etk}, and parameter estimation and statistical analyses were performed using \textit{MontePython}~\cite{Brinckmann:2018cvx}. N.S. also acknowledges the use of \textit{ChatGPT} (OpenAI) for assistance in language refinement, literature exploration, and Python scripting during the preparation of this manuscript. The author thanks the referee for their constructive comments, which have significantly improved the quality and clarity of the paper.
\end{acknowledgments}

\appendix

\section{List of 100 Randomly Selected Cosmological Models}
\label{ap:100param}

Table~\ref{tab:100param} summarizes the set of 100 cosmological models randomly selected from within the parameter space covered by \textit{Dark Emulator}, which were used throughout the main analysis.

\begin{table*}[t]
\centering
\footnotesize
\renewcommand{\arraystretch}{0.95}
\caption{100 rondomly selected cosmological parameters for wCDM models. Left: 01--50, Right: 51--100.}
\label{tab:100param}
\begin{tabular}{c|cccccc|c|cccccc}
\hline
Model & $\omega_b$ & $\omega_{\rm cdm}$ & $\Omega_\Lambda$ & $\ln(10^{10}A_{\rm s})$ & $n_s$ & $w$ &
Model & $\omega_b$ & $\omega_{\rm cdm}$ & $\Omega_\Lambda$ & $\ln(10^{10}A_{\rm s})$ & $n_s$ & $w$ \\
\hline
01 & 0.021971 & 0.130599 & 0.747911 & 3.216100 & 0.931323 & -1.137602 & 51 & 0.021252 & 0.120551 & 0.695524 & 3.264083 & 0.986307 & -0.809659 \\
02 & 0.021267 & 0.128574 & 0.712081 & 3.351511 & 0.918260 & -0.812036 & 52 & 0.022286 & 0.115558 & 0.765210 & 2.810382 & 0.958614 & -1.168617 \\
03 & 0.022990 & 0.112908 & 0.597296 & 2.702181 & 0.945619 & -0.990097 & 53 & 0.021194 & 0.130885 & 0.776378 & 3.336538 & 0.955719 & -1.130682 \\
04 & 0.022099 & 0.114798 & 0.715021 & 2.647838 & 0.944452 & -1.053455 & 54 & 0.021486 & 0.113816 & 0.697876 & 3.359584 & 0.979951 & -1.088026 \\
05 & 0.022152 & 0.126633 & 0.602183 & 3.111617 & 0.973413 & -1.181420 & 55 & 0.023262 & 0.125500 & 0.699280 & 3.232266 & 0.956745 & -1.100908 \\
06 & 0.022489 & 0.111906 & 0.565329 & 3.649541 & 1.009410 & -0.876641 & 56 & 0.021930 & 0.125978 & 0.551460 & 2.618851 & 0.920712 & -1.183708 \\
07 & 0.021815 & 0.110160 & 0.734836 & 3.019933 & 0.928046 & -1.001929 & 57 & 0.023041 & 0.124680 & 0.677330 & 2.596280 & 0.963691 & -1.010611 \\
08 & 0.021214 & 0.129607 & 0.618364 & 3.295138 & 0.946340 & -0.991973 & 58 & 0.021523 & 0.118215 & 0.656615 & 3.237376 & 0.977530 & -1.181878 \\
09 & 0.022354 & 0.112249 & 0.812953 & 3.434504 & 1.006890 & -0.842069 & 59 & 0.021971 & 0.122816 & 0.685259 & 3.535192 & 0.979806 & -1.134826 \\
10 & 0.022468 & 0.129908 & 0.571746 & 2.717748 & 0.920637 & -1.069868 & 60 & 0.021295 & 0.123212 & 0.554778 & 3.200156 & 1.006960 & -0.969810 \\
11 & 0.022002 & 0.114322 & 0.774395 & 2.916718 & 0.943371 & -0.982922 & 61 & 0.022001 & 0.123233 & 0.672971 & 3.150455 & 1.007079 & -1.045559 \\
12 & 0.021451 & 0.127041 & 0.567929 & 3.696571 & 0.990758 & -1.120514 & 62 & 0.023276 & 0.129512 & 0.601120 & 2.561042 & 0.925995 & -1.192711 \\
13 & 0.021150 & 0.127358 & 0.741029 & 3.377419 & 0.990664 & -1.170382 & 63 & 0.021348 & 0.124185 & 0.567009 & 2.869964 & 0.997763 & -1.190691 \\
14 & 0.021935 & 0.110596 & 0.783803 & 3.246594 & 0.948190 & -1.174577 & 64 & 0.022950 & 0.114573 & 0.579869 & 3.337482 & 0.976937 & -0.849011 \\
15 & 0.021829 & 0.115611 & 0.747257 & 3.264241 & 1.001847 & -1.011114 & 65 & 0.022773 & 0.127071 & 0.624730 & 2.694799 & 0.988672 & -0.877266 \\
16 & 0.021404 & 0.124909 & 0.755793 & 3.169837 & 0.990635 & -1.002482 & 66 & 0.023341 & 0.117706 & 0.649364 & 3.436089 & 0.949146 & -0.827697 \\
17 & 0.022301 & 0.118064 & 0.554479 & 2.608726 & 0.919306 & -0.945436 & 67 & 0.023047 & 0.118099 & 0.753078 & 3.409022 & 0.926221 & -0.838979 \\
18 & 0.021837 & 0.120005 & 0.795975 & 2.783724 & 0.955856 & -0.897780 & 68 & 0.022262 & 0.127622 & 0.635137 & 3.583500 & 0.953814 & -1.195665 \\
19 & 0.021647 & 0.109664 & 0.626842 & 2.674727 & 1.005944 & -0.876752 & 69 & 0.023152 & 0.110007 & 0.634935 & 3.650997 & 1.007961 & -0.970625 \\
20 & 0.022547 & 0.128700 & 0.767533 & 2.706099 & 1.002362 & -0.984263 & 70 & 0.022543 & 0.118565 & 0.627789 & 2.881955 & 0.981139 & -0.899050 \\
21 & 0.022934 & 0.129290 & 0.634577 & 2.611400 & 0.938259 & -1.029157 & 71 & 0.022899 & 0.126739 & 0.572489 & 3.087095 & 0.921827 & -0.980188 \\
22 & 0.022958 & 0.128443 & 0.549423 & 3.107301 & 0.956534 & -1.111157 & 72 & 0.022120 & 0.129089 & 0.643586 & 2.620082 & 0.930067 & -0.895396 \\
23 & 0.021404 & 0.115909 & 0.805651 & 2.875196 & 0.966312 & -0.918792 & 73 & 0.022513 & 0.110243 & 0.570545 & 3.342719 & 0.923293 & -0.871256 \\
24 & 0.021947 & 0.131104 & 0.811000 & 2.786806 & 0.964235 & -1.079649 & 74 & 0.022709 & 0.109769 & 0.570745 & 3.696265 & 0.952373 & -1.051743 \\
25 & 0.021771 & 0.108704 & 0.714394 & 3.097316 & 0.921240 & -1.088541 & 75 & 0.022946 & 0.130516 & 0.817448 & 3.407581 & 0.952565 & -1.166600 \\
26 & 0.023158 & 0.113560 & 0.587186 & 3.080947 & 1.011341 & -1.103178 & 76 & 0.022867 & 0.121199 & 0.663655 & 3.596904 & 0.927000 & -1.002950 \\
27 & 0.022633 & 0.126068 & 0.612576 & 3.376441 & 0.951748 & -0.947078 & 77 & 0.021163 & 0.119049 & 0.562934 & 2.622249 & 0.927610 & -0.940316 \\
28 & 0.022547 & 0.120657 & 0.572238 & 3.508970 & 0.947214 & -1.125393 & 78 & 0.022797 & 0.121798 & 0.810924 & 2.939140 & 0.943832 & -0.852560 \\
29 & 0.021228 & 0.121978 & 0.733010 & 2.495729 & 0.965666 & -1.109402 & 79 & 0.021635 & 0.130899 & 0.550847 & 3.675522 & 0.920438 & -0.843543 \\
30 & 0.022573 & 0.111998 & 0.736671 & 2.953824 & 1.006623 & -1.144992 & 80 & 0.022312 & 0.131611 & 0.567723 & 3.160650 & 1.009764 & -0.990761 \\
31 & 0.021896 & 0.110539 & 0.800664 & 3.560995 & 0.941153 & -0.936006 & 81 & 0.022538 & 0.124490 & 0.671955 & 3.251866 & 0.972632 & -0.839537 \\
32 & 0.022956 & 0.121123 & 0.692517 & 2.774516 & 0.925255 & -0.841114 & 82 & 0.021239 & 0.114552 & 0.807705 & 3.576990 & 0.960223 & -0.951947 \\
33 & 0.023141 & 0.122989 & 0.640333 & 2.907382 & 0.986293 & -0.841156 & 83 & 0.021755 & 0.112327 & 0.674462 & 2.912509 & 0.972569 & -1.168906 \\
34 & 0.023111 & 0.126506 & 0.723283 & 2.579332 & 0.931864 & -0.840578 & 84 & 0.023306 & 0.131450 & 0.738649 & 3.138673 & 0.946129 & -0.874482 \\
35 & 0.022487 & 0.108040 & 0.575299 & 3.296350 & 0.916763 & -1.135677 & 85 & 0.022661 & 0.111716 & 0.796895 & 3.493172 & 1.007883 & -0.909712 \\
36 & 0.022358 & 0.124398 & 0.726001 & 2.752756 & 0.984965 & -1.105100 & 86 & 0.022502 & 0.117841 & 0.802864 & 3.547041 & 0.920636 & -1.189453 \\
37 & 0.021862 & 0.125706 & 0.725364 & 3.526199 & 0.979702 & -0.972677 & 87 & 0.021975 & 0.127241 & 0.817797 & 2.661356 & 0.973579 & -1.047644 \\
38 & 0.021346 & 0.116630 & 0.620122 & 2.777162 & 1.010122 & -1.042761 & 88 & 0.023296 & 0.127997 & 0.777021 & 3.055255 & 0.956284 & -1.090637 \\
39 & 0.023122 & 0.122942 & 0.765108 & 3.097264 & 0.971917 & -1.002993 & 89 & 0.021263 & 0.128539 & 0.770060 & 3.712451 & 1.012401 & -0.977827 \\
40 & 0.021572 & 0.125130 & 0.624384 & 2.505293 & 0.978531 & -1.129156 & 90 & 0.022848 & 0.130457 & 0.780119 & 2.781318 & 0.959730 & -1.148336 \\
41 & 0.023230 & 0.130676 & 0.797973 & 2.933308 & 0.917766 & -0.828673 & 91 & 0.023260 & 0.122344 & 0.610113 & 3.306497 & 0.975893 & -1.056735 \\
42 & 0.022090 & 0.130981 & 0.811321 & 3.530885 & 0.944675 & -1.045961 & 92 & 0.021390 & 0.123911 & 0.689959 & 3.431021 & 0.966445 & -0.859127 \\
43 & 0.023031 & 0.115413 & 0.593920 & 3.164297 & 1.006567 & -0.921588 & 93 & 0.022365 & 0.121260 & 0.787513 & 2.974550 & 0.929201 & -1.188487 \\
44 & 0.022406 & 0.110148 & 0.715884 & 3.700491 & 0.929786 & -0.992668 & 94 & 0.022818 & 0.122683 & 0.740269 & 2.738764 & 0.929428 & -1.194182 \\
45 & 0.023090 & 0.125569 & 0.738335 & 3.344594 & 0.950948 & -1.082563 & 95 & 0.021918 & 0.121954 & 0.654901 & 3.016619 & 1.003481 & -1.060698 \\
46 & 0.022938 & 0.127230 & 0.784890 & 3.605427 & 0.965594 & -0.999393 & 96 & 0.022281 & 0.126596 & 0.656078 & 3.245095 & 0.999450 & -0.820192 \\
47 & 0.022914 & 0.123393 & 0.739690 & 3.460073 & 1.002116 & -1.064802 & 97 & 0.021465 & 0.130021 & 0.682242 & 2.794803 & 0.960559 & -0.807987 \\
48 & 0.021973 & 0.110072 & 0.705830 & 2.519682 & 0.961182 & -0.982942 & 98 & 0.022234 & 0.115697 & 0.720920 & 2.772404 & 0.923592 & -1.148448 \\
49 & 0.021775 & 0.121976 & 0.555870 & 2.521422 & 0.995615 & -1.055924 & 99 & 0.021422 & 0.111460 & 0.585525 & 3.268347 & 0.933817 & -1.061733 \\
50 & 0.021420 & 0.120333 & 0.758313 & 2.742300 & 0.976353 & -1.165861 & 100 & 0.023133 & 0.119176 & 0.730271 & 2.688463 & 0.934821 & -1.183653 \\
\hline
\end{tabular}
\renewcommand{\arraystretch}{1.0}
\end{table*}

\bibliography{ms}

\end{document}